\documentclass[twocolumn,showpacs,preprintnumbers,amsmath,amssymb]{revtex4}
\usepackage{graphicx}
\begin{document}
\title{Intra- and inter-shell  Kondo effects in carbon nanotube quantum dots}
\author{D. Krychowski and S. Lipi\'{n}ski}
\affiliation{%
Institute of Molecular Physics, Polish Academy of Sciences,\\M. Smoluchowskiego 17,
60-179 Pozna\'{n}, Poland
}%
\date{\today}
\begin{abstract}
The linear response transport properties of carbon nanotube quantum dot in the strongly correlated regime are discussed. The finite-U mean field slave boson approach is used to study many-body effects. Magnetic field can rebuilt Kondo correlations, which are destroyed  by  the effect of spin-orbit interaction or valley mixing. Apart from the field induced revivals of SU(2) Kondo effects of different types: spin, valley or spin-valley, also more exotic phenomena appear, such as SU(3) Kondo effect. Threefold degeneracy occurs due to the effective intervalley exchange induced by short-range part of Coulomb interaction or due to the intershell mixing. In narrow gap nanotubes the full spin-orbital degeneracy might be recovered in the absence of magnetic field opening the condition for a formation of SU(4) Kondo resonance.
\end{abstract}

\maketitle

\section{Introduction}
In the last two decades carbon nanotubes (CNTs) have attracted tremendous interest  from fundamental science and technological perspectives \cite{Saito,Cottet,Javey,Avouris,Shulaker,Laird}. The unique electronic properties of  these thin, hollow cylinders formed from graphene are due to confinement of electrons normal to the nanotube axis. In a very crude approach (zone-folding approximation \cite{Hamada,Saito2}), the bandstructure of nanotube  is obtained from the bandstructure of graphene  by imposing  periodic  boundary conditions along the circumference. Depending on the way graphene is rolled up, carbon nanotube can be either metallic or semiconducting.  In this simple picture bandgap depends  on the minimum separation of the quantization lines from Dirac points  and is  inversely proportional to diameter. The zone folding approximation breaks down for tubes of small diameter. In this case the  curvature induced corrections to the overlaps between adjacent orbitals cannot be neglected. In this way one can understand  experimentally observed opening of   a narrow gap ($\sim10$ meV)  even in nominally metallic tubes (these systems are sometimes called nearly metallic carbon nanotubes \cite{Steele}).  Similar effect occurs also due to strain. The resulting gaps are inversely proportional  to the square of  diameter, depend on chiral angles and are typically much smaller than the gaps resulting from circular quantization. The dispersion laws of narrow gap nanotubes differ considerably from the dispersions of wide gap nanotubes and consequently distinctly different is their behavior  in magnetic fields \cite{Steele}. Considerations of the present work are addressed to semiconducting and narrow gap carbon nanotubes. We focus on strong correlation effects, which are of importance in  quantum dots (CNTQDs) formed from a short part of nanotube e.g. by introducing a confining potential in the longitudinal direction. Electron correlations in CNTQDs  are strong due to the one-dimensional nature of confinement and due to the  low dielectric constant especially in suspended nanotubes \cite{Cao}.  As the size of the dot decreases the charging energy of a single excess charge on the dot increases. Transmission of the contacts determines the regime of charge transport. For very weak transparency charging effects dominate transport at low temperature and the electrons enter the dot one by one yielding the well known Coulomb blockade oscillations of conductance as a function of the gate voltage.  For more open contacts  the role of higher order tunneling processes (cotunneling)   increases   what results at low temperatures  in formation of many-body resonances at the Fermi level, and a new transport path opens in the valleys between Coulomb peaks. Until  the ground breaking paper of Kuemmeth et al. on spin-orbit coupling in CNTs \cite{Kuemmeth} it was believed that  spin and orbital (valley)  degrees of freedom  could be seen as independent in CNTQDs and  the electron spectrum of nanotubes in the absence of magnetic field  was considered as  fourfold spin-orbital  degenerate.  A conviction of orbital degeneration was based on  the presence of two equivalent dispersion cones ($K$ and $K'$) in graphene,  in nanotubes  this degeneracy can be intuitively viewed to originate from two equivalent ways electrons can circle the graphene cylinder, that is  clockwise and counter-clockwise \cite{Minot}. Experimentally the presence of quadruplets of degenerate levels  has been confirmed e.g. by   observation of fourfold periodicity of  the electron addition energy  (e.g. \cite{Cobden}). This degeneration  also manifests in  many-body effects by  the  enhanced symmetry of many body-resonances. The first report on  the appearance of the exotic many-body state in CNTs was the result  of  Jarillo-Herrero et al. indicating the occurrence of  Kondo effect of  spin-orbital SU(4)  symmetry \cite{Jarillo}. Later appeared similar reports also evidencing the occurrence of SU(4) Kondo effect in carbon nanotubes \cite{Makarovski,Grove,Wu}. The problem of  simultaneous screening of orbital and spin degrees of freedom has been also discussed from theoretical point of view in several publications \cite{Pohjola,Borda,Chudnovskiy,Choi,Lipinski,Lopez,Lim,Zarand,Galpin,Sakano,Mravlje,Le Hur,Busser,Fang,Anders,Krychowski,Makarovski2,Keller,Filippone,Schmid,Krychowski2,Jespersen}.  Whereas in conventional  spin Kondo effect (SU(2)),  a formation of many-body dynamical singlet  between localized spin and delocalized electrons is a consequence of spin flip cotunneling processes, in the case when spin and orbital degeneracies occur simultaneously (SU(4)),  a  formation of  many-body resonance results from spin, orbital isospin and spin-orbital  cotunneling.  The spin and orbital degrees of freedom are totally entangled  and Kondo resonance  is no longer peaked at the Fermi level ($E_{F}$) as in the standard spin Kondo state  and Kondo temperature is largely enhanced in comparison to SU(2) systems. Conclusive observation, confirming that  the phenomena reported by Jarillo-Herrero et al. was indeed SU(4) Kondo resonance was  detection of splitting of this resonance in parallel magnetic field into four lines. Parallel magnetic field  knocks out from degeneracies  both spin and orbital isospin and  sufficiently strong field  destroys Kondo  correlations. Some  other perturbations suppress the role of only one of the degrees of freedom and then  system is  left in SU(2) symmetry and a crossover from  Kondo effect of higher symmetry into Kondo effect of lower effect is observed \cite{Jarillo,Makarovski2}. Example of such a case is the action of perpendicular magnetic field, which breaks only spin degeneracy \cite{Makarovski2}. Similarly valley (orbital) coupling  caused by  local  perturbations  introduced e.g. by vacancies or substitution of atoms or by deformations of structure  in interatomic distances mix only orbital channels \cite{Grove2,Churchill}.  More recent experiments pointed out on the intrinsic source of breaking of the symmetries and lowering the degeneracy. The use of suspended CNTs  \cite{Kuemmeth} allowed to reduce both the disorder in the sample and the dielectric screening due to substrate. The single electron spectroscopy measurements  performed on  ultraclean nanotubes showed that even at zero magnetic field,  the spin and orbital motions are not independent and a level splitting into two Kramers doublets has been observed \cite{Kuemmeth}. This effect has been attributed as resulting from spin-orbit interaction (SO). Destruction of the full spin-orbital entanglement  is certainly disadvantageous for quantum computing applications, because storage capacities of qubit systems is smaller than for four-state bit systems associated with SU(4) symmetry. Also unfavorable effect of  SO interaction is opening of  a route for spin decoherence.  But there is also an advantage of SO interaction, it gives the way of  electrical manipulation of spin degrees of freedom \cite{Bulaev,Marcus}. The significant  SO interaction has been confirmed later by other authors \cite{Steele,Grove2,Jhang,Pei}, also in the many electron regime \cite{Cleuziou}. The origin of  spin-orbit coupling in CNTs is curvature, as was  already theoretically predicted by Ando \cite{Ando} and is described in more detail in \cite{Huertas,Izumida}.  In graphene this interaction is almost completely suppressed due to inversion symmetry of graphene plane. In nanotubes  inversion symmetry is broken  and in consequence the hopping between $p$ orbitals of different parity from different atoms is allowed. The combined effect of  curvature and intra-atomic SO coupling mixes spin and orbital channels and the four-fold degenerate manifolds of states are split into  doublets  by a few tenth of meV. One can look at the source of SO splitting as the result of effective radial electric field  arising out of curvature, which  in rotating electron frame is seen as an effective local magnetic field that has opposite direction for the two valleys. Depending on the sign of SO coupling constant it  introduces parallel, or antiparallel  alignment of spin and angular momentum. The energy of SO coupling  is comparable with the energy scale of Kondo effect and therefore the two effects interplay or compete. Several  interesting articles have been published  taking up this topic \cite{Galpin,Schmid,Fang2,Mantelli}. The first theoretical analysis showing how SO interaction significantly changes the low-energy Kondo physics in CNTQDs was the paper of Fang et al \cite{Fang2}, where the multipeak structure of many-body resonance resulting from the interplay of spin-orbit and Zeeman effect has been analyzed. The most comprehensive study of the impact of SO coupling on Kondo effect is the paper of Galpin et al. \cite{Galpin}, where applying numerical renormalization  group technique the differential conductance of CNTQD has been calculated as a function of gate voltage and magnetic field.  Similar calculations, but additionally taking into account the effect of valley mixing have been presented in \cite{Schmid,Mantelli}. In \cite{Schmid} also nonlinear transport has been discussed and universality of conductance curves  with the energy scale determined by Kondo temperature  has been tested.

Our present publication generalizes and supplements the earlier  studies in  three aspects:

(i) We extend the discussion of the interplay of SO interaction and valley mixing in many-body processes by considering additional to disorder,  other  important sources of valley scattering,  not discussed earlier.   We analyze the impact of  indirect  intervalley mixing resulting from the coupling to the leads, where due to interference appears  e-h asymmetry within the shell, fact often observed in experiment (e.g.  \cite{Cleuziou}).   We also discuss the effect of  intervalley mixing induced by short -range  Coulomb interactions (valley backscattering - VBS). VBS changes the splitting between different valley and spin states, what can be interpreted as a result of the effective spin and intervalley  exchange. In effect of this interaction the tendency to form entangled two-electron states competes with Kondo screening of spin or orbital isospin. The reconstruction of dot states resulting from VBS  leads in some cases to enhanced degeneracy, what  opens  the path for  the occurrence of high symmetry  Kondo effects.

(ii) Discussing   the effect  of magnetic field in CNTQDs, we draw special attention on the role of  field induced reconstruction  of single electron dispersion curves.  This effect  is particularly important  for narrow gap nanotubes, where the details of the band structure are decisive for  the response on the field.  This manifests most visibly   in the vicinity of the gap. Close to the gap effective spin-orbit coupling changes dramatically and often exhibits different sign for electrons and holes \cite{Jespersen}. All these changes strongly modify the conditions for the occurrence of  Kondo resonances, what reflects in  the gate and field dependencies of conductance and polarization of conductance. We show that for dots in nearly metallic nanotubes the effective spin-orbit splitting can even vanish for some specific values of gate voltages  and full  SU(4) symmetry could be recovered with all consequences for many-body physics, including possibility of  occurrence of  SU(4) Kondo effect.

(iii) We generalize the analysis of correlation effects by taking into account  the shell mixing processes, that are not negligible for long CNTQDs or high magnetic fields.   For specific cases magnetic field partially  recovers degeneracy broken by SO interaction, what leads to a  formation of the  many body resonances of different origins and symmetries. For vanishing intervalley mixing  only  spin and valley Kondo effects are induced within the single shell.  Intershell effects  or intervalley mixing  additionally allow  formation of  spin-valley  Kondo resonance. Another interesting observation is that for strong Coulomb induced intervalley coupling, or due to intershell coupling, magnetic field can in some cases  lead  not only two, but also  three states to degeneracy. It is interesting from the point of view of quantum computing, because it would allow to operate in these systems not only on qubits, but also on qutrics (tree state information units).   Threefold degeneracy allows also for a formation of  Kondo state of exotic SU(3) symmetry, problem previously discussed only in a few works \cite{Carmi,Moca,Lopez2}.

To complement the discussion of the crossovers between Kondo states of different symmetries we present  in some cases,  apart from evolution of total conductance  also the corresponding dependencies of other transport characteristics, like thermopower or noise Fano factor. This is especially useful for analysis of SU(4) to SU(2) crossover, because total conductances coincide for both symmetries.  Special attention is also paid to spin dependent characteristics important for spintronic applications.

This paper is organized as follows.  In Sec. II, we describe the model of  CNTQD and the many body technique we use in analysis - slave boson approach.  Numerical results  and  analysis are given in sections III and IV, where we successively present  the effect of SO interaction on transport characteristics, discuss the  impact of different types of intervalley mixing on Kondo physics and analyze the effect of magnetic field and intershell mixing. Section V contains summary and a short discussion.

\section{Model And Slave Boson Mean-Field Formulation}
Our considerations are addressed to the low temperature range and it is enough therefore to restrict to the lowest quadruplet of states  labeled by spin ($\sigma=\pm1=\uparrow,\downarrow$)  and the orbital pseudospin ($m=\pm1=K,K'$) commonly referred to as valley. Although the CNTQD  states are in principle  additionally numbered by  longitudinal momentum $k_{\parallel}$ and  circumferential momentum $k_{\perp}$ we do not write them explicitly, because the level separations corresponding to  the quantizations of these quantities are much larger than the  thermal energy and the energy scale of the processes engaged in many-body effects discussed. Fixed values of   $k_{\perp}$ and  $k_{\parallel}$  corresponding to the lowest occupied states are assumed, with only one exception in section IV, where intershell effects are discussed and in this case  apart from the lowest also the  first excited  longitudinal mode is considered. The basic Hamiltonian describing CNTQD is extended  two orbital Anderson model:
\begin{eqnarray}
&&{\cal{H}}={\cal{H}}_{L}+{\cal{H}}_{R}+{\cal{H}}_{d}+{\cal{H}}_{t}
\end{eqnarray}
The first two terms describe noninteracting electrons in the leads:
\begin{eqnarray}
&&{\cal{H}}_{\alpha}=\sum_{km\sigma}E_{k\alpha m\sigma}c^{\dag}_{k\alpha m\sigma}c_{k\alpha m\sigma}
\end{eqnarray}
($\alpha =L, R$) for the left (right) electrode and $E_{k\alpha m\sigma}$ is the energy of an electron in the lead $\alpha$. The dot Hamiltonian is given by:
\begin{eqnarray}
&&{\cal{H}}_{d}=\sum_{m\sigma}E_{m\sigma}n_{m\sigma}+{\cal{U}}\sum_{m}n_{m\uparrow}n_{m\downarrow}+\nonumber\\
&&+{\cal{U}}'\sum_{\sigma\sigma'}n_{1\sigma}n_{-1\sigma'},
\end{eqnarray}
with dot energies
\begin{eqnarray}
&&E^{\pm}_{m\sigma}=\pm\sqrt{(-m \frac{E_{g}}{2}+\mu_{o}B_{\parallel}+\sigma\Delta^{1}_{so})^{2}+E^{2}_{0}(V_{g})}+\nonumber\\
&&+m\sigma\Delta^{0}_{so}+\frac{\sigma\mu_{s}B_{\parallel}}{2},
\end{eqnarray}
the upper or the lower sign refer to conduction or valence band states, $E_{g}$ is the bandgap at zero field without spin-orbit coupling, $E_{0}$ is gate dependent dot energy for the fixed value of $k_{\parallel}$, $B_{\parallel}$ is magnetic field directed along the nanotube axis. SO interaction sets as spin quantization axis the nanotube axis and locks spin and valley degrees of freedom. $\Delta^{0}_{so}$ and $\Delta^{1}_{so}$ parameterize the strength of  Zeeman and orbital contributions to the spin-orbit coupling \cite{Ando,Huertas,Izumida}. The former gives rise to a vertical shift of the Dirac cones that is opposite for two spin directions and $\Delta^{1}_{so}$ gives rise to horizontal shift of Dirac cones, what is equivalent to spin dependent  magnetic flux. As it is seen  we have included the SO corrections already in the single particle energies (4). They result from  SO perturbation of the form:
\begin{eqnarray}
{\cal{H}}_{so}=\Delta^{1}_{so}s_{z}\mathbf{\tau_{x}}+\Delta^{0}_{so}m s_{z}
\end{eqnarray}
where $s_{z}$ is the spin component along the nanotube axis and $\mathbf{\tau_{x}}$ is Pauli matrix in the A-B graphene sublattice space. For large bandgap tubes i.e. in the limit $E_{g}\gg \Delta^{0}_{so}$, $\Delta^{1}_{so}$ the SO splitting can be described  by one parameter $\Delta=2(\Delta^{0}_{so}\mp\frac{\Delta^{1}_{so}}{\sqrt{1+(E_{0}/E_{g})^{2}}})\approx2(\Delta^{0}_{so}\mp\Delta^{1}_{so})$.
The terms parameterized by ${\cal{U}}$, ${\cal{U}}'$ describe intra and interorbital Coulomb interactions and the last term ${\cal{H}}_{t}$ describes electron tunneling from the leads to the dot (or vice versa) and takes the form:
\begin{eqnarray}
{\cal{H}}_{t}=\sum_{k\alpha m\sigma}t(c^{\dag}_{k\alpha m\sigma}c_{m\sigma}+h.c.)
\end{eqnarray}
Coupling of the dot to the electrodes can be parameterized by $\Gamma_{m\sigma}(E)=2\pi\sum_{k\alpha}|t|^{2}\delta(E-E_{k\alpha m \sigma})$.  We assume that  $\Gamma_{m\sigma}(E)$ is constant within the energy band, $\Gamma_{m\sigma}(E)=\Gamma$. The full spin-orbital rotational SU(4) symmetry occurs only  for $B=0$, $\Delta=0$ and ${\cal{U}}={\cal{U}}'$. Apart from magnetic field and SO interaction we will also discuss the effect of other symmetry breaking perturbations: direct and indirect valley mixing and valley and spin exchange. The forms of these additional perturbations will be given in the sections, where these problems will be discussed.

To analyze correlation effects, we use  finite U slave boson mean field approach (SBMFA)  of Kotliar and Ruckenstein \cite{Kotliar} and introduce a set of boson operators for each electronic configuration of the single shell of  CNTQD.  These operators act as projectors onto empty state $e$, single occupied state $p_{m\sigma}$, doubly occupied $d$, triply occupied $t_{m\sigma}$ and fully occupied $f$. The $e$ operators are labeled by orbital index, $p$ operators by indices specifying the corresponding single-electron states, $t$ by indices of state occupied by a hole, and the six $d$ operators denote projectors onto double occupied states $d_{m=1,-1}$ ($\uparrow\downarrow$,$0$),($0$,$\uparrow\downarrow$) and $d_{\sigma\sigma'}$ ($\uparrow$,$\uparrow$),($\uparrow$,$\downarrow$),($\downarrow$,$\uparrow$) and ($\downarrow$,$\downarrow$).
To eliminate unphysical states, the completeness relations for the slave boson operators  and  the conditions for the correspondence between fermions and bosons have to be imposed. These constraints can be enforced by introducing Lagrange multipliers $\lambda$, $\lambda_{m\sigma}$ and supplementing the effective slave boson Hamiltonian by corresponding terms in (7). The  SB Hamiltonian then  reads:
\begin{widetext}
\begin{equation}
\begin{split}
{\mathcal{H}}^{K-R}=\sum_{m\sigma}E_{m\sigma}n^{f}_{m\sigma}+{\mathcal{U}}\sum_{m}d^{\dag}_{m}d_{m}
+{\cal{U}'}\sum_{\sigma\sigma'}d^{\dag}_{\sigma\sigma'}d_{\sigma\sigma'}+\sum_{m\sigma}({\cal{U}}+2{\cal{U}'})t^{\dag}_{m\sigma}t_{m\sigma}
+(2{\cal{U}}+4{\cal{U}'})f^{\dag}f\\+\sum_{m\sigma}\lambda_{m\sigma}(n^{f}_{m\sigma}-Q_{m\sigma})
+\lambda({\cal{I}}-1)+t\sum_{k\alpha m\sigma}(c^{\dag}_{k\alpha m\sigma}z_{m\sigma}f_{m\sigma}+h.c)
+\sum_{k\alpha m\sigma}E_{k\alpha m\sigma}n_{k\alpha m\sigma}
\end{split}
\end{equation}
\end{widetext}
with $Q_{m\sigma} = p^{\dag}_{m\sigma}p_{m\sigma}+d^{\dag}_{m}d_{m}+d^{\dag}_{\sigma\sigma}d_{\sigma\sigma}
+d^{\dag}_{\sigma\overline{\sigma}}d_{\sigma\overline{\sigma}}+t^{\dag}_{m\sigma}t_{m\sigma}
+t^{\dag}_{\overline{m}\sigma}t_{\overline{m}\sigma}+t^{\dag}_{\overline{m}\overline{\sigma}}t_{\overline{m}\overline{\sigma}}
+f^{\dag}f$, ${\cal{I}}=\sum_{m\sigma\sigma'}(e^{\dag}e+p^{\dag}_{m\sigma}p_{m\sigma}
+d^{\dag}_{m}d_{m}+d^{\dag}_{\sigma\sigma'}d_{\sigma\sigma'}
+t^{\dag}_{m\sigma}t_{m\sigma}+f^{\dag}f)$  and $z_{m\sigma}=(e^{\dag}p_{m\sigma}+p^{\dag}_{m\overline{\sigma}}d_{m}
+p^{\dag}_{\overline{m}\overline{\sigma}}(\delta_{m,1}d_{\sigma\overline{\sigma}}+\delta_{m,-1}d_{\overline{\sigma}\sigma})
+p^{\dag}_{\overline{m}\sigma}d_{\sigma\sigma}+
d^{\dag}_{\overline{m}}t_{m\sigma}+d^{\dag}_{\overline{\sigma}\overline{\sigma}}t_{\overline{m}\overline{\sigma}}
+(\delta_{m,-1}d^{\dag}_{\sigma\overline{\sigma}}+\delta_{m,1}d^{\dag}_{\overline{\sigma}\sigma})t_{\overline{m}\sigma}
+t^{\dag}_{m\overline{\sigma}}f)/\sqrt{Q_{m\sigma}(1-Q_{m\sigma})}$. $z_{m\sigma}$ renormalize interdot hoppings and dot-lead hybridization (6). In the mean field approximation the slave boson operators are replaced by their expectation values. In this way the problem is formally reduced  to  the effective free-particle model with renormalized hopping integrals and renormalized dot energies. The stable mean field solutions are found from the saddle point of the partition function i.e. from the minimum of the free energy with respect to  the mean values of boson operators  and Lagrange multipliers.  SBMFA best works close to the unitary Kondo limit, but it gives also reliable results of linear conductance for systems with weakly broken symmetry in a relatively wide dot level range, being in a reasonably agreement with experiment and with renormalization group calculations \cite{Krychowski,Dong,Bulka,Lim,Trocha}.
Current flowing through CNTQD in the ($m\sigma$) channel can be expressed as ${\cal{I}}_{m\sigma}=\int [f_{L}(E)-f_{R}(E)]{\cal{T}}_{m\sigma}(E)dE$,
where transmission is given by ${\cal{T}}_{m\sigma}=\frac{4 \pi \Gamma_{L}\Gamma_{R}}{\Gamma_{L}+\Gamma_{R}} A_{m\sigma}(E)$ and $f_{\alpha}$ are the Fermi distribution functions.
The dot spectral weight $A_{m\sigma}(E)$ is obtained from the retarded Green's function $G^{R}_{m\sigma,m\sigma}$ by $A_{m\sigma}=(-1/\pi)Im[G^{R}_{m\sigma,m\sigma}]$.
The linear conductance is defined by ${\cal{G}}_{m\sigma}=\frac{d{\cal{I}}_{m\sigma}}{dV}|_{V\rightarrow0}$ and spin polarization of conductance  by  $PC_{s}=(\sum_{m}{\cal{G}}_{m\uparrow}-{\cal{G}}_{m\downarrow})/{\cal{G}}$, ${\cal{G}}=\sum_{m\sigma}{\cal{G}}_{m\sigma}$. In an analogous way defined are also polarizations associated with other degree of freedom, orbital $PC_{o}=(\sum_{\sigma}{\cal{G}}_{1\sigma}-{\cal{G}}_{-1\sigma})/{\cal{G}}$ or spin-orbital Kramers polarizations  $PC_{K}=({\cal{G}}_{1\uparrow}+{\cal{G}}_{-1\downarrow}-{\cal{G}}_{1\downarrow}-{\cal{G}}_{-1\uparrow})/{\cal{G}}$.
To get closer insight into evolution of many-body correlations under symmetry breaking perturbations we also  analyze  other transport quantities. Thermopower (TEP) acts as an excellent tool to describe the crossover between SU(4) and SU(2) Kondo states. It is defined by $S=-(1/|e|T)(\sum_{m\sigma}L_{1m\sigma}/\sum_{m\sigma}L_{0m\sigma})$, where $L_{nm\sigma}=-(1/h)\int (E-\mu)^{n}\frac{df}{dE} {\cal{T}}_{m\sigma}(E)dE$.
Also the  shot noise is strongly affected by a change of correlations and we analyze  the noise Fano factor $F$ defined as the ratio between the actual shot noise ${\cal{S}}$ and the Poissonian noise of uncorrelated carriers ($2|e|{\cal{I}}$). Due to the fact, that SBMFA  is an  effective noninteracting particle picture, in the linear regime Fano factor reads $F=\sum_{m\sigma}{\cal{T}}_{m\sigma}(1-{\cal{T}}_{m\sigma})/(\sum_{m\sigma}{\cal{T}}_{m\sigma})$.
We parameterize the unperturbed fully spin-orbital rotationally invariant CNTQD Hamiltonian by three parameters: charging energy $\cal{U}$, tunnel rate between the dot and the reservoirs $\Gamma$, and the half-bandwidth $D$.  The numerical results discussed below are presented with the use of energy unit defined by its relation to the bandwidth $2D=100$. The assumed energy unit corresponds to energy of 1 to 4 meV. If not specified differently  the results are presented for ${\cal{U}} = 3$ and $\Gamma = 0.05$. Typically charging energy for semiconducting CNTs is of order of tens of meV \cite{Herrero2,Babic2} and $\Gamma$ in the assumed weak-coupling regime is of order of several meV \cite{Jarillo,Makarovski,Kouvenhoven2}.

\section{Single-Shell Many-Body Effects}
\subsection{Breaking of SU(4) symmetry by spin-orbit interaction}
The fully symmetric system characterized by  SU(4) unitary group of spin-orbital rotations  (${\cal{U}}={\cal{U}}'$, $\Delta=0$ and $B=0$) exhibits  the fourfold degeneracy  for odd occupancies ($N =1,3$) and sixfold degeneracy at half filling ($N =2$).
\begin{figure}
\includegraphics[width=0.48\linewidth,bb=0 0 439 438,clip]{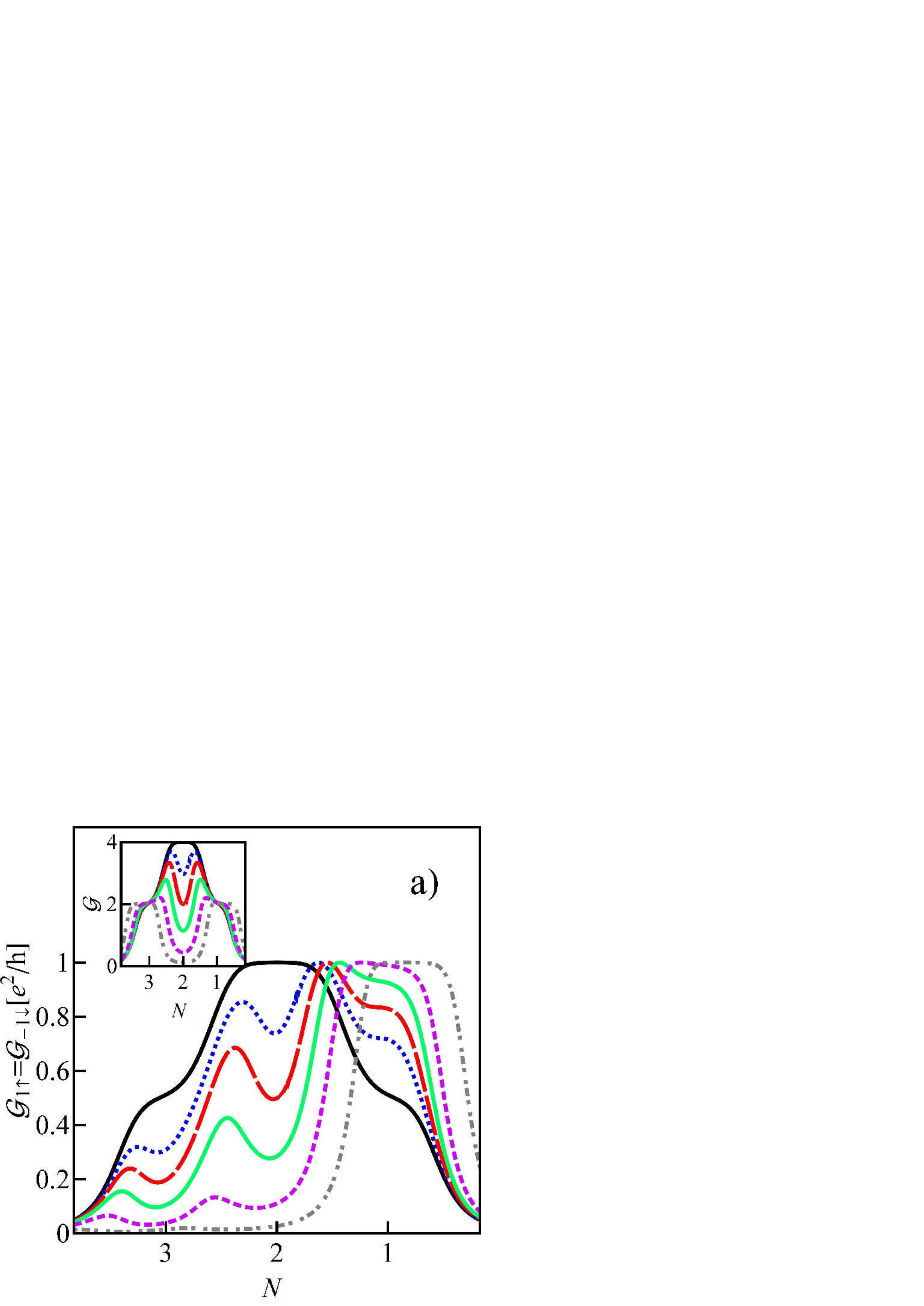}
\includegraphics[width=0.48\linewidth,bb=0 0 439 451,clip]{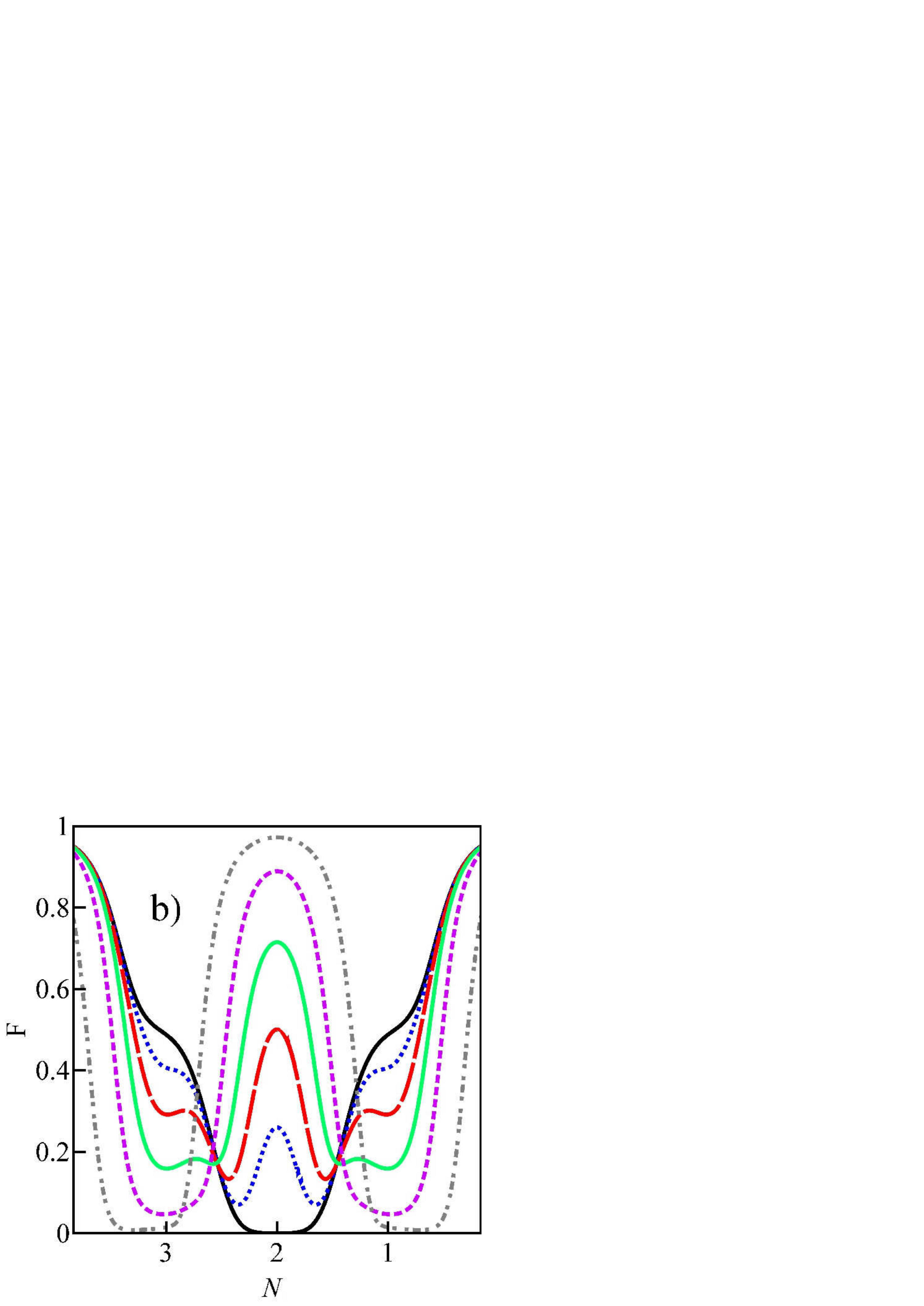}
\includegraphics[width=0.48\linewidth,bb=0 0 439 452,clip]{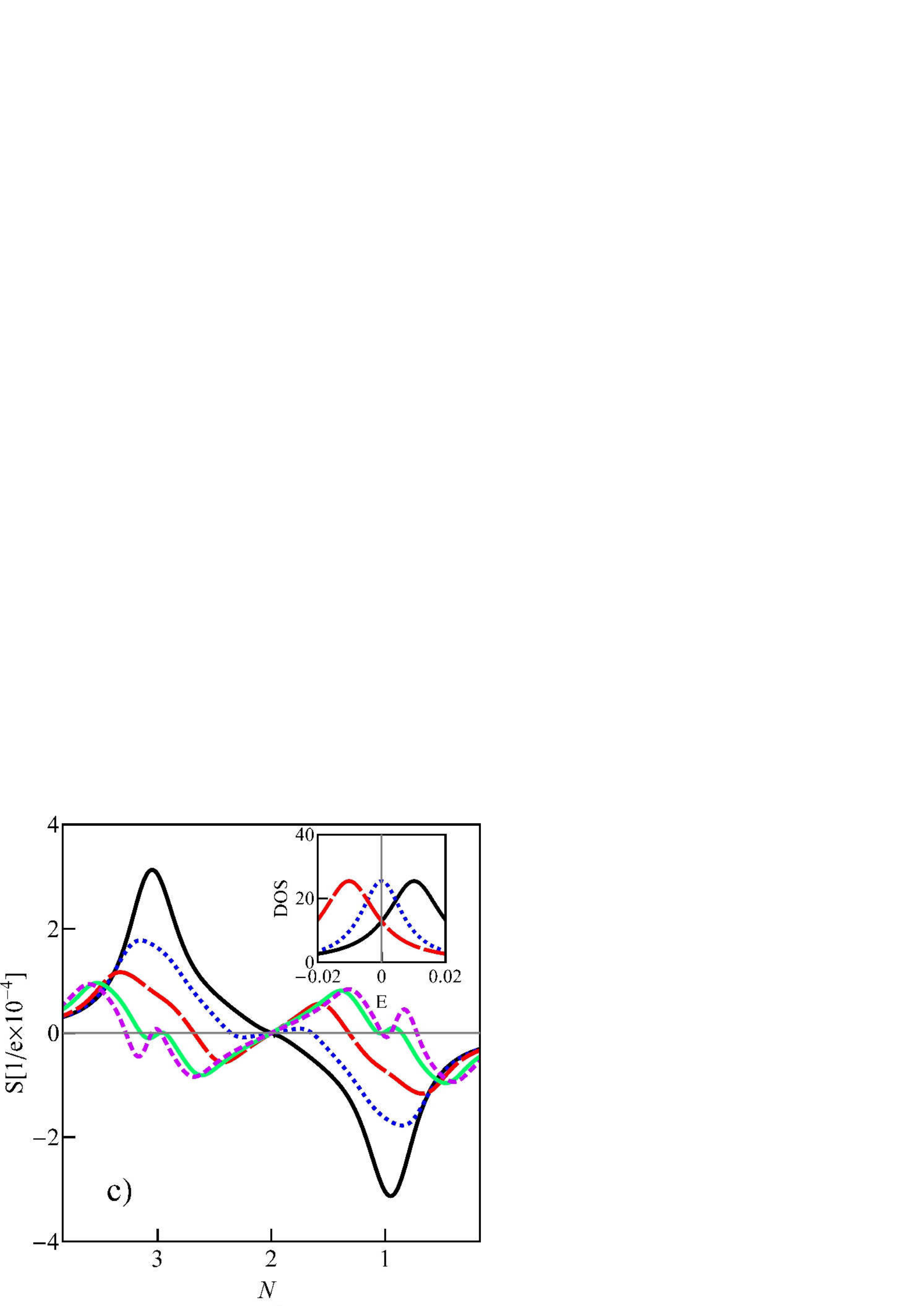}
\includegraphics[width=0.48\linewidth,bb=0 0 439 422,clip]{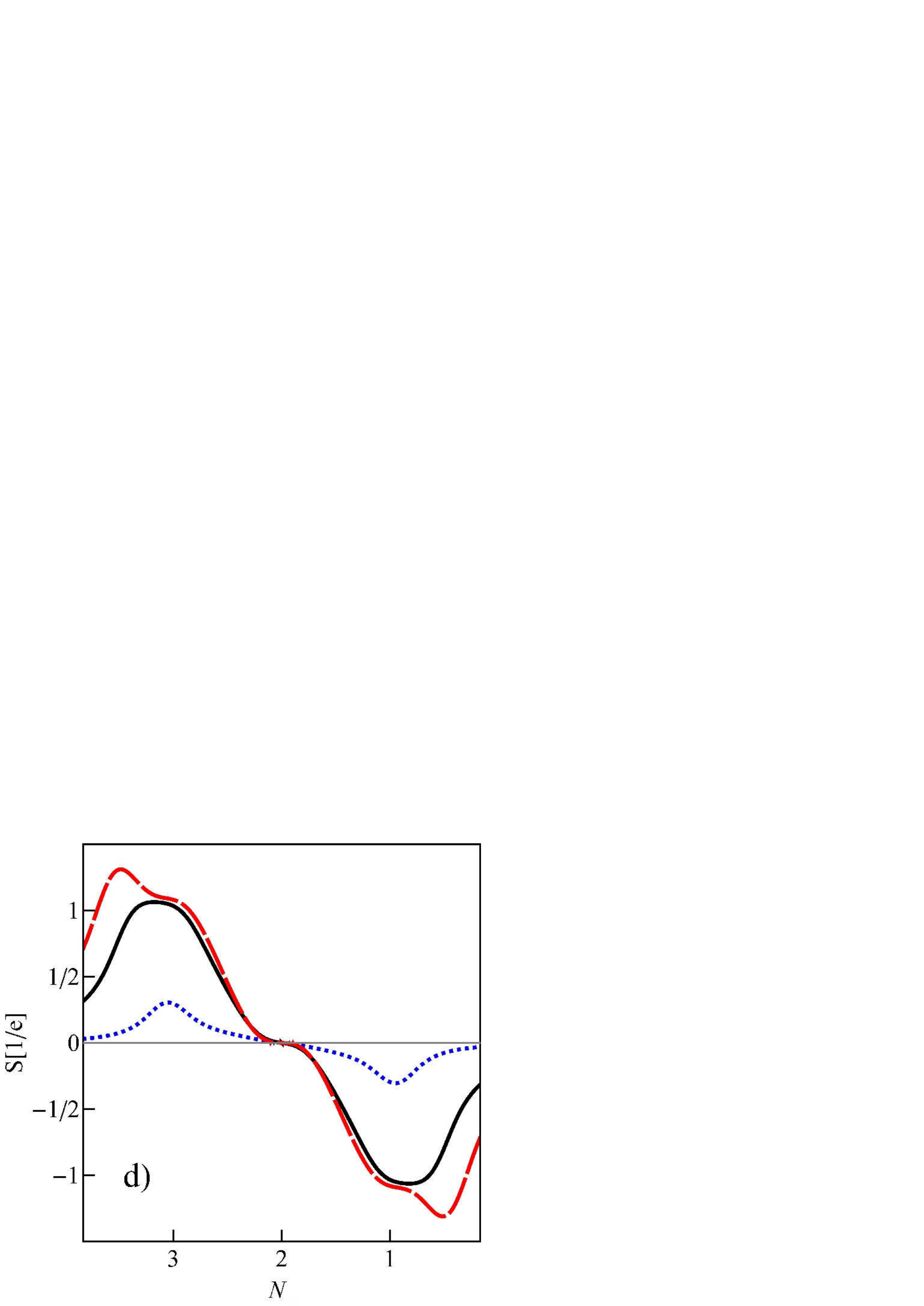}
\includegraphics[width=0.48\linewidth,bb=0 0 439 434,clip]{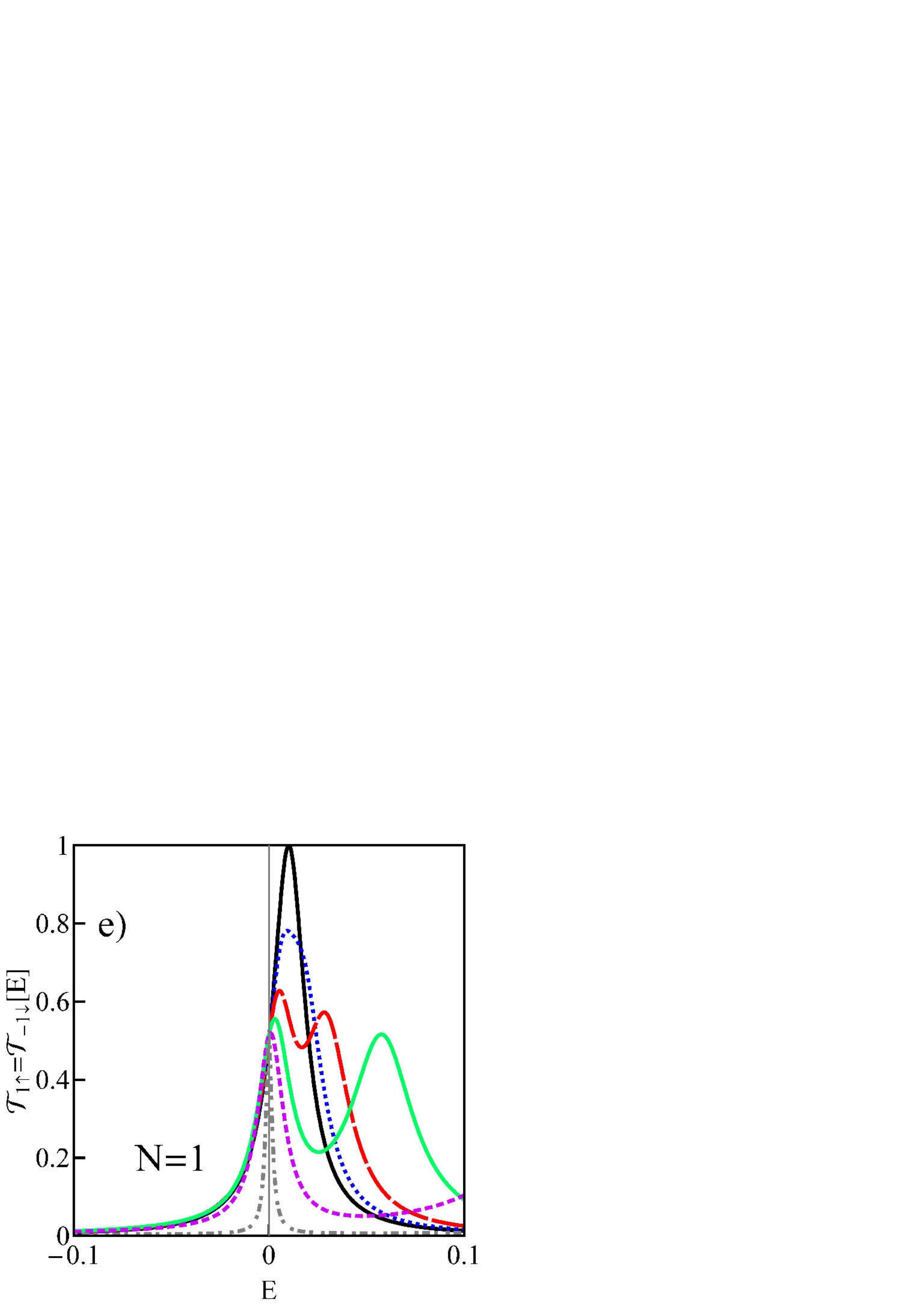}
\includegraphics[width=0.48\linewidth,bb=0 0 439 434,clip]{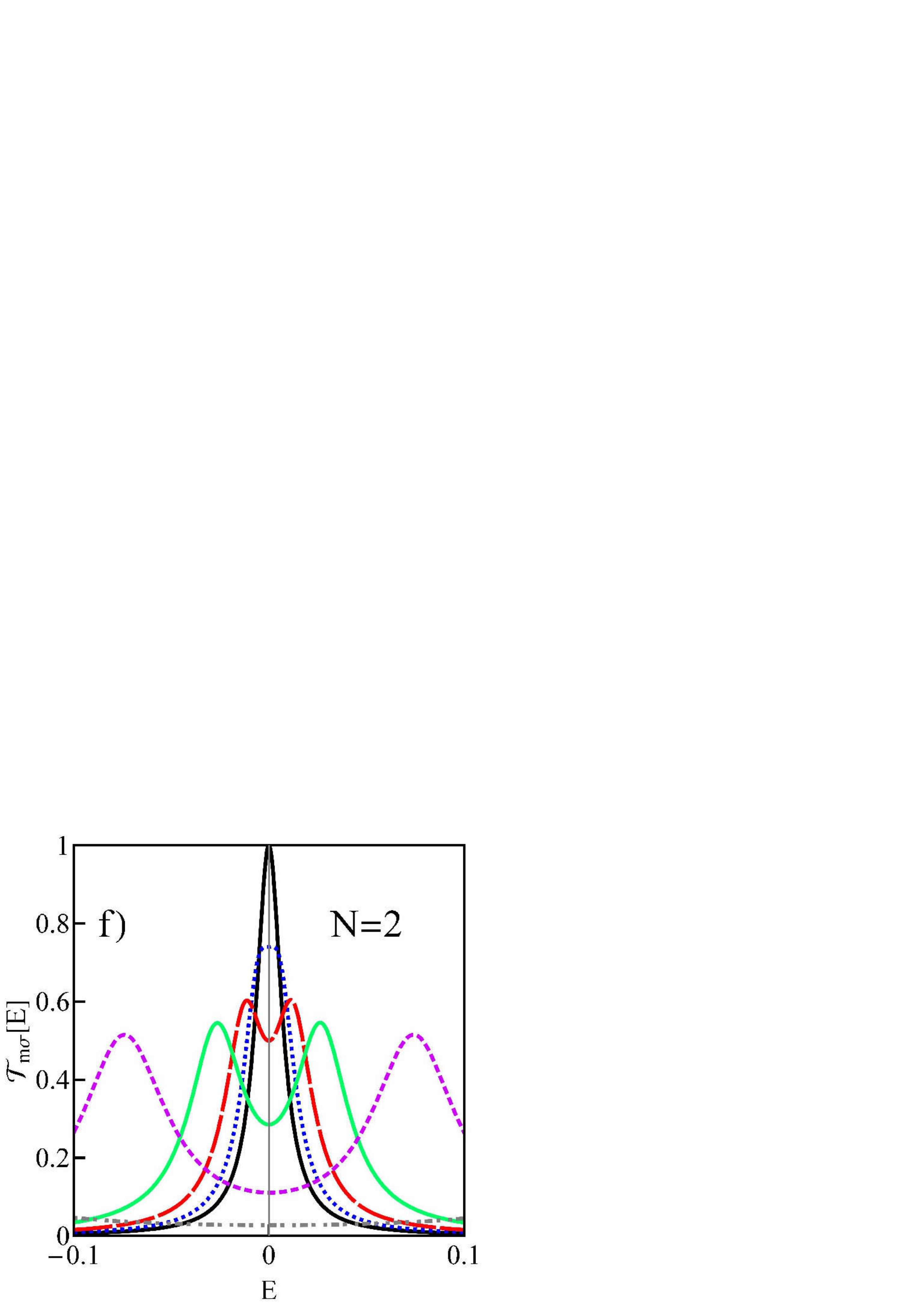}
\caption{\label{fig1} (Color online) Transport characteristics of  CNTQD with SO interaction.  (a) Spin and orbital resolved conductances of CNTQD  vs dot occupation $N=(1/2)(1-2E_{0}/{\cal{U}})$ plotted for several values of SO splitting $\Delta$: $\Delta=0$ (solid black line), $\Delta=0.025$ (dotted blue), $\Delta=0.05$ (dashed red), $\Delta=0.1$ (solid green), $\Delta=0.25$ (short dashed purple) and $\Delta=0.75$ (dashed dotted grey). The same assignment of the lines is valid also for the inset and figures b, e, f. Inset presents corresponding total conductances. Parameters:  ${\cal{U}}=3$ and $\Gamma=0.05$ ( if not specified differently the same parameters apply also to other pictures).  (b) Noise Fano factors. (c) Thermopower for $\Delta=0$   (solid black), $\Delta=0.08$   (dotted blue), $\Delta=0.2$  (dashed red), $\Delta=0.5$  (solid green), $\Delta=0.65$   (short dashed purple) ($V=0.005$ and $T=10^{-6}$). (d) Temperature evolution of thermopower for $\Delta=0$ and  $T=2\times10^{-2}$ (dashed line), $T = 10^{-2}\approx T_{K}$ (solid black), $T = 10^{-3}$ (dotted blue) ($V=0.005$). (e,f) Partial transmissions for $N =1$ and $N = 2$ plotted for different values of SO splitting.}
\end{figure}
CNT Hamiltonian is in this case invariant under time-reversal and valley-reversal.  For $N=1$ ($1e$ valley) the degenerate states are ($|m\sigma\rangle$, $m=K,K'$, $\sigma=\uparrow,\downarrow$) and  for $N=3$ ($|\underline{m}\underline{\sigma}\rangle$, $\underline{m}=K,K'$, $\underline{\sigma}=\uparrow,\downarrow$), where  the underlined quantum numbers refer to the unoccupied states i.e. $|\underline{K\uparrow}\rangle$ for instance  denotes the state $|K\downarrow K'\uparrow K'\downarrow\rangle$. In the following we will also interchangeably use instead of $K$, $K'$ quantum numbers $1,-1$. The six  degenerate two-electron states for $N=2$ are $|K\uparrow K'\downarrow\rangle$, $|K\uparrow K\downarrow\rangle$, $|K'\uparrow K'\downarrow\rangle$, $|K\uparrow K'\uparrow\rangle$, $|K\downarrow K'\downarrow\rangle$,  $|K\downarrow K'\uparrow\rangle$.
For the Kondo effect to occur apart from degeneracy it is also required that  spin and valley quantum number are conserved.
It  may be approximately satisfied if the contacts constitute parts of the same  CNT as the dot and the carriers dwell for some time before moving into the metallic electrodes. One can think then that before and after tunneling process electron share the same degrees of freedom \cite{Choi,Schmid}.  Kondo effect in odd electron valleys results from effective spin and valley isospin fluctuations (transitions  between four degenerate states) occurring due to cotunneling processes. Both the total spin $S_{Z}=(n_{\uparrow}-n_{\downarrow})/2$ and orbital pseudospin $T_{Z}=(n_{1}-n_{-1})/2$ are quenched by these  fluctuations.  The spin-valley many-body peak is shifted above ($1e$) or below ($3e$)  the Fermi level corresponding to phase shifts $\delta=\pi/4$ and $\delta=3\pi/4$ respectively. Increased degeneracy reflects in broadening of Kondo resonance in comparison to the standard spin Kondo peak, what means exponential  enhancement of  Kondo temperature.
In 2e valley for $\Delta=0$ transitions between all six states lead to a formation of Kondo  resonance  centered at $E_{F}$, the corresponding phase shift is  $\delta=\pi/2$ and conductance is doubled in comparison to SU(4)  Kondo effect for odd number of electrons. Spin-orbit coupling depends on chirality and diameter. The observed values of SO splitting are of order of $0.1-0.4$ meV \cite{Kuemmeth,Cleuziou,Jespersen}, but also splitting of up to $3.4$ meV has been reported \cite{Steele}. Fig. 1 presents  partial conductances, noise Fano factor, transmissions and thermoelectric power plotted for different values of SO coupling. In all these quantities the crossover between different symmetries induced by SO interaction is  evidently reflected.
\begin{figure}
\includegraphics[width=0.48\linewidth,bb=0 0 439 432,clip]{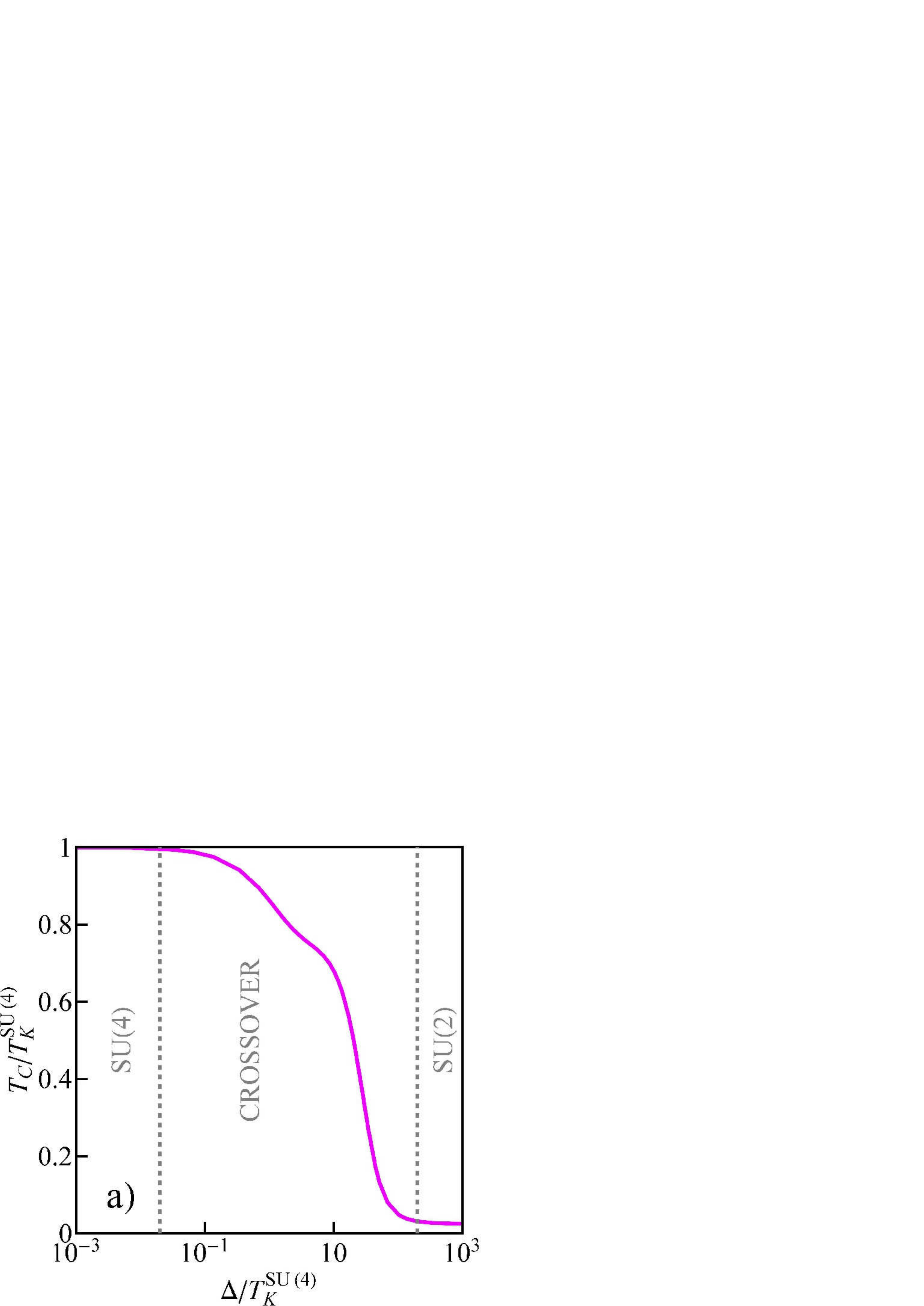}
\includegraphics[width=0.48\linewidth,bb=0 0 439 421,clip]{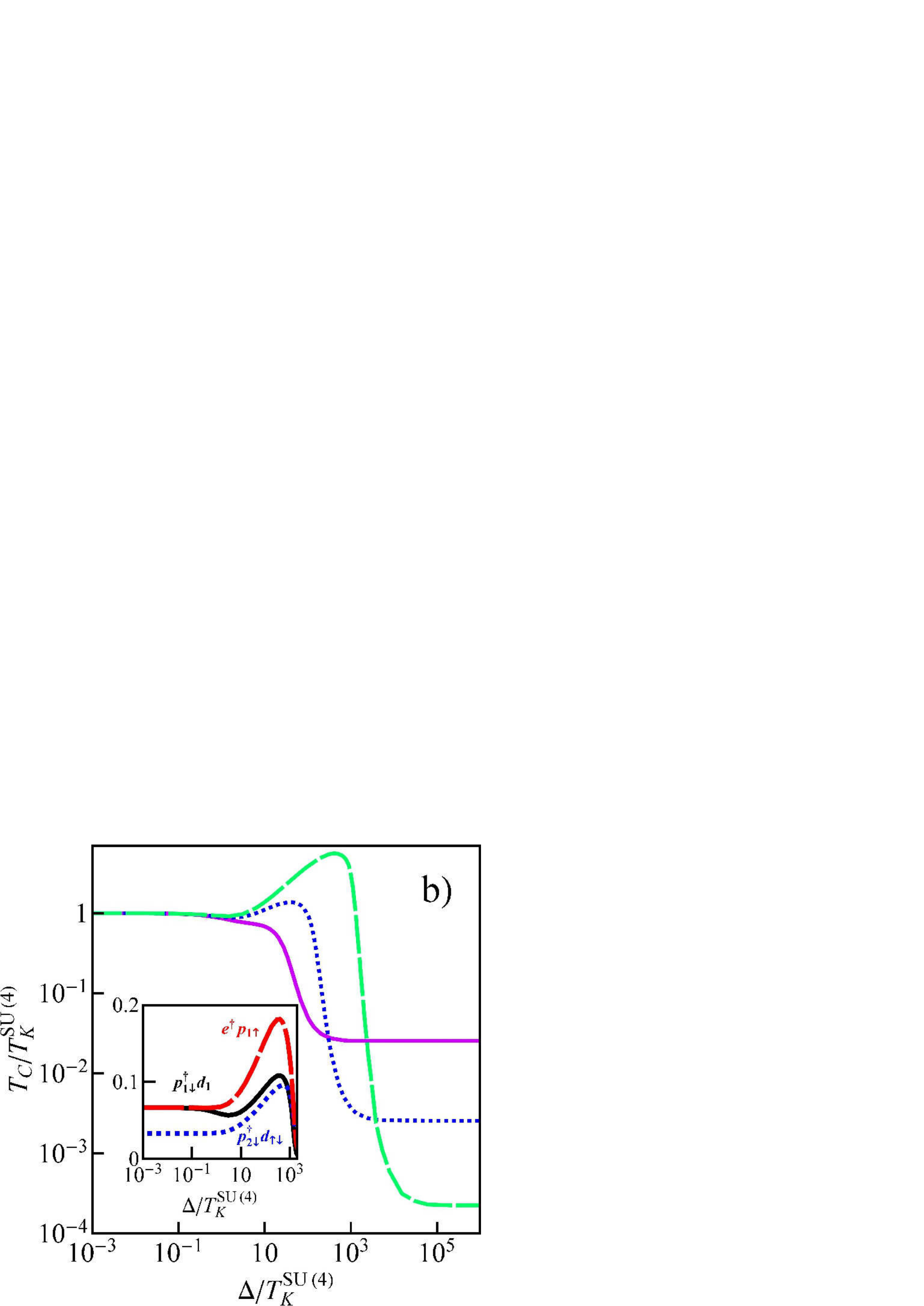}
\includegraphics[width=0.48\linewidth,bb=0 0 439 402,clip]{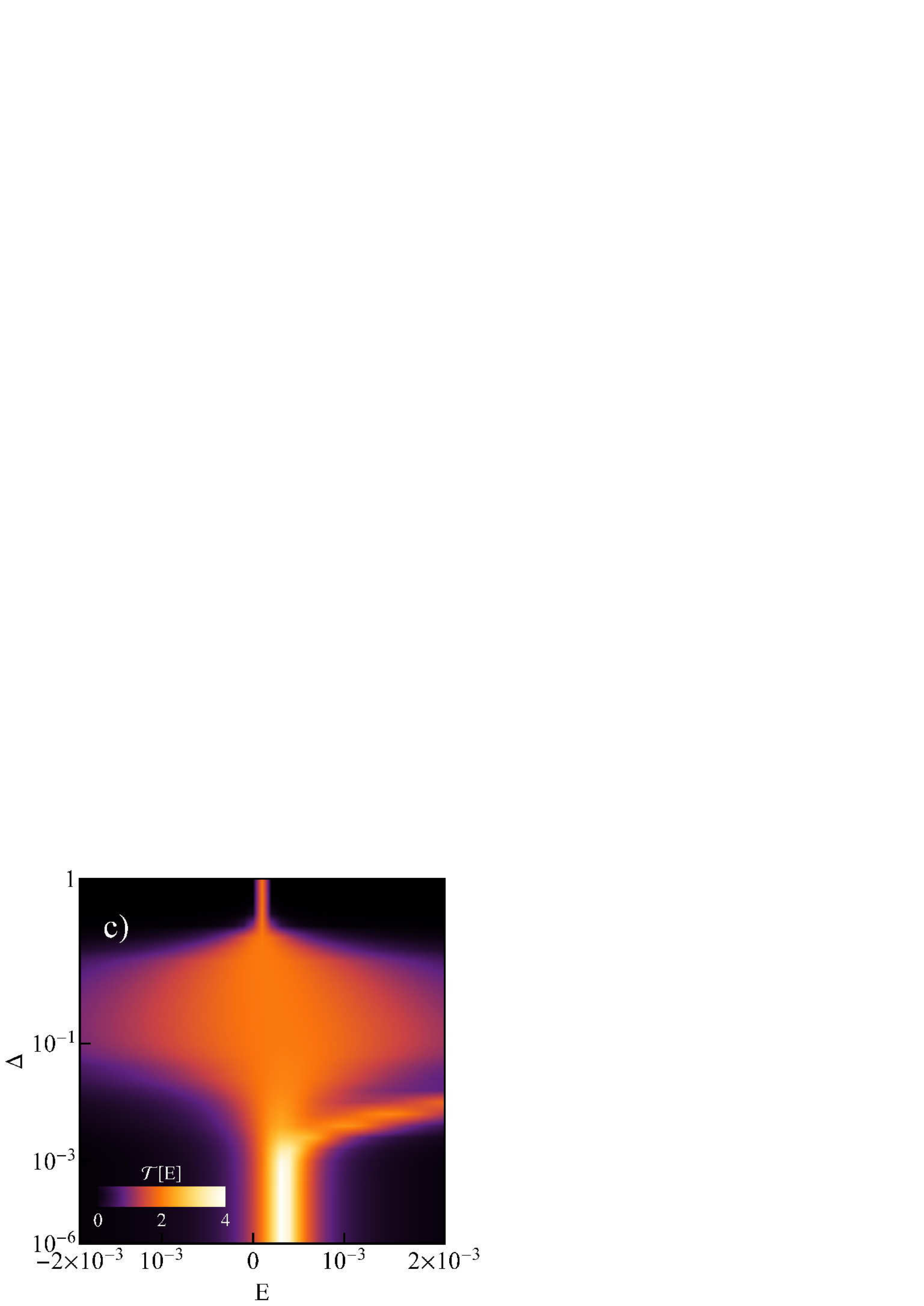}
\includegraphics[width=0.48\linewidth,bb=0 0 439 414,clip]{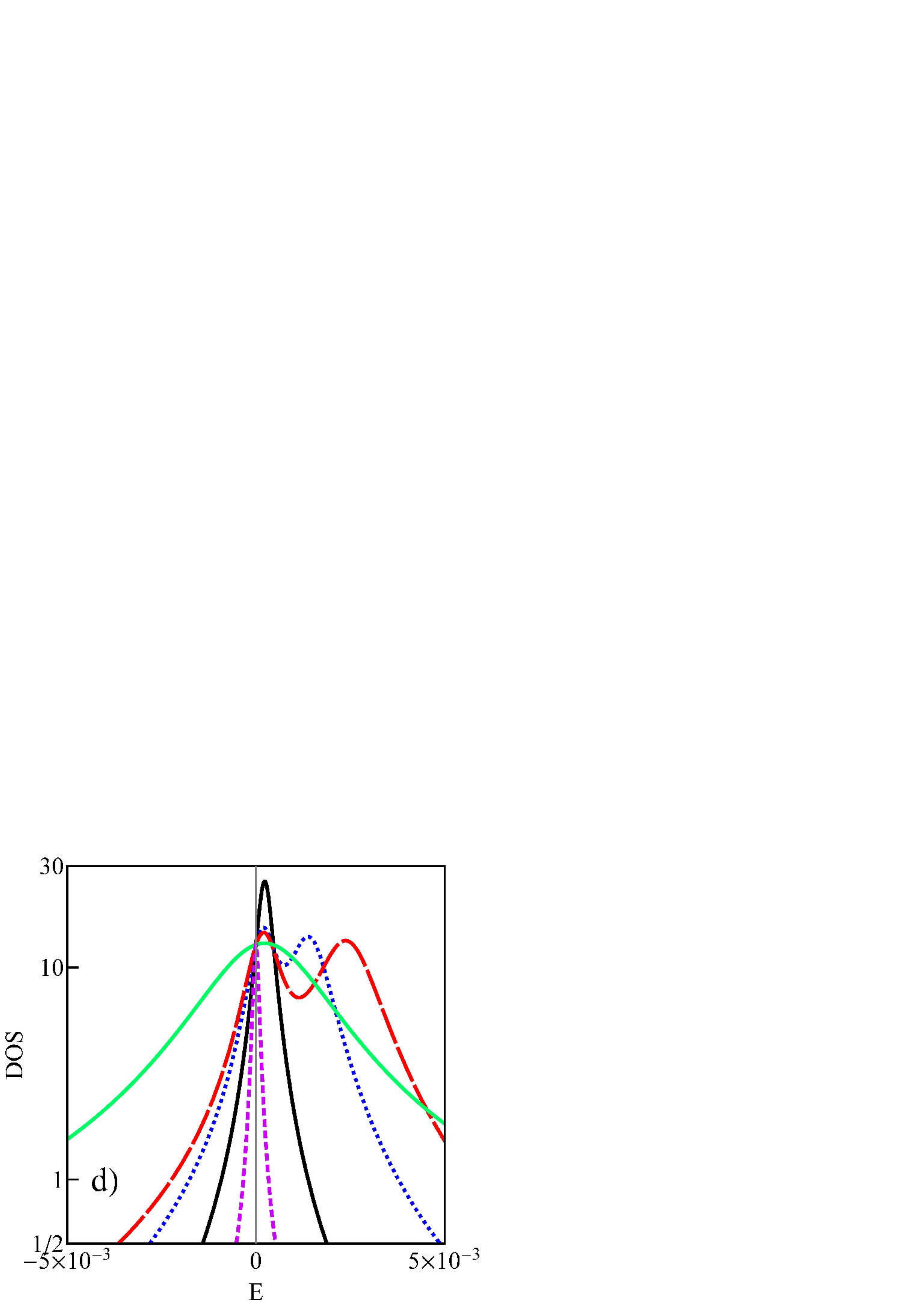}
\caption{\label{fig2} (Color online) Spin-orbit interaction induced SU(4)$\rightarrow$ SU(2) crossover in the  single electron valley ($N = 1$).  (a) Characteristic temperatures $T_{C}$ of many-body resonances vs SO splitting $\Delta$ ($U =3$). (b)  $T_{C}(\Delta)$ for ${\cal{U}} = 3$ (solid line), ${\cal{U}} = 4$ (dotted), ${\cal{U}} = 5$ (dashed). Inset shows mean values  of the product of slave boson operators plotted vs SO splitting parameter. (c)  Transmission map  ${\cal{T}}(E,\Delta)$ for ${\cal{U}} = 5$. (d) Densities of states of many-body resonances for $\Delta = 0$  (solid black line), $\Delta = 0.003$  (dotted blue), $\Delta = 0.005$  (dashed red), $\Delta = 0.1$  (solid green) and $\Delta = 0.52$  (short dashed purple).}
\end{figure}
Total conductance curves are given as a point of reference in the inset of Fig. 1, but we skip an analysis of their  evolution   with the increase of SO coupling, because this problem has been discussed earlier in \cite{Galpin,Mantelli}.  It is also  known that the total conductance  is not  a proper quantity to track the crossover between the considered symmetries, because  both symmetries SU(4) and SU(2)  achieve  the same unitary limit   in  the  Kondo range for odd electron occupations. For $N=2$  conductance is doubled in comparison to the odd case, what is a consequence of  half transmission of each channel in odd valleys and full transmission for even occupation (Fig. 1c,d).  In the latter case tunnel induced  fluctuation between six two-electron states lead to  a formation of   Kondo resonance centered at the Fermi level. Different opening of the  many-body transmission channels is clearly seen  in different limits of spin and orbital resolved conductances in odd and even valleys (Fig. 1a). This is also reflected in noise. Perfect transmission of all channels in $2e$ valley corresponds to no noise ($F=0$, Fig. 1b).
In odd valleys  on the other hand, the observed  half transmission of spin-orbital channels corresponds to  strong partition, what  means strong shot noise  ($F=1/2$).   Total conductance curve and noise Fano factor are symmetric vs $E_{0}=-(3/2){\cal{U}}$ line (symmetry of 1e and 3e regions),  as they reflect  transmission values at $E_{F}$ for $T=0$ or transmissions averaged over the thermal energy $k_{B}T$ for higher temperatures. Thermopower complements this information. This quantity provides information on the possible  asymmetry of the  transmission line  ${\cal{T}}(E)$ near the Fermi energy in the range of the thermal broadening.  In  the limit of $T\rightarrow0$  the sign of TEP reflects the slope of the spectral function at the Fermi level. For $N=1$ Kondo resonance lies above $E_{F}$ (inset of  Fig. 1c), the slope is positive, resulting in a negative TEP. For $N=3$ the peak is located below $E_{F}$ and the opposite sign is observed.
Zero value of thermopower for $N=2$ corresponds to the central position at  the Fermi energy of  Kondo resonance for even occupation. Now let us look at the broken symmetry case induced by   spin-orbit interaction ($\Delta\neq0$). For $N=2$  the sextet is  split into three groups of states, of degeneracy 1, 4 and 1, with relative energies $-\Delta$, $0$, and $\Delta$  respectively   (Fig. 5f).  Since the state of the lowest energy is singlet, Kondo correlations are gradually destroyed by SO interaction,  what results in a drop of conductance.
In transmission,  the top of the peak is replaced by a valley, deepening with the increase of $\Delta$ and since this effect occurs around the Fermi level and it manifests  by a change of the sign  of TEP   in the $N=2$ region. Spin-orbit  interaction lowers the symmetry, but it does not break time-inversion symmetry. Consequently in odd valleys it splits the four degenerate levels into  the pairs of Kramers doublets separated by an energy $\Delta$ (Fig. 5e) and the system   gradually falls into SU(2) symmetry class with the increase of  SO coupling.  Only  lower doublet in the single electron range ${|K\downarrow\rangle, |K'\uparrow\rangle}$ or the  higher  doublet ${|K\uparrow\rangle, |K'\downarrow\rangle}$ for  triple occupancy remains active in   Kondo fluctuations. Evolution of  partial conductances is illustrated in Fig.1a, taking as the examples  spin and orbital resolved conductances for the channels corresponding to the lower Kramers doublet. In the SU(4) limit ($\Delta=0$)  ${\cal{G}}_{1-}={\cal{G}}_{-1+}=1/2$ for $N=1$  corresponding to a quarter occupancy of  each spin-orbital. Increase of $\Delta$ results in the increase of the partial conductances, for  $\Delta\gg T$  they reach the unitary SU(2)  Kondo limit.    The curves  ${\cal{G}}_{1+}={\cal{G}}_{-1-}$, not presented here (higher Kramers doublet),  are symmetric under shell electron-hole symmetry line  and these transport channels are  almost completely closed in $N=1$ range for strong SO coupling, they are  open however  in  $3e$ valley.  For this filling, in turn, as seen in Figure 1a,   ${\cal{G}}_{1-}$  and  ${\cal{G}}_{-1+}$  are negligible. Reduction of symmetry with the  increase of SO coupling manifests itself  in a tendency to total suppression of  TEP at the points  $N=1,3$, which indicates the location of the SU(2)  Kondo resonances at the Fermi energies. Local minima of maxima of thermopower curves visible for small deviations from these integer occupations reflect the influence of  many-body fluctuations in Kondo active doublets.  Similar local extremes  are also seen  closer to the Coulomb border lines and in this case they signal the impact of  charge fluctuations.  Fig. 1d illustrates temperature evolution of thermoelectric power  shown only for the full SU(4) symmetric case. Apart from an obvious increase of TEP,  a shift of extremes  to the borders of Coulomb blockade regions is observed for temperatures above $T_{K}$, i.e. in the range, where Kondo correlations are destroyed.  Destruction of the symmetry or the   crossover between the symmetry classes can be also monitored  by shot noise (Fig. 1b). Killing of Kondo correlations at half filling for strong enough SO coupling causes that the  noise gets Poissonian.  For odd occupations the  SU(4) symmetry transmission  is distributed evenly across all channels and they are only partially open. When  system evolves into  SU(2) class  two of the channels  are closed and  two other  become transparent. In consequence   Fano factors change from $F=1/2$  for SU(4) symmetry  to $F=0$ for SU(2). In the latter case electrons pass through the dot with probability one, and the linear conductance is noiseless. Aforementioned direct signature of the symmetry change   in noise  obtained within effective  non-interacting quasiparticle picture of Landauer-Buttiker is consistent with the spirit of SBMFA. This approach is satisfactory for the discussion of linear noise. In nonlinear range, not discussed here, also other factors distinguish  noise  between different symmetry classes. Important contribution to the  nonlinear  noise give  two-particle scattering processes  induced by interactions out of equilibrium.   A clear shot noise enhancement compared with noninteracting case has been reported for SU(4)  CNTQD \cite{Vituschinsky,Mora}. This  is a distinct feature of the SU(4) Kondo effect, distinguishing it from the SU(2) case, for which a shot noise reduction is expected instead \cite{Mora}.   These effects are however outside the formalism we use in the present paper. Fig. 2a   shows characteristic temperature $T_{C}$ of many-body resonances in $N=1$ region, vs.   SO splitting $\Delta$ and  Fig. 2b presents $T_{C}(\Delta)$ for several values of  Coulomb interaction. $T_{C}$  has been estimated from  the temperature dependence of conductance (${\cal{G}}(T_{C})={\cal{G}}(0)/2$)  or from the position of the center of the peak ($\widetilde{\epsilon}$) and half width of the resonance peak ($\widetilde{\Delta}$), $T_{C}=\sqrt{\widetilde{\epsilon}^{2}+ \widetilde{\Delta}^{2}}$, and both estimations are qualitatively consistent and show the same tendency.  The stronger  enhancement of SU(4) Kondo temperature with respect to SU(2) Kondo temperature is observed for larger values of ${\cal{U}}$, what is consistent with the well known exponential dependencies of $T_{K}$ on ${\cal{U}}$ combined with the dependence on the fold of degeneracy \cite{Hewson}.  The crossover region also  extends with increasing ${\cal{U}}$ and interestingly, the characteristic temperatures do not always evolve monotonically. For high ${\cal{U}}$ values, where pure SU(4) resonances  are relatively narrow,  the effect of symmetry breaking  induced broadening is more significant than for wider SU(4) Kondo   resonances  and this effect dominates in a broad range of crossover  region over the tendency for extracting the additional peak associated with the processes  encountered in   the  doublet inactive in SU(2) Kondo  fluctuations (Fig. 2d). The region of enhanced characteristic temperature corresponds to mixed valence range, where  charge fluctuations become also of importance. In SBMFA formalism it manifests for  discussed here  $1e$ valley   in  enhancement of  mean values of $p$ and $e$ SB operators  in the range of enhanced $T_{C}$. The products of operators presented in the inset of Fig. 2b are the dominant contributions to SB parameters $z_{m\sigma}$, which determine the widths of the resonance.  The enormous broadening of  many-body resonance in the crossover region is also  visualized  on the ${\cal{T}}(\Delta,E)$  transmission  map (Fig. 2c).

\subsection{Direct and indirect valley mixing}

Apart from lifting of the fourfold degeneracy by  SO interaction, some experiments indicate  that  a similar role can be played by valley mixing \cite{Schmid,Kuemmeth}. To scatter from one valley to the other, a scattering vector of the order of the Brillouin zone vector is needed.
\begin{figure}
\includegraphics[width=0.6\linewidth,bb=0 0 439 436,clip]{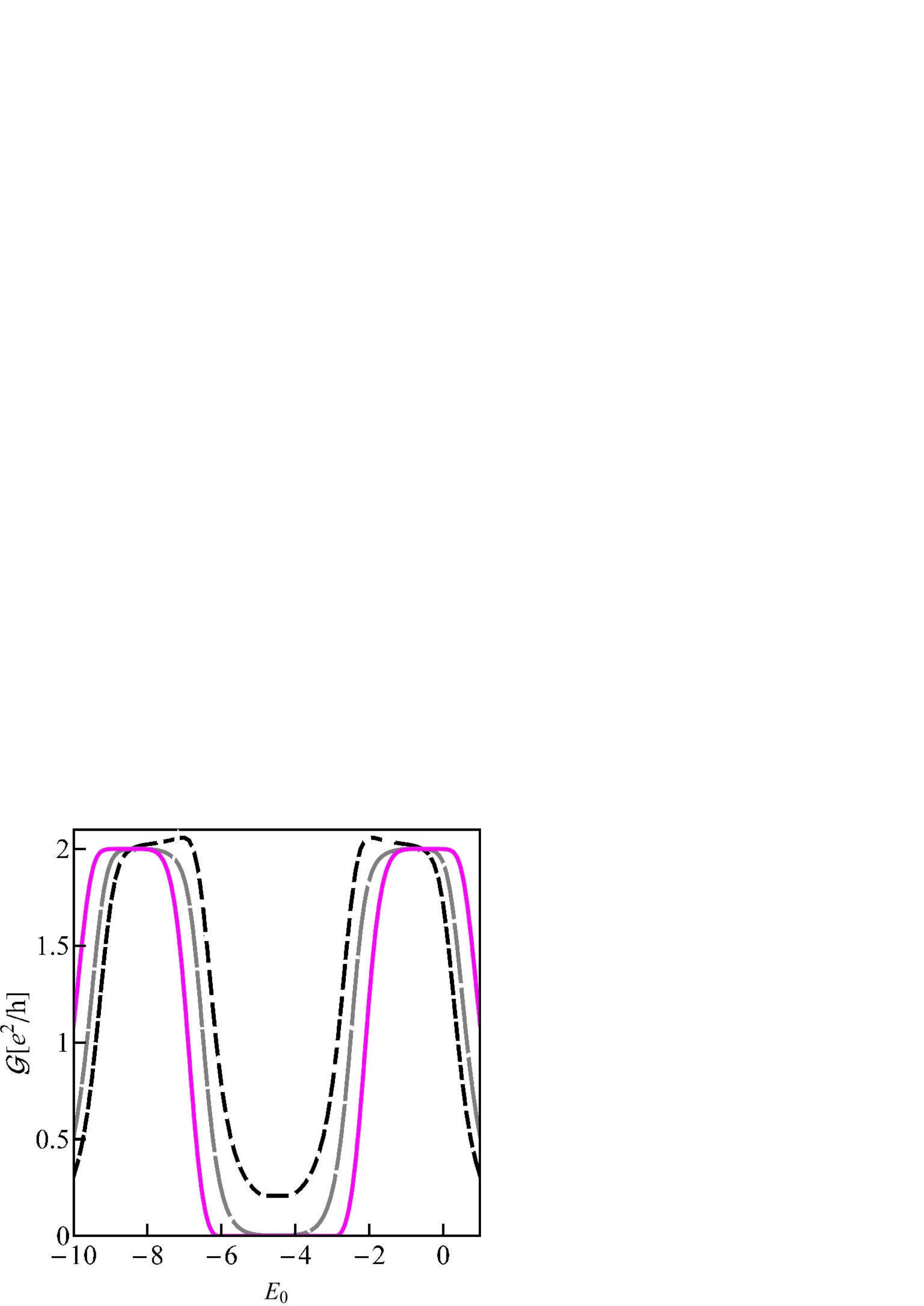}
\caption{\label{fig3} (Color online) Direct valley mixing.  Conductances of CNTQD vs unperturbed dot energy level $E_{0}$ for $\Delta=0.5$ and valley mixing parameters: $\Delta_{KK'} = 0$ (short dashed line), $\Delta_{KK'} = 0.3$ (dashed) and $\Delta_{KK'} = 0.8$ (solid).}
\end{figure}
This can be caused by  scattering on  nonmagnetic impurities, structure defects or deformations of structure e.g. by putting CNT on a rough substrate.  Valley mixing can be modeled by $\sum_{m}\Delta_{m,-m}c^{\dagger}_{m\sigma}c_{-m\sigma}$. Spin-orbit  interaction reduces degeneration, but does not change the quantum numbers, valley mixing preserves only spin.  Due to the time inversion symmetry double degeneracy is preserved in odd valleys. Splitting between the doublets caused by SO interaction and valley mixing is given by  $\Delta_{eff}=\sqrt{\Delta^{2}+\Delta_{KK'}^{2}}$.
Fig. 3 shows the  gate voltage dependence of conductance for fixed value of SO coupling presented for several values of  $\Delta_{KK'}$. Valley mixing enhances the  SO destruction of  Kondo correlations  at half filling. The borders of odd occupation  valleys and Kondo conductance  plateaus move symmetrically  away from shell e-h symmetry point with the increase of  $\Delta_{KK'}$. Since valley scattering has been  detected  also in ultraclean  nanotubes \cite{Kuemmeth,Schmid,Jespersen,Churchill}, it is worth considering also the role of other mechanisms that cause valley mixing. One of them  is local Coulomb scattering, which we will discuss in  section D. Also electrical contacts can induce valley scattering due to valley mixing during tunneling.
\begin{figure}
\includegraphics[width=0.48\linewidth,bb=0 0 439 468,clip]{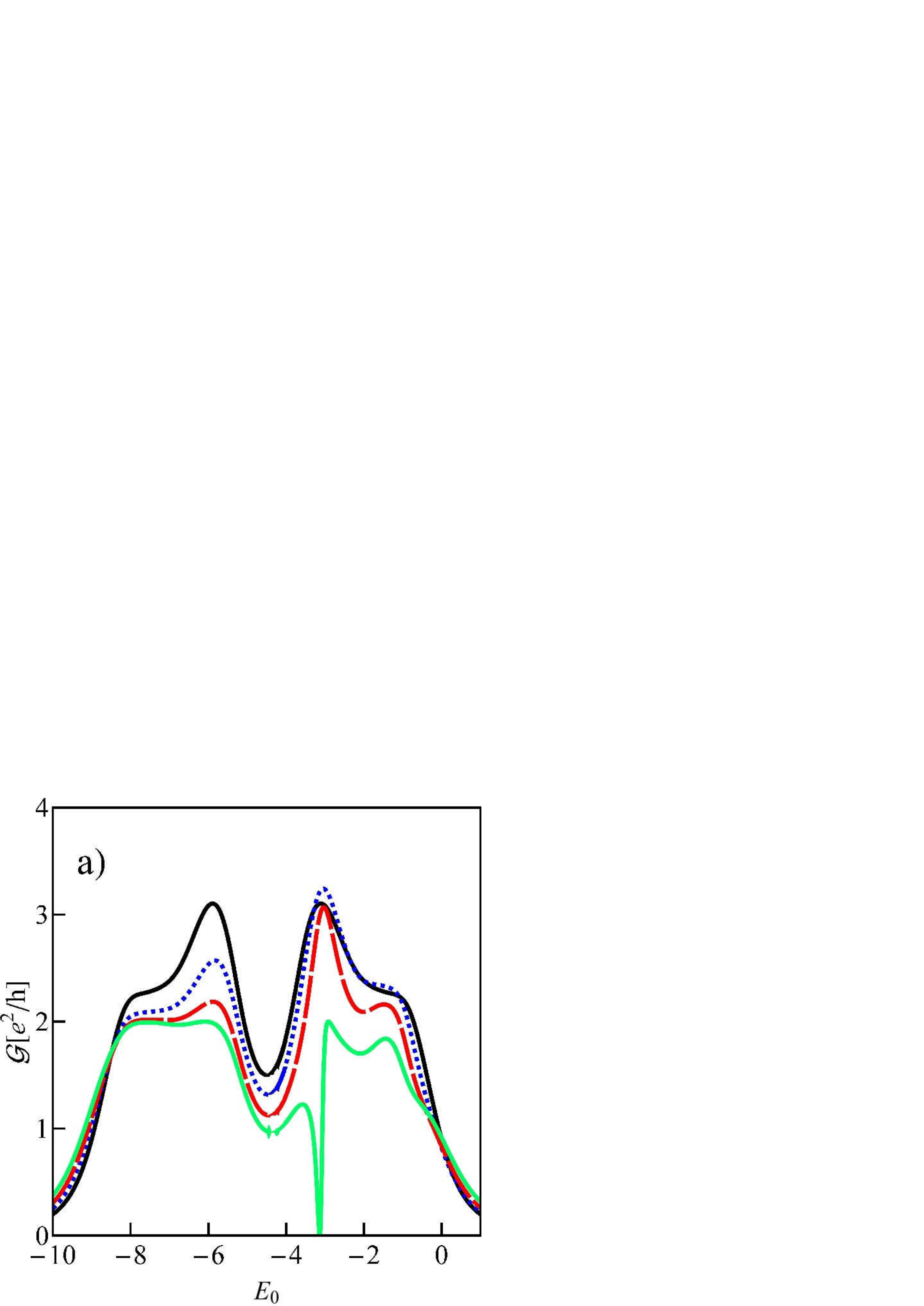}
\includegraphics[width=0.48\linewidth,bb=0 0 439 408,clip]{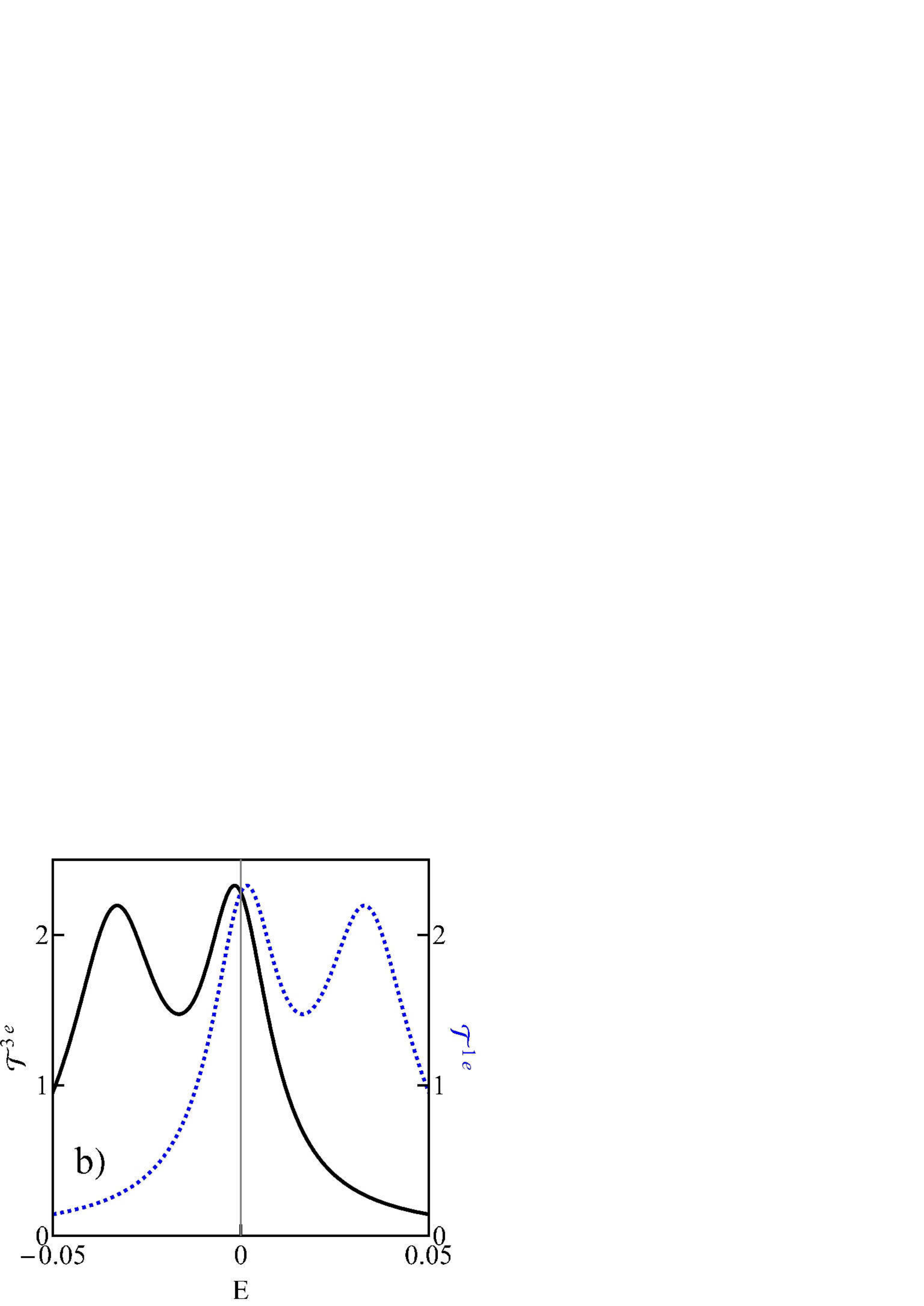}
\includegraphics[width=0.48\linewidth,bb=0 0 439 408,clip]{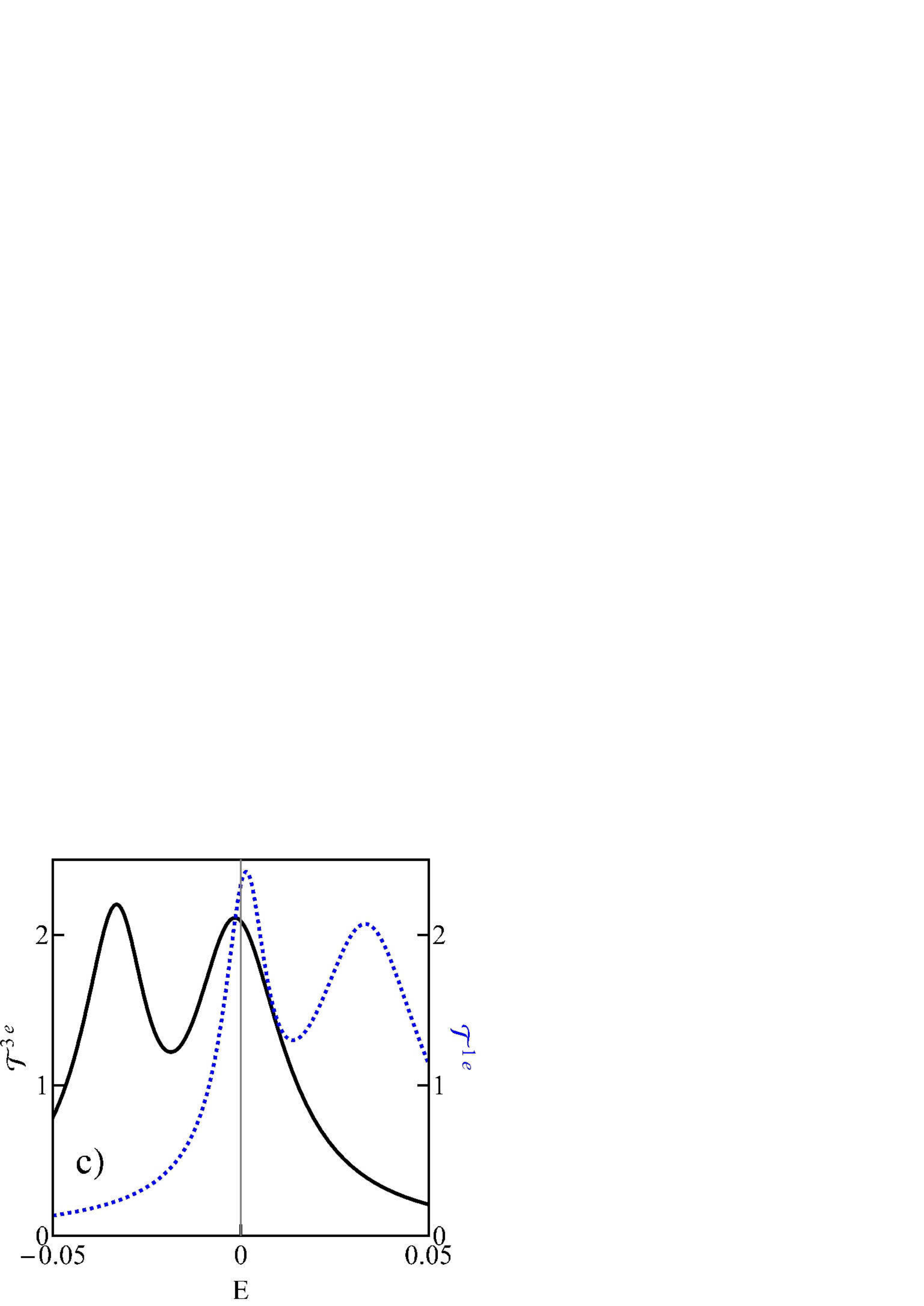}
\includegraphics[width=0.48\linewidth,bb=0 0 439 408,clip]{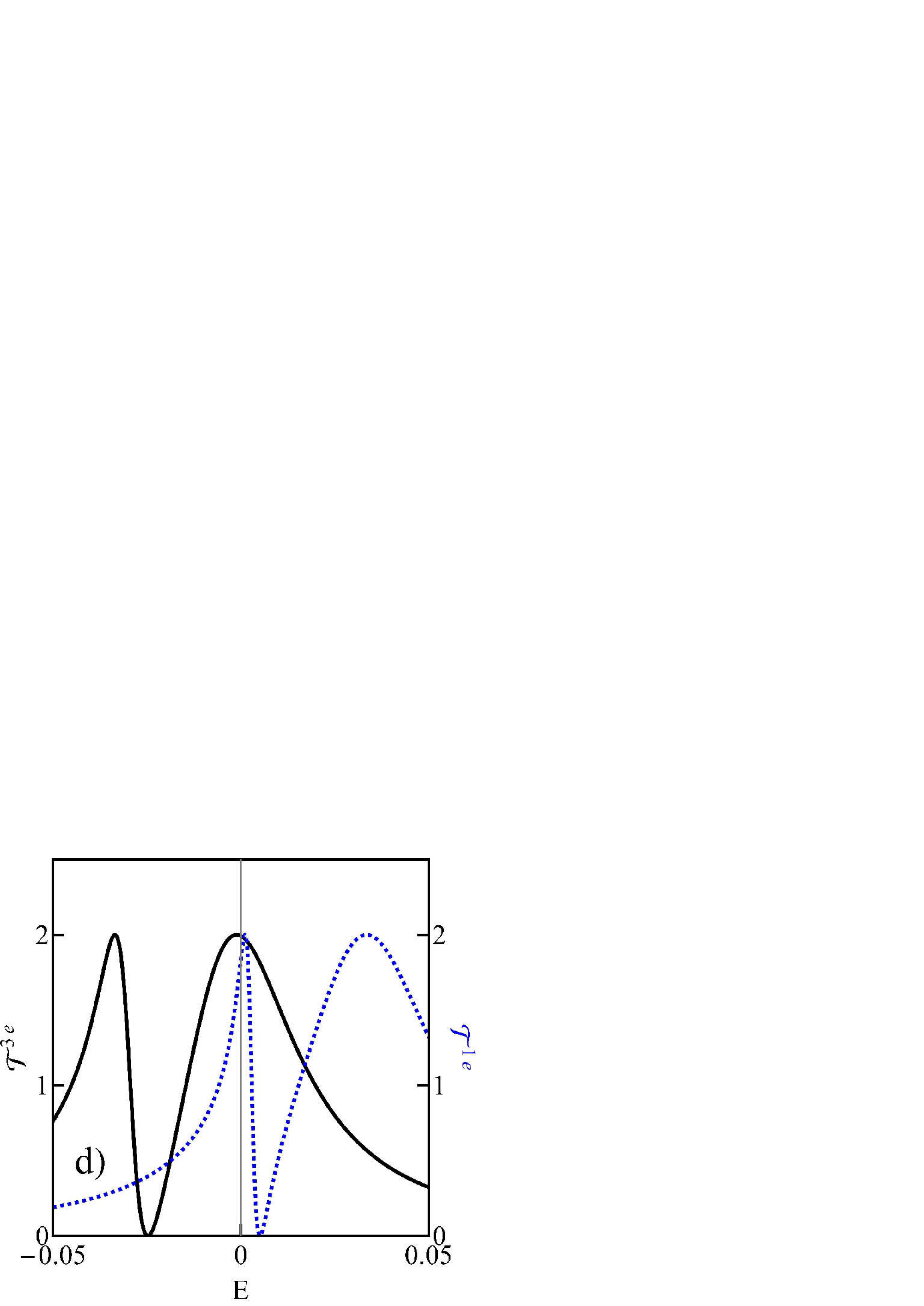}
\includegraphics[width=0.48\linewidth,bb=0 0 439 455,clip]{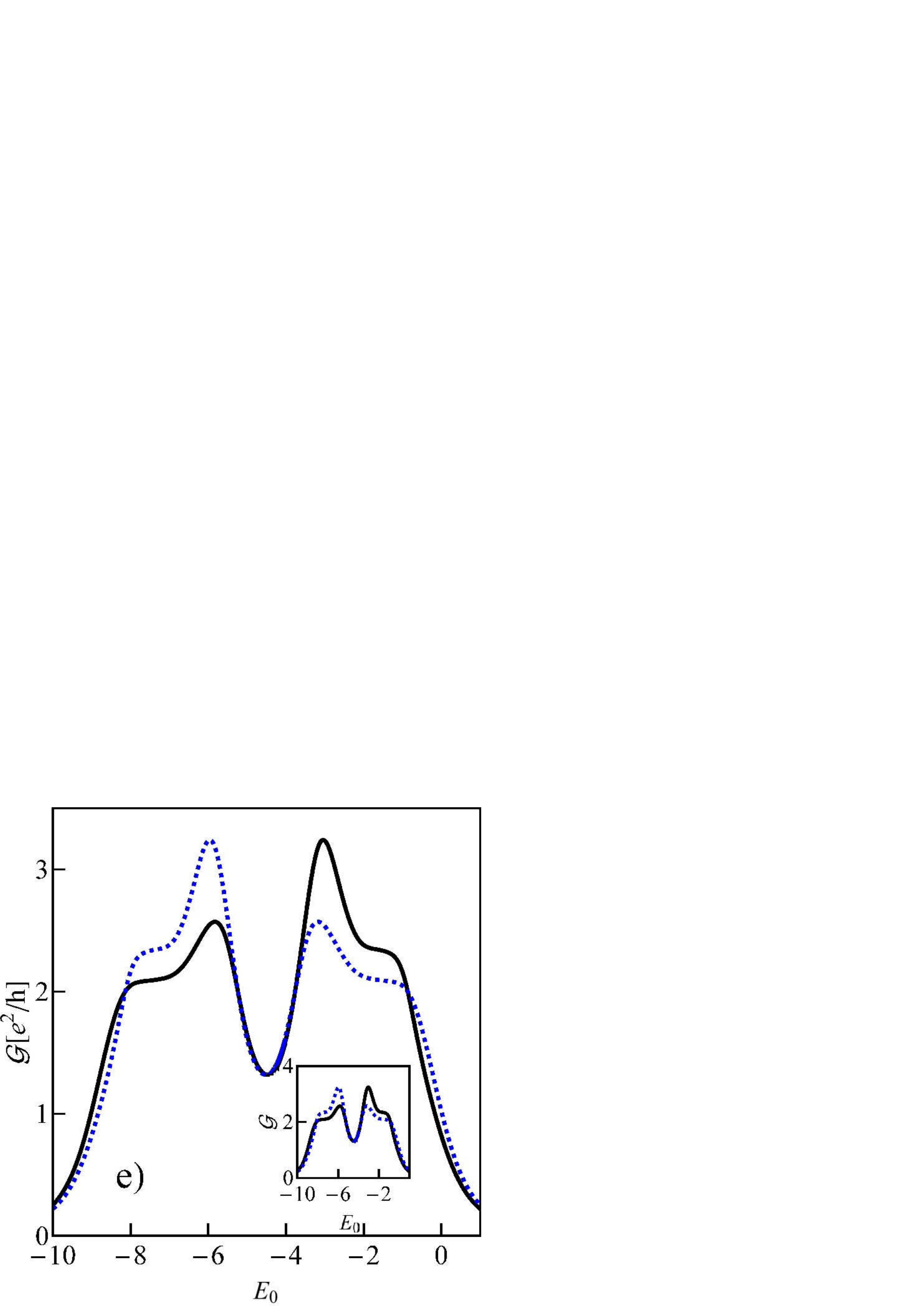}
\includegraphics[width=0.48\linewidth,bb=0 0 439 468,clip]{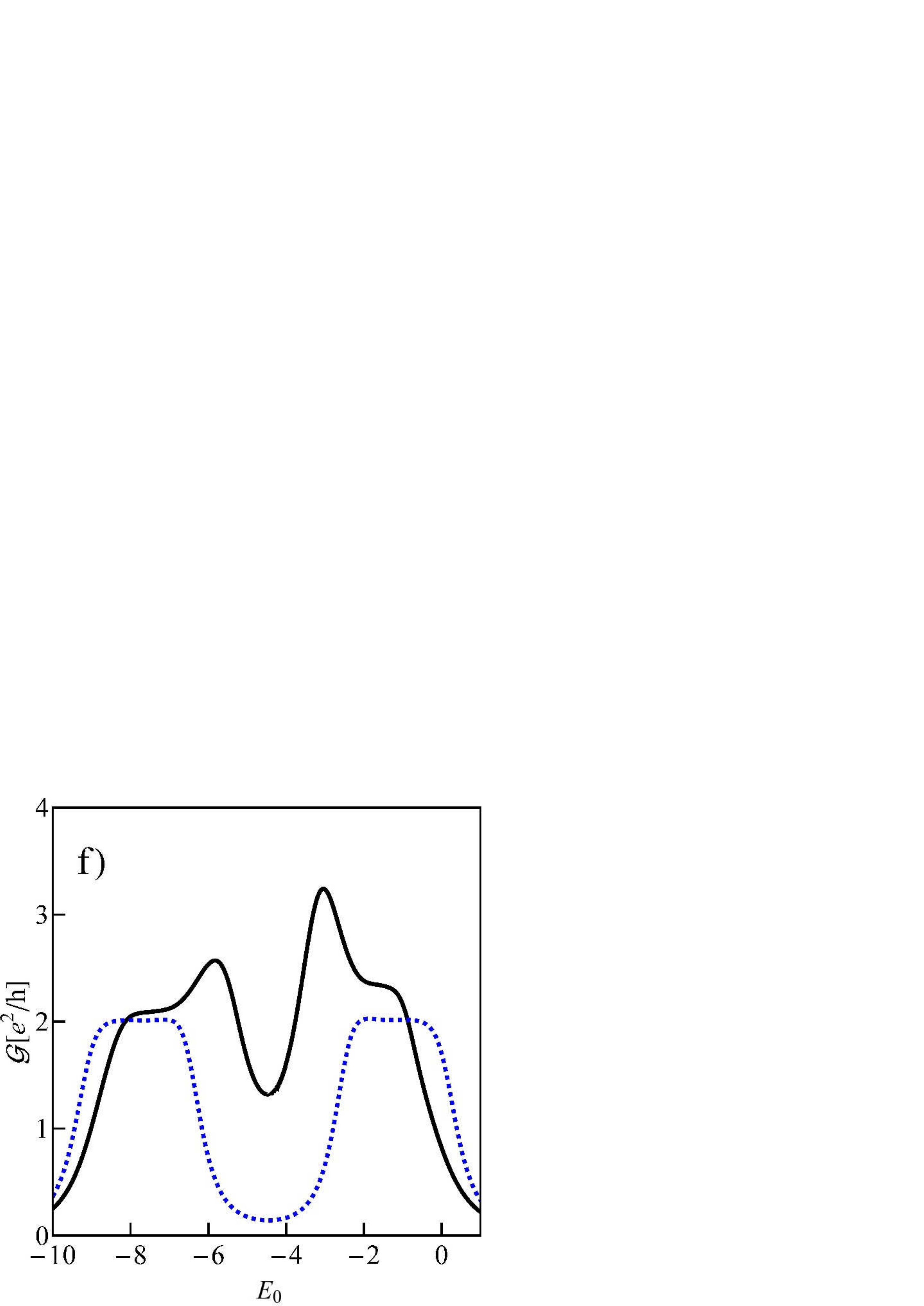}
\caption{\label{fig4} (Color online) Indirect valley mixing.  (a) Conductance of CNTQD disturbed by direct and indirect valley mixing, $\Delta =0.1$ , $\Delta_{KK'}=0.01$ and $q = 0$ (solid black line), $q =1/2$  (dotted blue), $q =3/4$ (dashed red), $q=1$  (solid green). (b,c,d) Transmissions for $N = 1$ (dotted line) and $N = 3$ (solid line) $\Delta =0.1$ , $\Delta_{KK'}=0.01$  and $q = 0$ (b), $q =1/2$   (c), $q =1$  (d). (e)  Intershell e-h asymmetry: conductance for $\Delta=0.1$   and $\Delta_{KK'} =0.01$, $q =-1/2$  (dotted blue line), $\Delta_{KK'} =0.01$  , $q =1/2$   (solid).  Inset presents conductances for $\Delta_{KK'} =-0.01$, $q =1/2$  (dotted blue line), $\Delta_{KK'}=0.01$, $q=1/2$   (solid). (f) Comparison of conductances for $\Delta<T_{K}$ ($\Delta=0.1$ , solid line) with the case  $\Delta\gg T_{K}$ ($\Delta=1$, dotted line),  $\Delta_{KK'}=0.01$, $q=1/2$.}
\end{figure}
Commonly accepted conviction on valley conservation during tunneling, which is also adopted in this work, is based on the view, that the leads to the dot are formed within the same nanotube as the dot,  or that electrons from   metallic leads  enter the nanotube segment before tunneling. If, however, these requirements are not fully met,  some mixing in the orbital channels may occur. The effect of contacts which may mix orbital numbers is thoroughly discussed in  \cite{Schmid,Jespersen}. Indirect valley scattering  via the electron states of electrodes can be represented by the off diagonal terms of  electrode-dot coupling matrix  $\Gamma_{m,-m}$, and   following \cite{Kubo,Karwacki} they can be approximated by $\Gamma_{m,-m}=iq\Gamma$.
The processes represented by off diagonal terms  of $\Gamma$ result from   various interference effects and are the consequence of indirect  transitions  between  dot orbitals by states in the electrodes. Parameter $q$ describes the strength of this indirect mixing, in general $q$ can be a complex number, but in our considerations is assumed  real  $|q|<1$. In the case discussed at present, besides the cotunneling processes, which  preserve valley also cotunneling, which flips  isospin comes into play in formation of many-body resonances. The partially  separated peaks of  transmissions presented in Figures 4b, c, d correspond  to resonances set up primarily by   cotunneling processes within only one of the Kramers doublets for each of the resonances, although  for small SO splitting  also the states from the second doublet play some role. The peaks located around $E_{F}$  correspond to Kondo like resonances.  In $1e$ valley the Kondo peak is lower in energy than the peak associated  with tunneling into the second doublet and in $3e$  valley the opposite energy ordering of the peaks occurs. Due to the electrode mediated destructive interference between  the Kondo state and the renormalized states from the second doublet the asymmetric Fano resonance is formed  between the peaks. As it is seen Fano resonance differently perturbs Kondo peak in $1e$ and $3e$ valleys and this results in intrashell e-h asymmetry ($1e-3e$) of conductance  (Fig. 4a). Formally changing the sign of one of the parameters either $\Delta_{KK'}$ or $q$ on the opposite results in reversing of  the asymmetry, conductances of  $1e$ and $3e$ valleys change the role. Similar interference induced asymmetry  effects have been earlier reported e.g.  for double dot  Kondo systems \cite{Karwacki}. In carbon nanotubes the single shell  e-h  conductance asymmetry has been observed in many systems \cite{Jarillo,Schmid,Makarovski3}. Of course experimentally observed conductance asymmetry between $1e$ and $3e$ valleys   can be also  caused by many other reasons, which we do not discuss here, e.g. asymmetry in coupling of the  leads to  different  SO Kramers doublets.
Also the gate dependence of dot-lead coupling, or gate dependence of  spin-orbit interaction can introduce intrashell $1e-3e$  asymmetry, but for the narrow energy interval of the single shell, the latter effects are expected to be of minor importance at least for wide gap tubes. As we discuss in section E they might  be of importance in nearly metallic nanotubes. In the limit of strong indirect intervalley mixing ($q =1$) the destructive interference leads to a zero transmission dip  (Fig. 4d). For  certain values of gate voltage this  dip locates  at the Fermi energy and  linear conductance vanishes in this case. The largest asymmetry caused by indirect valley mixing  is observed in the crossover  region between SU(4) and SU(2) symmetries, for $\Delta\gg T^{SU(4)}_{K}$ conductance again becomes approximately symmetric with respect to $E_{0} =-(3/2){\cal{U}}$ line  (Fig. 4f).

\subsection{Magnetic field}
Analysis of transport through strongly correlated CNTQD in magnetic field, which takes  into account SO coupling has been already discussed in  literature \cite{Steele,Schmid,Jespersen,Galpin,Mantelli}.
\begin{figure}
\includegraphics[width=0.48\linewidth,bb=0 0 439 430,clip]{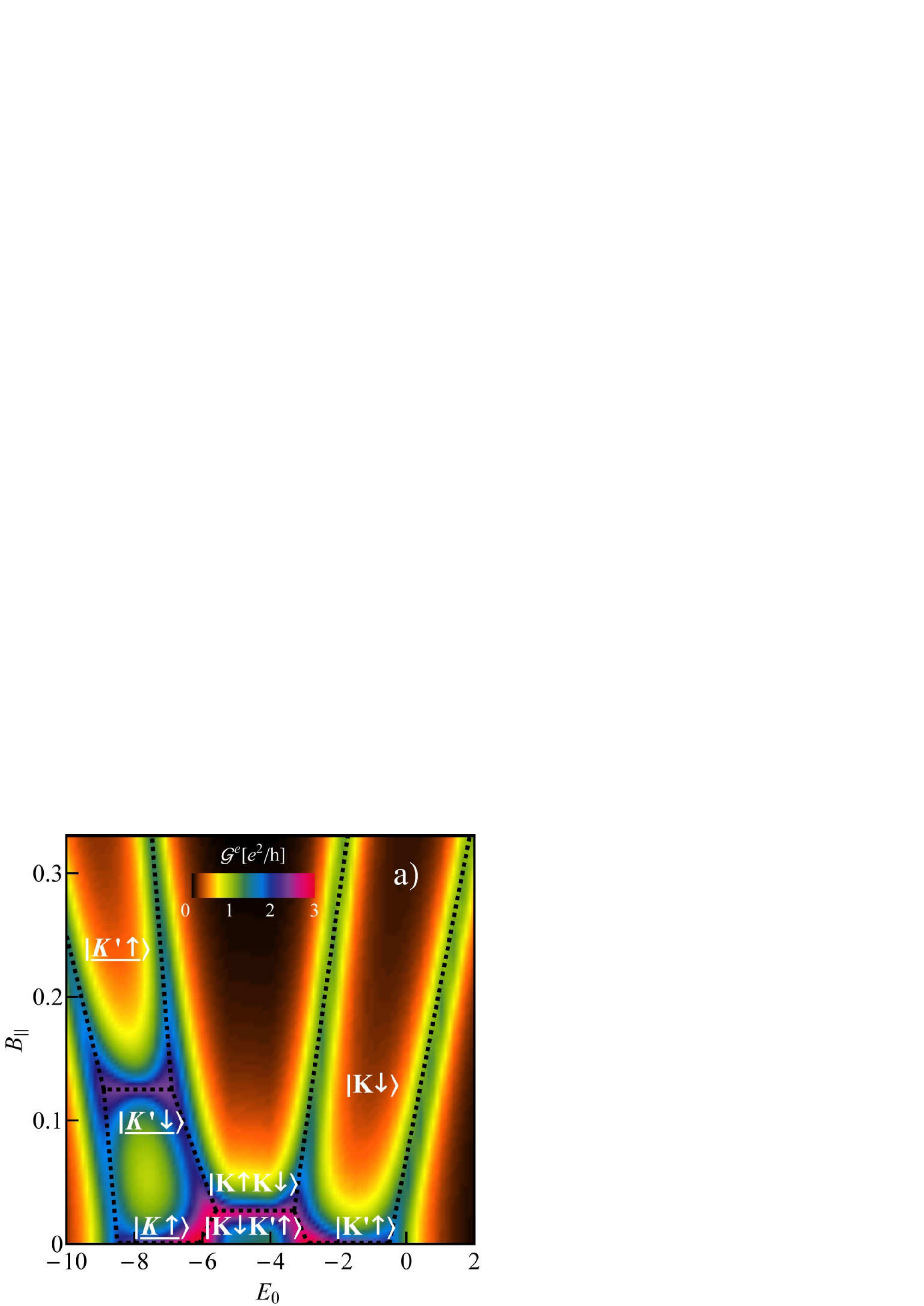}
\includegraphics[width=0.48\linewidth,bb=0 0 439 430,clip]{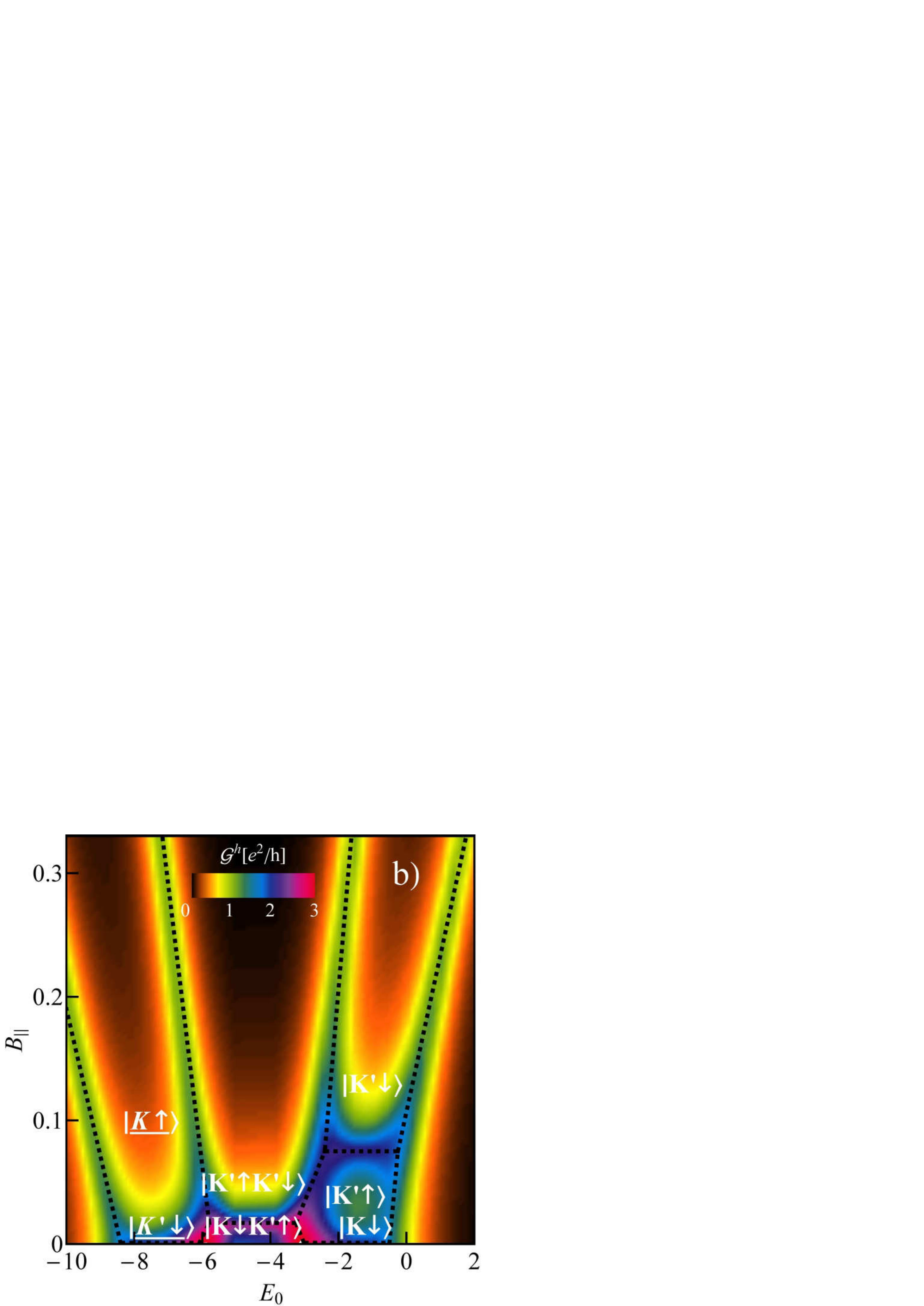}
\includegraphics[width=0.48\linewidth,bb=0 0 439 430,clip]{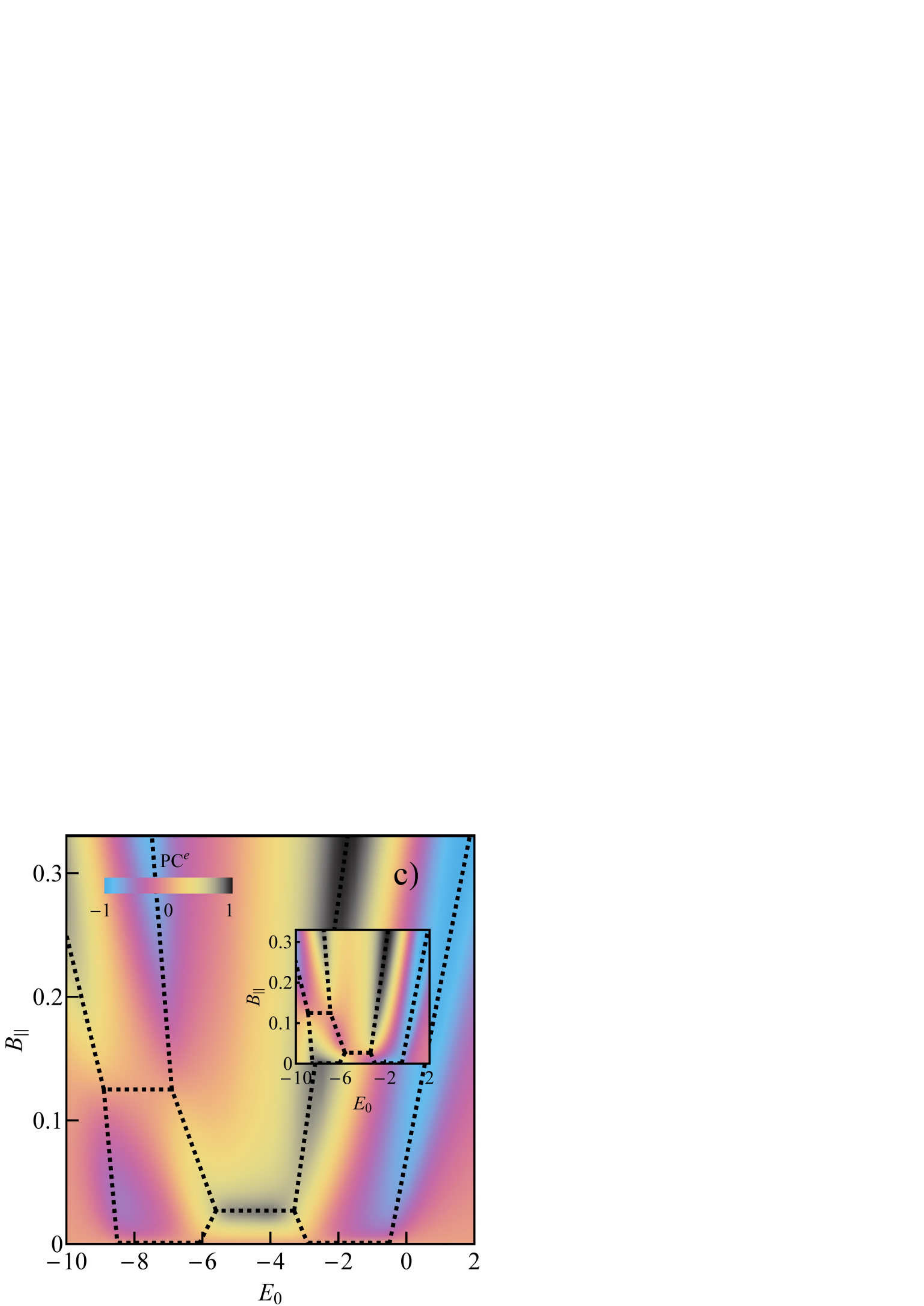}
\includegraphics[width=0.48\linewidth,bb=0 0 439 430,clip]{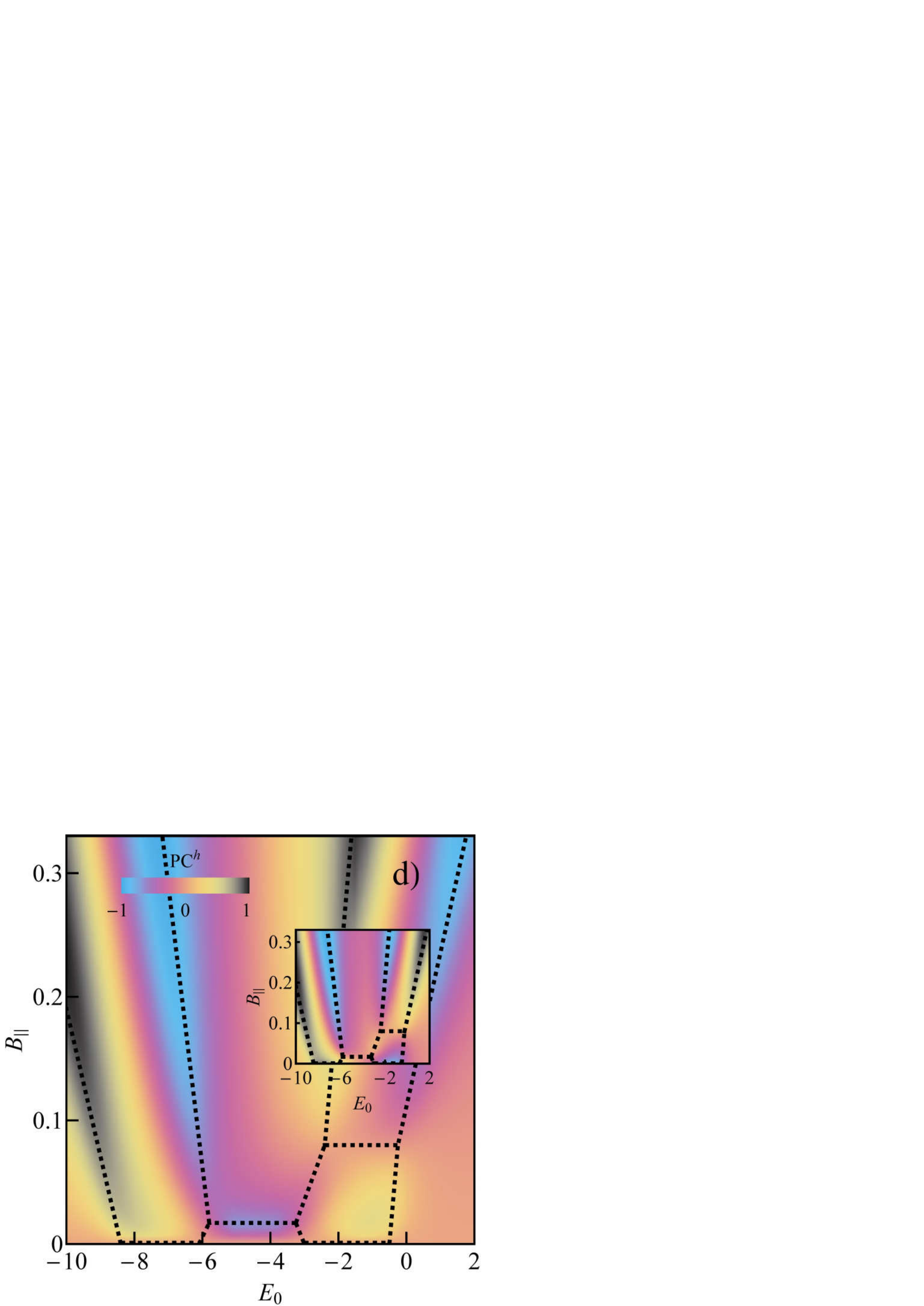}
\includegraphics[width=0.48\linewidth,bb=0 0 439 451,clip]{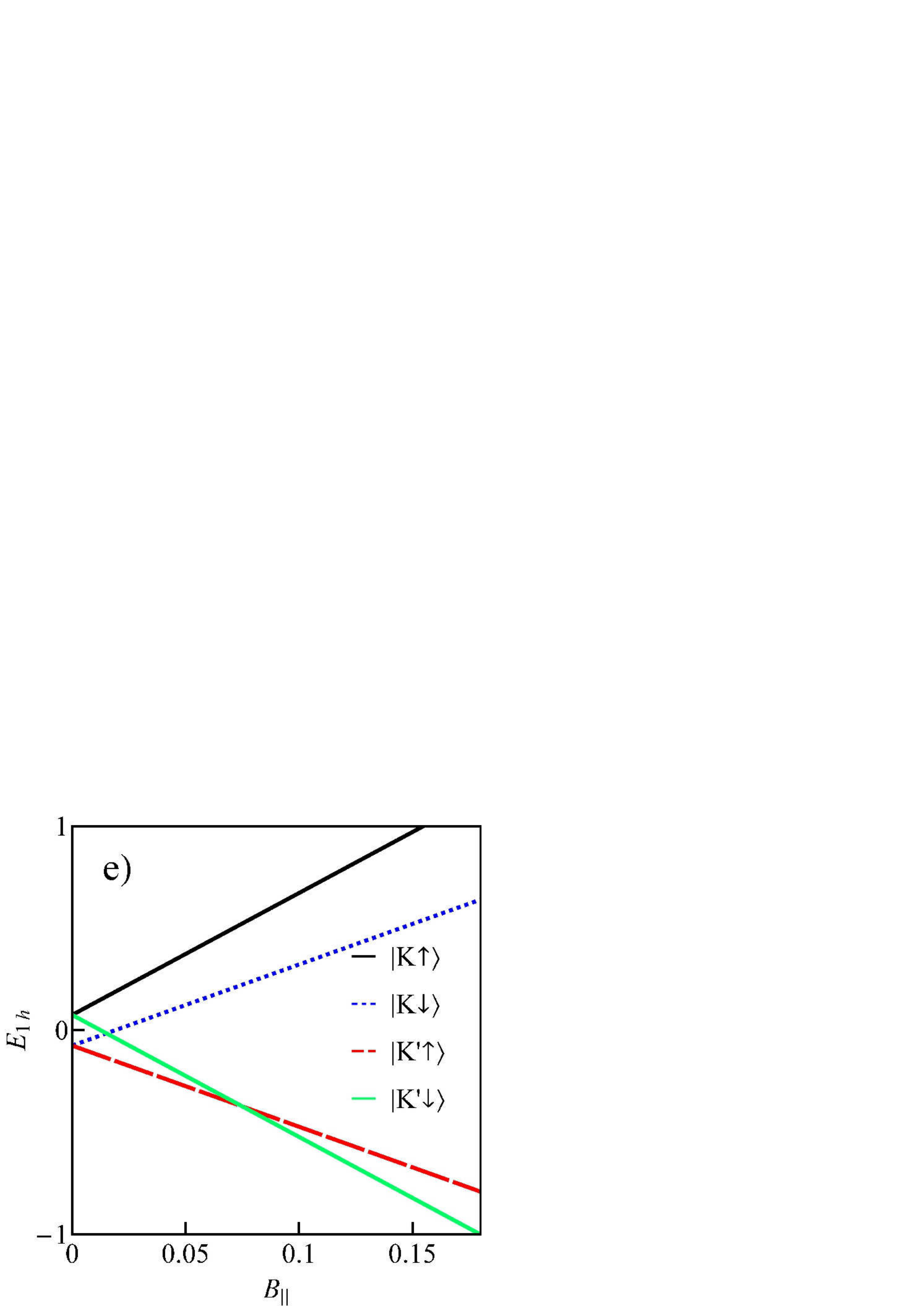}
\includegraphics[width=0.48\linewidth,bb=0 0 439 469,clip]{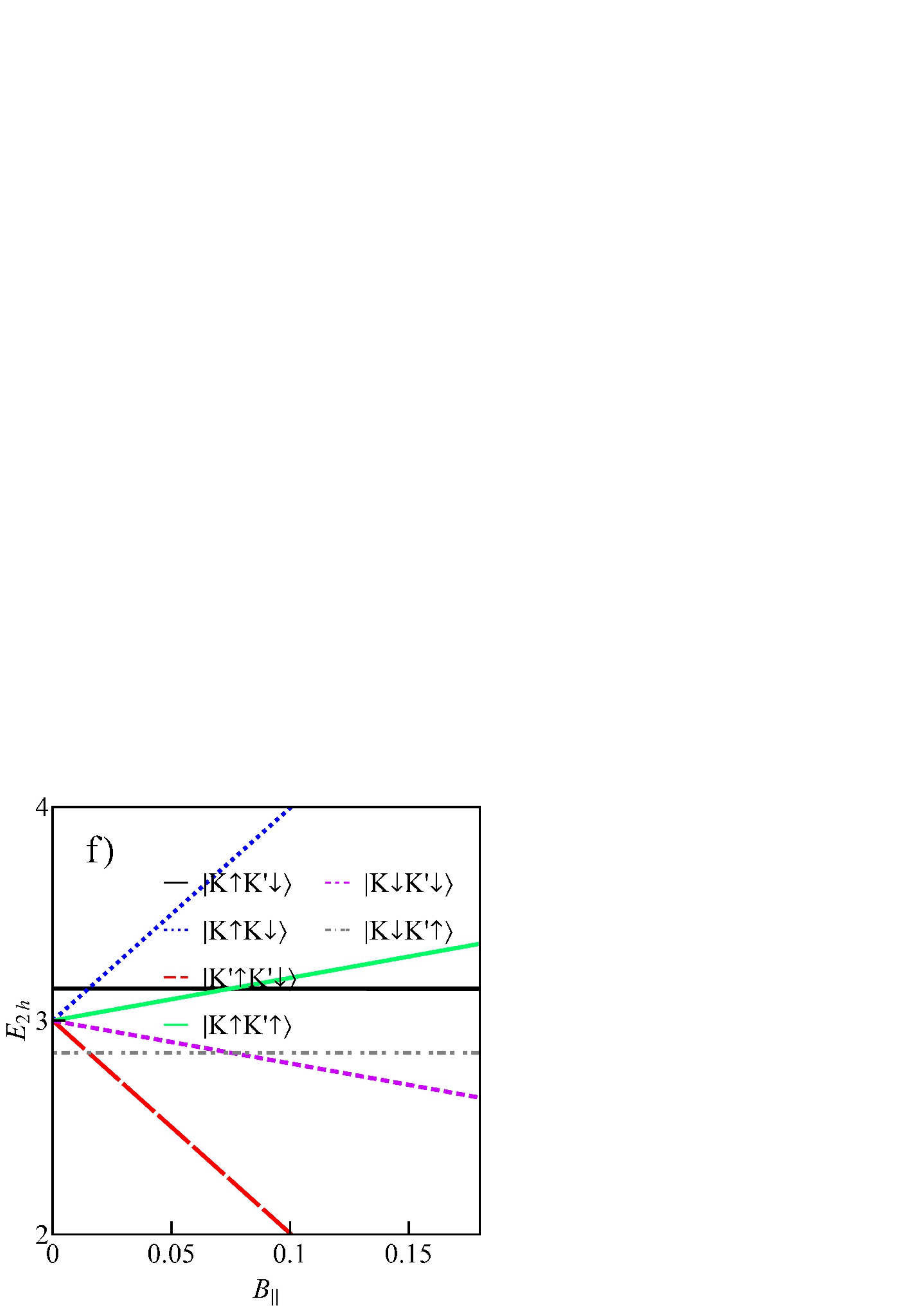}
\caption{\label{fig5} (Color online) CNTQD in parallel magnetic field ($\Delta^{1}_{so}=-0.01$, $\Delta^{0}_{so}=0.04$, $\mu_{o}/\mu_{s}=5$). (a) Electron  conductance map  ${\cal{G}}^{e}(E_{0},B_{\parallel})$ with the ground states stability regions on it. (b) The same as in (a) but for holes. (c) Spin polarization of electron conductance and valley conductance polarization in the inset (d) The same as in (c), but for holes. (e) Field dependence of single hole states. (f) Field dependence of two-hole states.}
\end{figure}
Here we only  supplement  the earlier reports by presenting  conductance map also for  holes and discussing polarization. In order to facilitate the latter discussion  we complement the conductance maps by the  ground state configuration diagram of the system (Figs. 5a, b).
Magnetic field breaks time-inversion symmetry, parallel magnetic field  splits the Kramers doublets in odd valleys and splits the quartet in $2e$ valley  by both spin and orbital Zeeman effects (Figs. 5e, f).  For magnetic field of energy exceeding Kondo energy the corresponding many-body  correlations are destroyed and as a consequence the  linear conductance is suppressed. Let us focus on the hole case, for electrons the picture is analogous, replaced is only the role of single occupied and triple occupied valleys. Fig. 5e shows single-hole energies. Hereafter presenting electron or hole energies we use the shifted energies defined by $E_{1e(1h)}(m\sigma)=E^{\pm}_{m\sigma}-E^{\pm}_{0}$, where $E^{\pm}_{0}=\pm\sqrt{E^{2}_{g}+E^{2}_{0}}$. As it is seen in Fig. 5e a competition of  SO interaction and Zeeman effect in $1h$ valley leads to level crossing in the ground state. For the field $B_{s}=\Delta^{h}/g\mu_{s}$ ($N=1$) (Fig. 5e), the energy of the  state $|K'\uparrow\rangle$ is crossed by energy line of  one of the  states from excited Kramers doublet $|K'\downarrow\rangle$  i.e. degeneracy is recovered.
\begin{figure}
\includegraphics[width=0.6\linewidth,bb=0 0 439 427,clip]{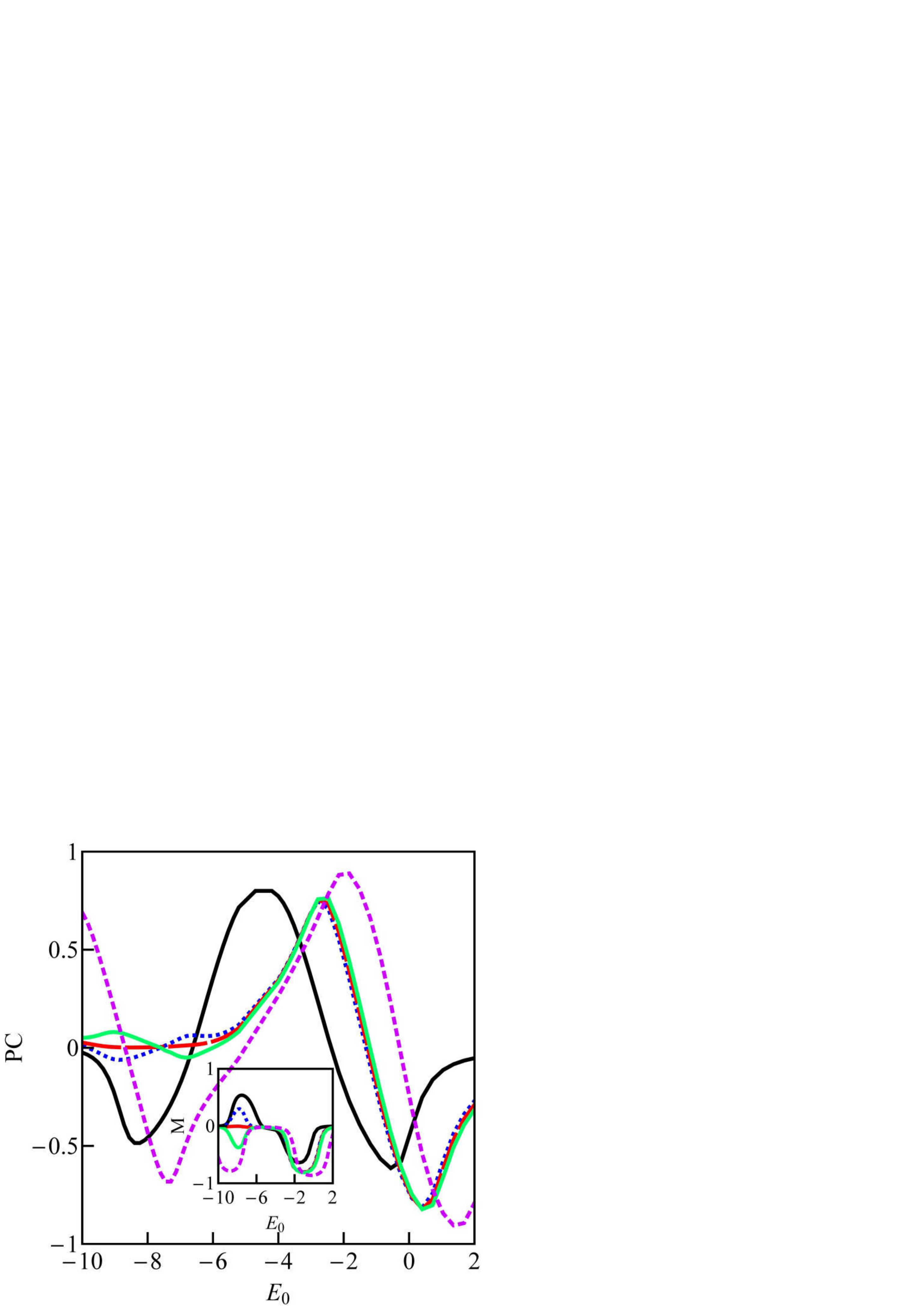}
\caption{\label{fig6} (Color online) Spin polarization of conductance for the chosen values of magnetic field: $B_{\parallel} = 0.025$  ($B_{o}$) (solid black line), $B_{\parallel} = 0.118$ (dotted blue), $B_{\parallel} = 0.125$ ($B_{s}$) (dashed red), $B_{\parallel} = 0.13$ (solid green), $B_{\parallel} = 0.278$ (short dashed purple). Inset presents magnetic moments at the dot for the same choice of the fields.}
\end{figure}
This allows revival of the  Kondo effect. Cotunneling induced  spin fluctuations are responsible for the creation of the many-body resonance in this case (spin SU(2) Kondo effect). Similar field induced  recovery of degeneracy  is observed in the $N=2$ sector (Fig. 5f), where at the  field  $B_{o}=\Delta^{h}/g\mu_{o}$ ground state $|K\uparrow K'\downarrow\rangle$  crosses $|K'\uparrow K'\downarrow\rangle$  state.
The corresponding quantum fluctuations leading to Kondo effect are now valley isospin fluctuations and spin is preserved (valley SU(2) Kondo effect). In the wide gap limit here discussed, $E^{0}_{G}\gg\Delta^{0}_{so}$, $\Delta^{1}_{so}$ and for $\Delta_{KK'}=0$ the Kondo lines of enhanced conductance are parallel to the  gate voltage axis, because the corresponding characteristic fields $B_{o}$  and  $B_{s}$ are determined by SO splitting alone. Comparing conductance maps for electrons and holes (Figs. 5a, b) it is seen that electron-hole symmetry around the band gap is broken.  The reason is that orbital-like contribution to SO coupling $\Delta^{1}_{so}$ contributes with opposite signs for electrons and holes. Looking at the areas of stability of states presented on Figures 5a,b one can  see that neighboring Coulomb blockade regions differ in  magnetic moments and this reveals by an increase of polarization of conductance at the borders of  Coulomb blockade regions (Figs. 5c, d). The corresponding lines for electrons and holes are characterized by polarizations of opposite signs. Conductance on the Kondo lines is unpolarized in odd valleys and polarized in even.  In odd valleys it is the  spin, which due to cotunneling processes, despite applied magnetic field,  freely fluctuates between two orientations and orbital pseudospin is fixed. Transport occurs  effectively only  in one orbital channel with equal probability for both spin orientations and thus conductance is spin unpolarized  and orbitally (valley) polarized in this case (compare a map of valley polarization presented in the insets of Figs. 5c, d).  For half filling on the other hand, orbital pseudospin fluctuates and spin orientation is fixed, what results in spin polarization of  conductance. The orientations of net  spin magnetic moments of Kondo active  doublets are opposite for  $2e$ and $2h$  occupations and correspondingly opposite are also  spin polarizations of conductance in these regions. Conductance is orbitally unpolarized in this case. Fig. 6 presenting spin polarization vs. gate voltage for  several values of magnetic field highlights the possible spintronic applications.  Polarization of conductance can be changed both by magnetic field and gate voltage. The former is a consequence of the field dependence of gate voltages at which the Coulomb borders occur  and electrical control is due to subsequent crossing  of different Coulomb borders with the change of  the gate.
Finally  let us look at the impact of  valley scattering on  transport characteristics in magnetic field. When both SO coupling and valley mixing are taken into account the field dependencies  of dot energies become nonlinear. The characteristic fields of  crossing of the states $B_{c}$  are modified by  valley mixing, for $\Delta_{KK'}=0$ $B_{c}=B_{s}=0.5$ (Fig. 7a) and for $\Delta_{KK'}=0.3$ $B_{c}=0.47$ (Figs. 7b, c). For strong mixing no ground state crossing is observed (inset of Fig. 7c).  Field  induced SU(2) Kondo effect results from the  effective fluctuation of isospin characterizing Kramers doublet. The isospin in this case  is neither  the pure  spin nor valley pseudospin, but a combination of both and its screening does not mean quenching  of pure spin nor valley pseudospin.
\begin{figure}
\includegraphics[width=0.48\linewidth,bb=0 0 439 445,clip]{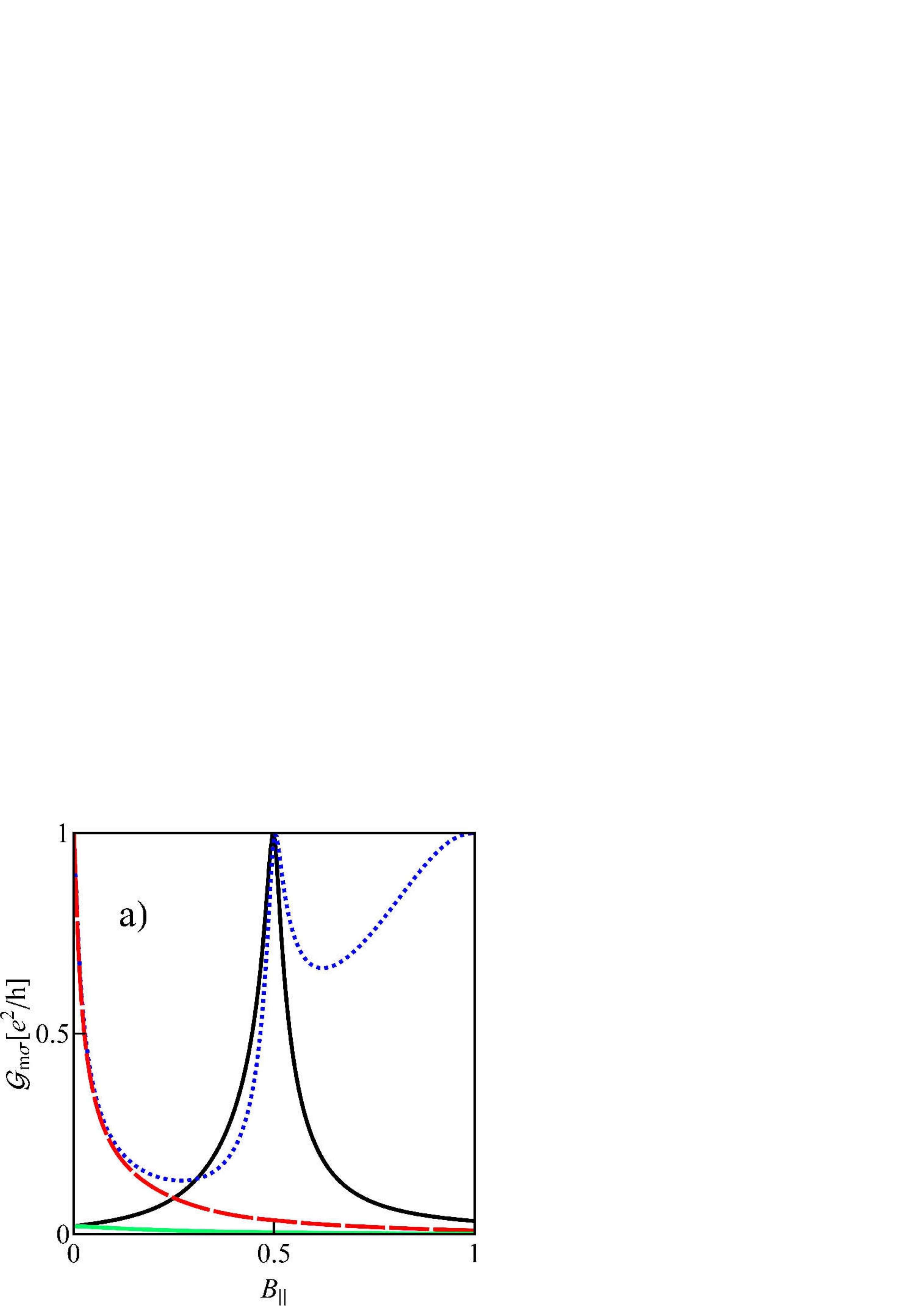}
\includegraphics[width=0.48\linewidth,bb=0 0 439 445,clip]{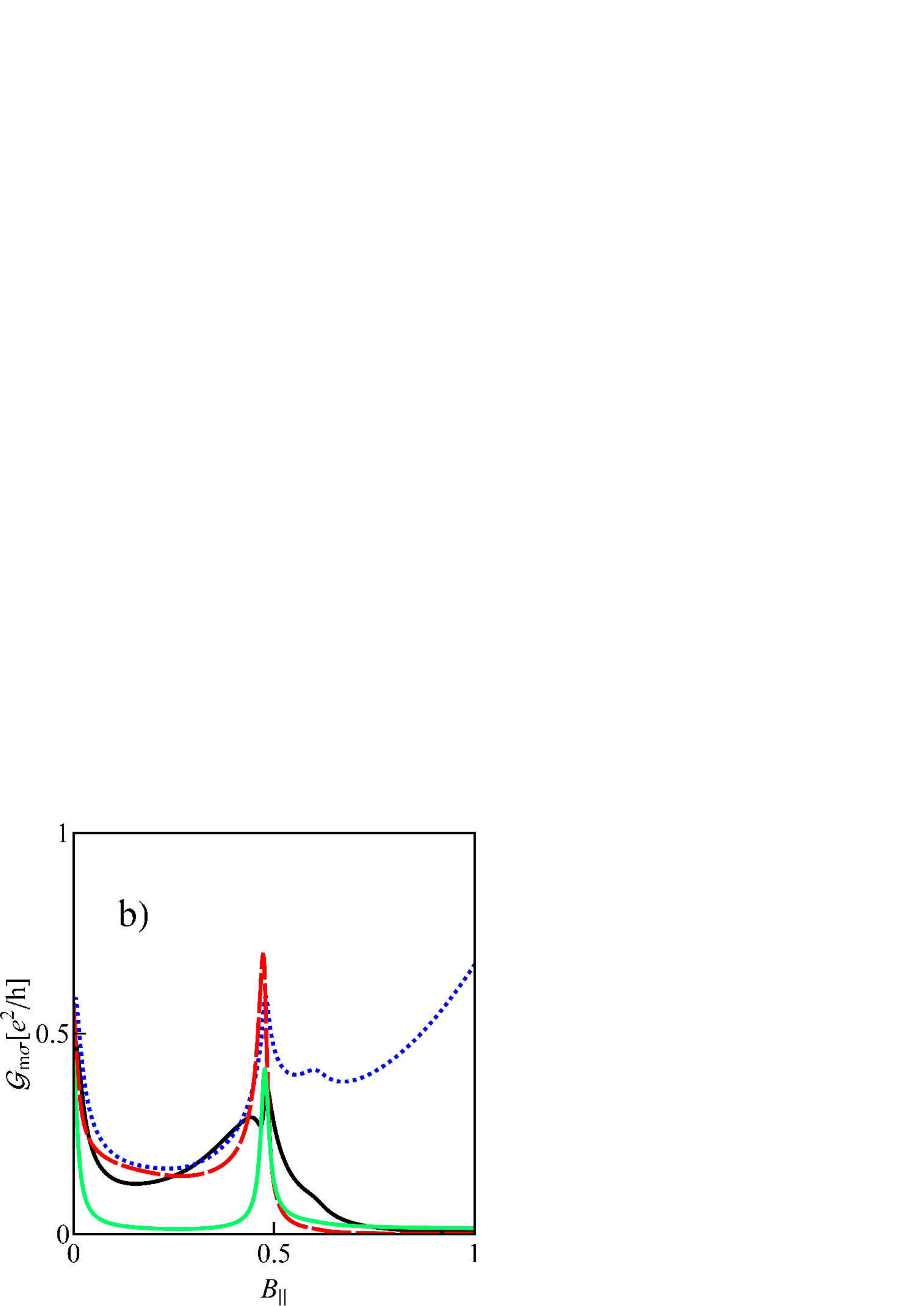}
\includegraphics[width=0.48\linewidth,bb=0 0 439 438,clip]{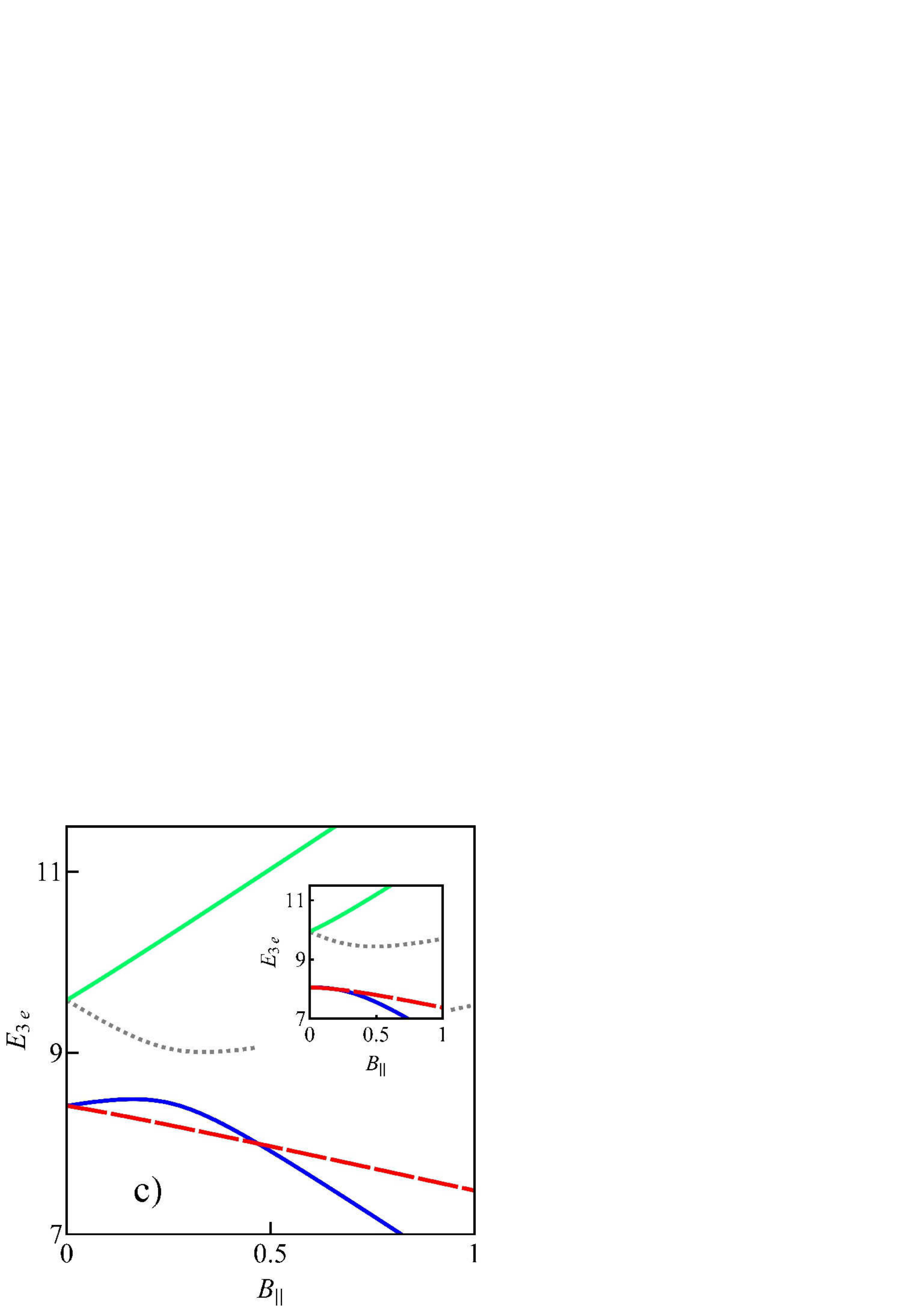}
\includegraphics[width=0.48\linewidth,bb=0 0 439 433,clip]{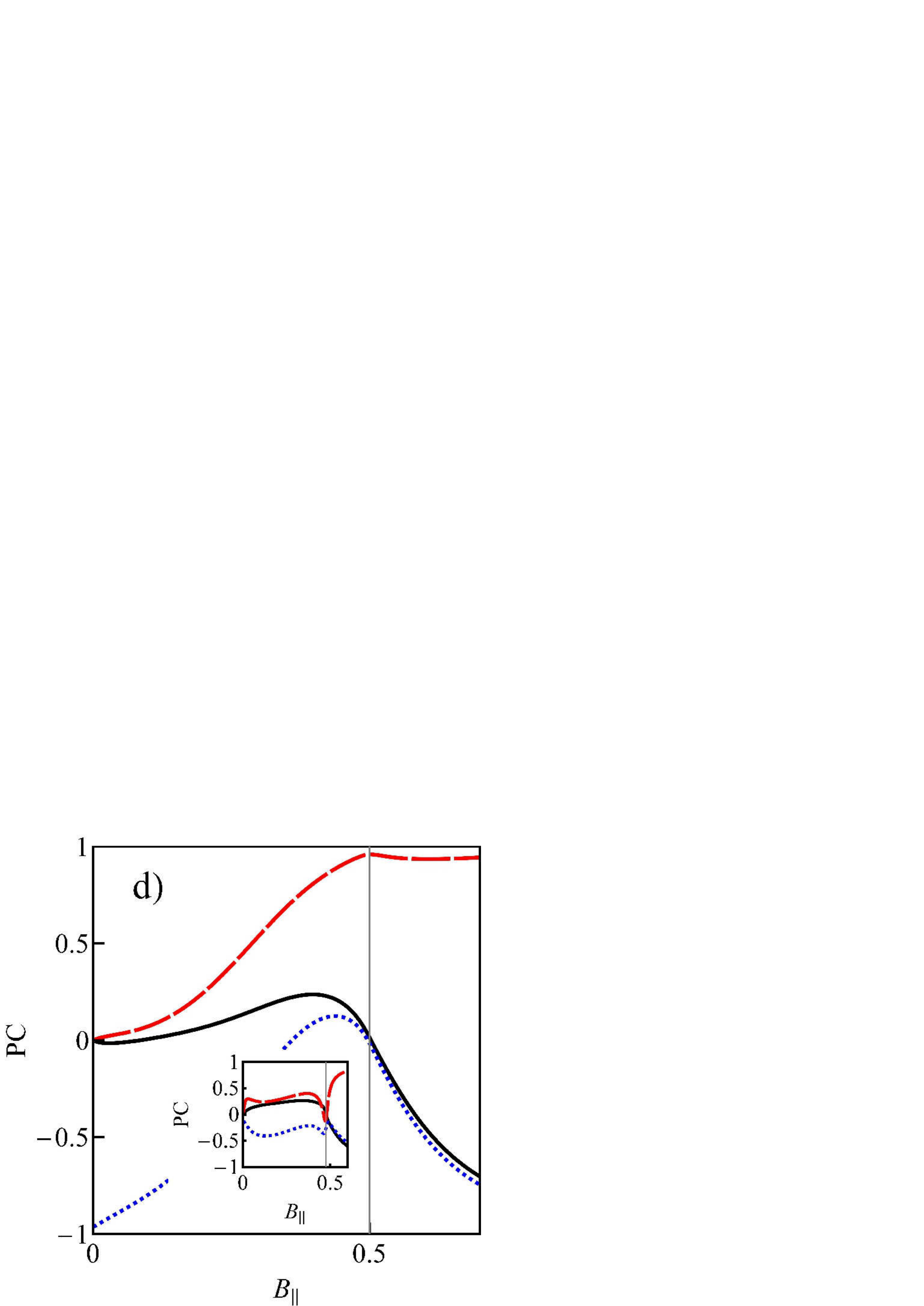}
\caption{\label{fig7} (Color online) Impact of valley scattering on field dependencies (a) Field dependencies of spin-orbital resolved conductances of CNTQD for $N = 3$. (a) $\Delta_{KK'} = 0$,  ${\cal{G}}_{1\uparrow}$   (solid black line) , ${\cal{G}}_{1\downarrow}$    (dotted blue) , ${\cal{G}}_{-1\uparrow}$    (dashed red) , ${\cal{G}}_{-1\downarrow}$    (solid green). (b) The same as in (a), but for $\Delta_{KK'} = 0.3$. (c) Field dependencies of three-electron states for $\Delta_{KK'} = 0.3$ and  for $\Delta_{KK'} = 0.8$   in the inset. (d) Polarizations of conductance for  $N = 3$: spin conductance polarization (solid line), valley polarization (broken), Kramers polarization (dotted) for $\Delta_{KK'}=0$ and for $\Delta_{KK'}=0.3$ in the inset.}
\end{figure}
It reflects in significant and almost equal contribution of all four  spin-orbital channels to total conductances, both at zero field and for  the field of revival of Kondo correlations (Fig. 7a). When valley mixing is absent and valley quantum number is preserved, only two  channels are opened (Fig. 7b). To elucidate the role of  spin and valley pseudospin in  Kondo fluctuations we present spin, valley and Kramers polarizations of conductance  for $\Delta_{KK'} = 0$ (Fig. 7d)  and in the inset for finite valley mixing.  For $B =0$  spin and valley polarizations vanish both for finite and zero valley mixing.   Kramers polarization is finite and indicates which Kramers doublet is active in Kondo processes, for $\Delta_{KK'}\neq0$  pure spin-orbital Kramers polarization considerably decreases. For the fields of Kondo revivals,  spin conductance polarizations vanish, but  Kondo states become valley polarized, opposite in two cases.

\subsection{Intervalley scattering induced by Coulomb interaction}
In section B we have discussed intervalley scattering  resulting from disorder on the scale of interatomic spacing and indirect valley  mixing, where interference processes with the electrode states played the essential role.   Here we analyze mixing between valleys caused by local part of electron-electron interaction.
The dominant part of Coulomb interaction  -  long range interaction is diagonal in valley and spin degrees of freedom. The corresponding scattering processes, called forward scattering (FS), are associated with  small momentum transfer. For short quantum dots also local interactions, which can exchange isospin come into play (backward scattering term VBS \cite{Ando2,Wunsch,Secchi,Pecker}).  This  term is described by $(1/2)V\sum_{\sigma\sigma'}(c^{\dagger}_{m\sigma}c^{\dagger}_{-m\sigma'}c_{m\sigma'}c_{-m\sigma'})$ \cite{Secchi}.
\begin{figure}
\includegraphics[width=0.48\linewidth,bb=0 0 439 450,clip]{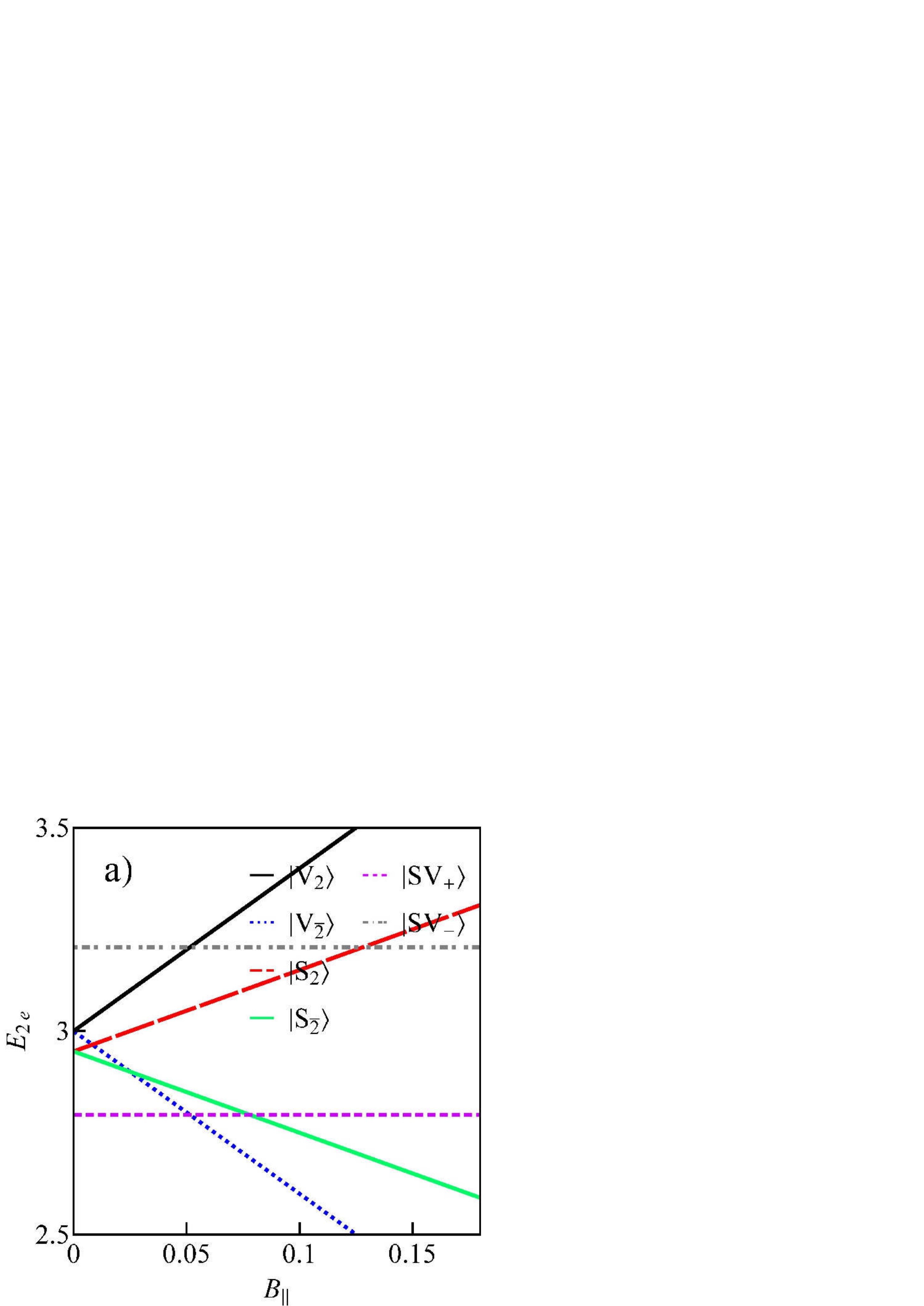}
\includegraphics[width=0.48\linewidth,bb=0 0 439 450,clip]{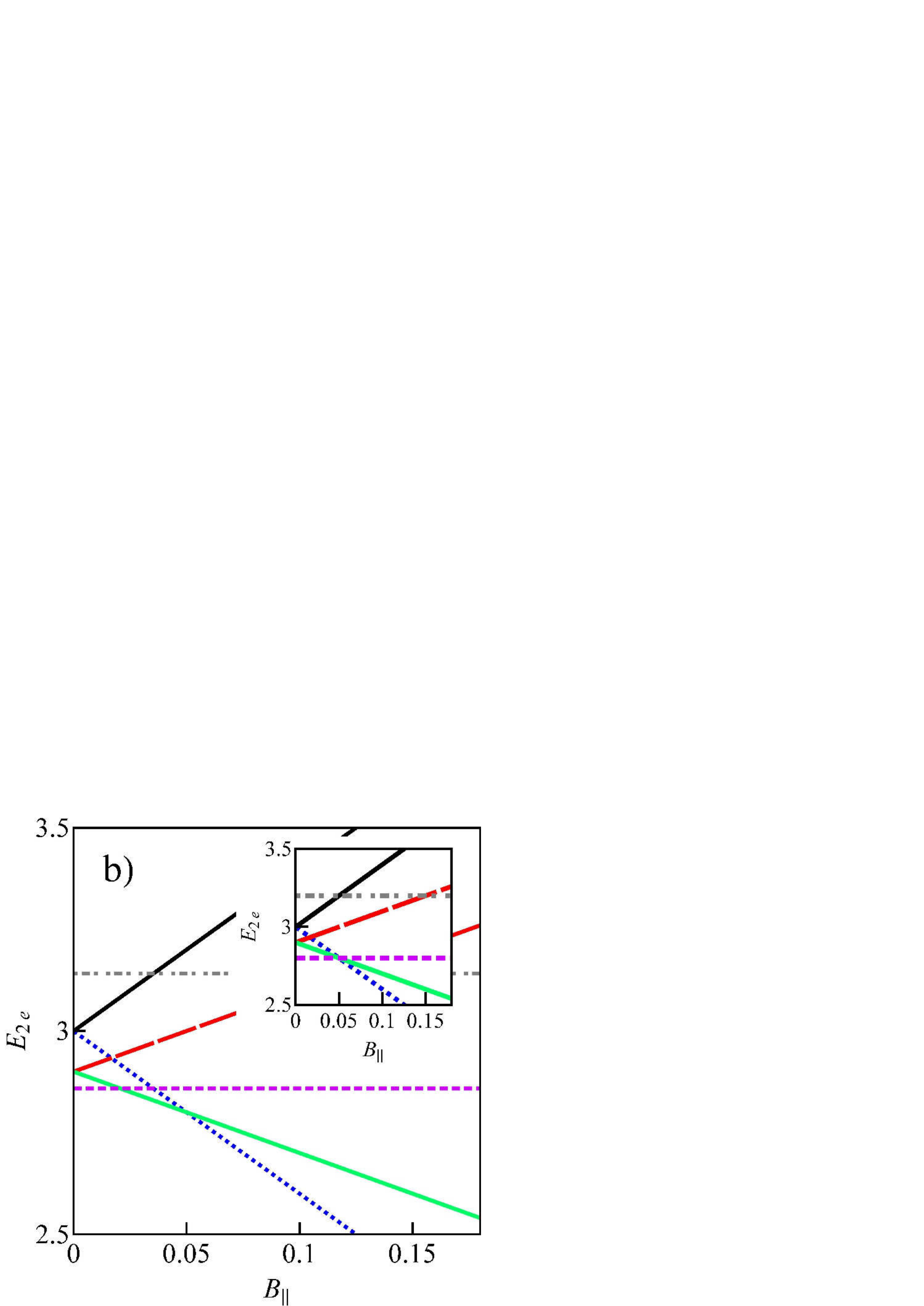}
\includegraphics[width=0.48\linewidth,bb=0 0 439 455,clip]{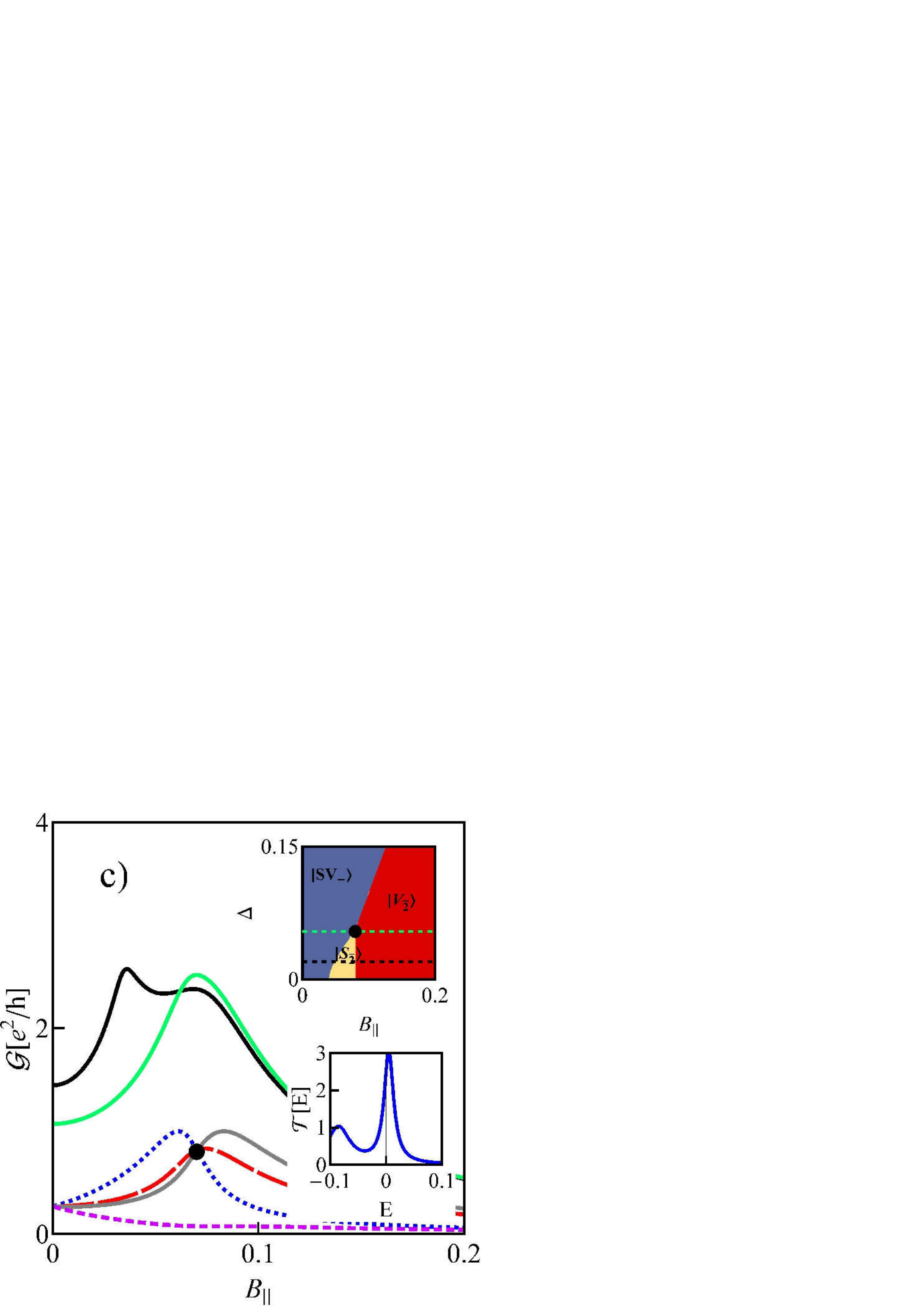}
\includegraphics[width=0.48\linewidth,bb=0 0 439 455,clip]{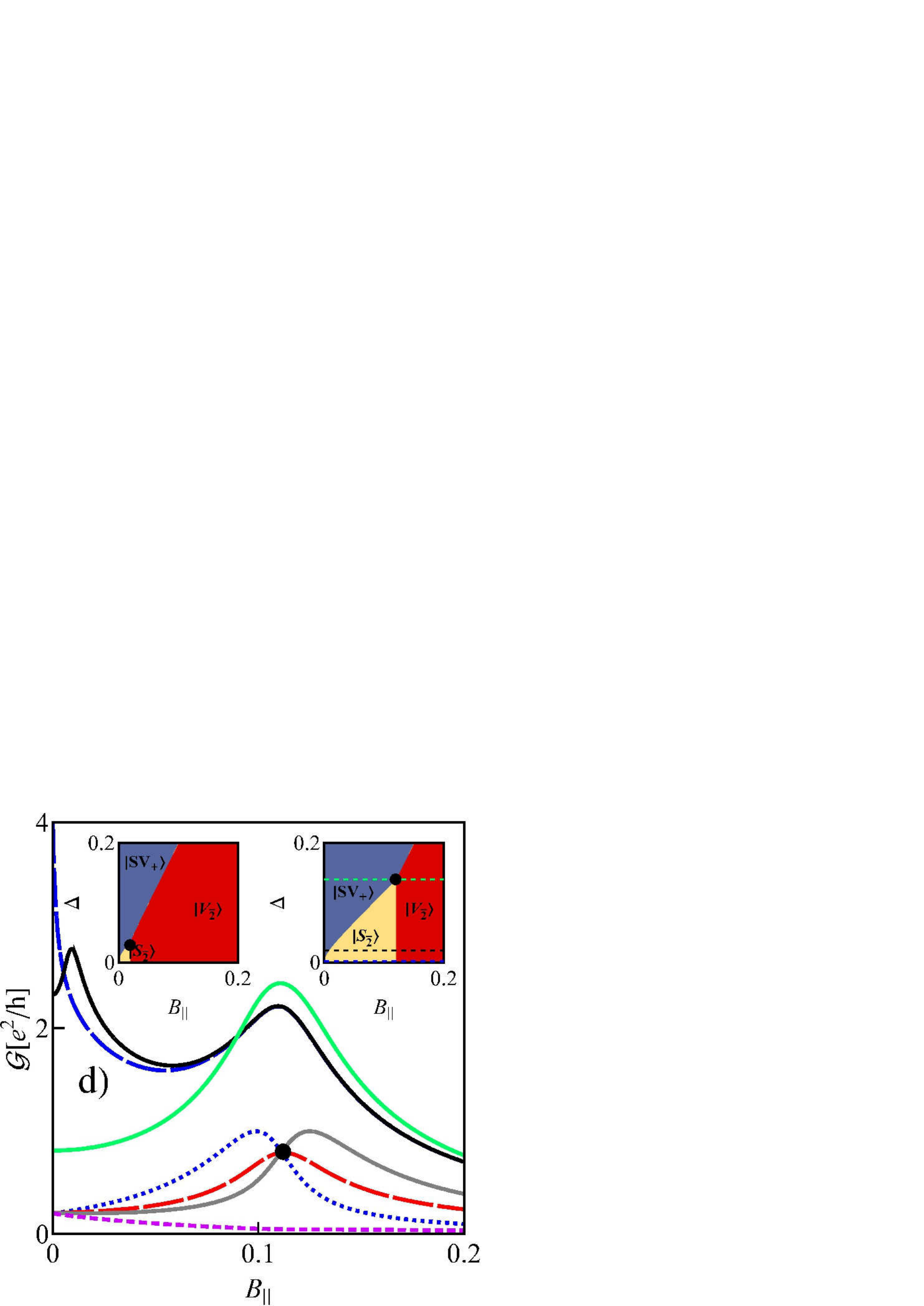}
\caption{\label{fig8} (Color online) Coulomb Intervalley scattering.  (a) Field dependencies of two-electron states of CNTQD  in the case of weak intervalley scattering ($V=0.05$, $\Delta=0.1$) and (b) for strong intervalley scattering ($V=0.1$, $\Delta=0.05$) (${\cal{U}}=3$ and ${\cal{U}}'=2.8$). Inset of Fig. b  illustrates the  field induced threefold degeneracy of two-electron states occurring for $V =0.1$ and $\Delta=0.087$. (c) Total conductance for $V=-0.04$ and $\Delta=0.02$ (solid black) and for $\Delta=0.053$ (solid green) with the spin-orbital resolved contributions (solid gray line ($-1\uparrow$), dashed red ($1\downarrow$), dotted blue ($1\uparrow$) and dashed magenta ($-1\downarrow$)). Upper inset shows the ground-state diagram. Dashed lines on the inset mark $\Delta$ values, for which the conductances are plotted and black point marks the threefold degeneracy. Lower inset on fig (c) presents transmission in the SU(3) Kondo state. (d) Conductance for $V=0.04$ $\Delta=0$ (dashed blue line) and $\Delta=0.02$ (solid black) and for $\Delta=0.127$ (solid green). Insets show the ground-state diagrams for (left: ${\cal{U}}={\cal{U}}'=3$) and (right: ${\cal{U}}=3$ and ${\cal{U}}'=2.8$). Lower curves present the spin-orbital resolved conductances for the SU(3) symmetry.}
\end{figure}
Isospin flip implies a momentum transfer of the order $1/a$ ($a$ - lattice constant) and it  occurs  when electrons have non-negligible probability of being closer than a distance a.  VBS term can be viewed as an effective intervalley exchange, since  it lifts the degeneracy of the spin polarized and valley polarized states.  The confirmation of this view was the  observation of huge enhancement of effective spin-orbit splitting at half filling of  hole shell \cite{Cleuziou}.  A similar observation was also reported for electrons \cite{Pecker}. The difference between odd and even filling of the nanotube multiplets   was attributed to the effective intervalley backscattering. VB scattering acts on a much smaller energy scale than FS scattering,  but these processes  are relevant for the problems discussed by us, because their  energy  is  comparable with SO interaction energy and Kondo temperature. We focus on two electron case.  Due to large separation of longitudinal modes, when  one can restrict to a good approximation  to the single longitudinal mode, the CNTQD  wave function can be decomposed   into  a spatial and  a spin-isospin components. Since the total wave function of two electrons has to be antisymmetric under exchange of electrons, the antisymmetric function in real space implies symmetric form in spin-isospin sector and  vice versa. The ground state of the two-electron system has a symmetric orbital wave function with corresponding six antisymmetric  spin-valley functions. They can be labeled by total helicity   $\kappa=\sigma_{1}m_{1} + \sigma_{2}m_{2}$ , total spin $\sigma = \sigma_{1}+\sigma_{2}$ and isospin $m=m_{1}+m_{2}$, ($|\kappa m\sigma\rangle$).
SO interaction acts only on the states   $\kappa\neq0$, VB scattering on the states $m=0$, and total spin remains a good quantum number. As has been discussed in section A, in the absence of magnetic field  SO interaction splits the sixfold multiplet into two singlets ($\kappa=\bar{2}, 2$ and quartet $\kappa=0$). In effect of backscattering $\kappa=0$ quartet is further split by $\Delta_{VBS}$ with the corresponding lower states, which are spin polarized ($S_{2}=|002\rangle$, $S_{\bar{2}}=|00\bar{2}\rangle$) and higher states, valley polarized ($V_{2}=|020\rangle$, $V_{\bar{2}}=|0\bar{2}0\rangle$). The valley polarized doublet is raised with respect to spin polarized doublet. VB scattering does not conserve helicity and spin-valley unpolarized states ($\kappa=\bar{2}, 2$) become mixed  by VBS perturbation ($SV_{\pm}=1/\sqrt{2}(|\bar{2}00\rangle\pm|200\rangle)$). The corresponding shifts of energies are $\pm\Lambda=\pm\sqrt{\Delta^{2}+\Delta^{2}_{VBS}}$ \cite{Secchi}. The values of $\Delta_{VBS}$ calculated from the observed effective SO splitting $\Lambda$ are $\Delta_{VBS}=0.2$ meV \cite{Pecker} and $\Delta_{VBS}= 1.56$ meV \cite{Cleuziou}. Estimates from uncorrelated states based on first-principles perturbation theory predict $\Delta_{VBS}$  to be of hundreds of  $\mu eV$ \cite{Secchi,Laird} and other theoretical estimates  give much smaller values of  $\Delta_{VBS}\sim 1-10$  $\mu eV$ \cite{Pecker,Secchi,Wunsch}. There are suggestions that neglect of superexchange  with participation of the  state from different shell is responsible for an underestimation of valley exchange \cite{Laird}. Since there is no clear view  in literature on the relative value of VBS parameter $V$ in relation to SO splitting, and even the sign of $V$  is the subject of discussion \cite{Laird} we analyze  both cases $V>0$ and $V<0$.
The examples of  field evolution of the mentioned  two-electron dot eigenstates are shown for weak VBS case ($|V/\Lambda|\ll1$) on Fig. 8a, and in the  strong VB scattering limit ($|V/\Lambda|\sim1$) on Fig. 8b. Interesting observation shown in the inset of Fig. 8b  is that apart from field induced recovery of double degeneracy, for strong  VBS, also triple degeneracy may appear.
Figures 8c, d show examples of the field dependencies of conductance in the presence of VB scattering and insets  present   the ground state maps of the dot  with lines, which mark  cross sections, for which conductances have been drawn. The results  are presented for   ${\cal{U}}'<{\cal{U}}$ , because no triple degeneracy point appears on the ground state map for ${\cal{U}} = {\cal{U}}'$ for   $V<0$. For $V>0$  triple degeneracy occurs also for ${\cal{U}} = {\cal{U}}'$ (left inset of Fig 8d), but the degeneracy region on the ($\Delta$, $B$) map  is considerably  smaller than for ${\cal{U}}'<{\cal{U}}$ (right inset). Conductance curves with two maxima reflect two consecutive field induced  Kondo effects, maxima for lower fields  correspond to spin SU(2) resonances and for  higher fields spin-valley  Kondo effects.  Conductance curves with the  single maximum correspond to SU(3) Kondo effect.  The corresponding transmission peak of this  many-body resonance of enhanced symmetry is asymmetric and is  shifted away from the Fermi energy  and is characterized by a phase shift $\delta\sim \pi/3$ (lower inset of Fig. 8c).  We also present partial conductances for SU(3) case, and it is seen that practically  only three channels participate  in transport and they equally contribute to the  total conductance (${\cal{G}}\sim(9/4)(e^{2}/h)$).  The residual value derived from the fourth channel only slightly enhances the total conductance.

\subsection{Small bandgap carbon nanotube quantum dots}
The small gaps  occur in the nanotubes which  in the simple zone folding picture  should be nominally metallic.  These gaps do not result from quantization of momentum along circumference (gaps of order of few hundred meV), but are induced by curvature or strain (band gaps  $\leq10$ meV) \cite{Steele}.
Nearly metallic nanotubes exhibit crossing of the Dirac points at anomalously low magnetic fields ($B_{Dirac}\sim2$ T \cite{Steele,Deshpande}), what indicates the small shifts of circumference quantization lines from the Dirac point and thus confirm the small values of the bandgaps.
The small gaps are comparable with the energies of  SO splitting or valley mixing and therefore the commonly used  large gap expansion of single electron energies cannot be applied in this case. Using the full expression (4) the parabolic field dependencies of single electron energies result  (see the inset of Fig. 9a).  These dependencies are  determined not only  by the response of  orbital and spin magnetic moments, as in the case of large gaps,  but crucially depend  also on the value of   the gap and gate voltage. Consequently neither Coulomb lines nor the Kondo lines are   straight lines on gate voltage- magnetic field conductance maps (Fig. 9a). The fields of Kondo recovery  depend on voltage and this manifests in that   the  Kondo lines are not parallel to the gate axis. As it is seen the maps do not exhibit intrashell e-h symmetry. Apart from $2e$ valley,   recovery of Kondo effect is observed  not only in one, but in both  odd valleys (Fig. 9b). This is in contrast to what is observed in wide gap nanotubes, where depending on the sign of SO coupling Kondo effect appears in $1e$ or $3e$ valley (Section C). The  ability to restore degeneration in both odd valleys  is again a consequence of nonlinear field dependence of energies. The surprising result is observed also at zero magnetic field.  For some values of the gap, different for different SO couplings, Kondo effect of SU(4) symmetry can appear. Due to nonlinear gate dependence of the dot energies characteristic  for   small gap carbon nanotubes, the effective spin-orbit splitting  tends to zero for some values of the bandgap and SO parameters.
\begin{figure}
\includegraphics[width=0.48\linewidth,bb=0 0 439 447,clip]{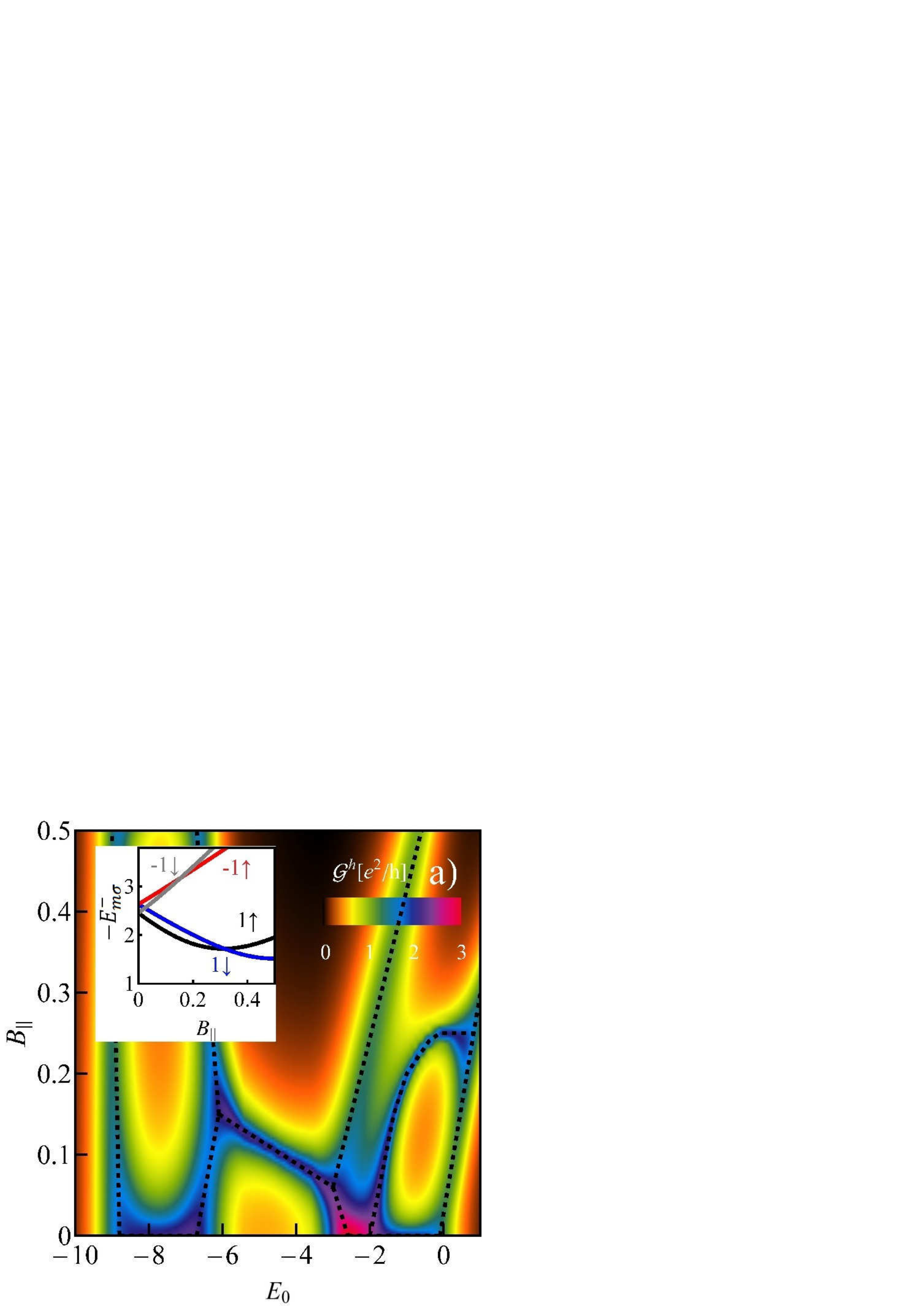}
\includegraphics[width=0.48\linewidth,bb=0 0 439 448,clip]{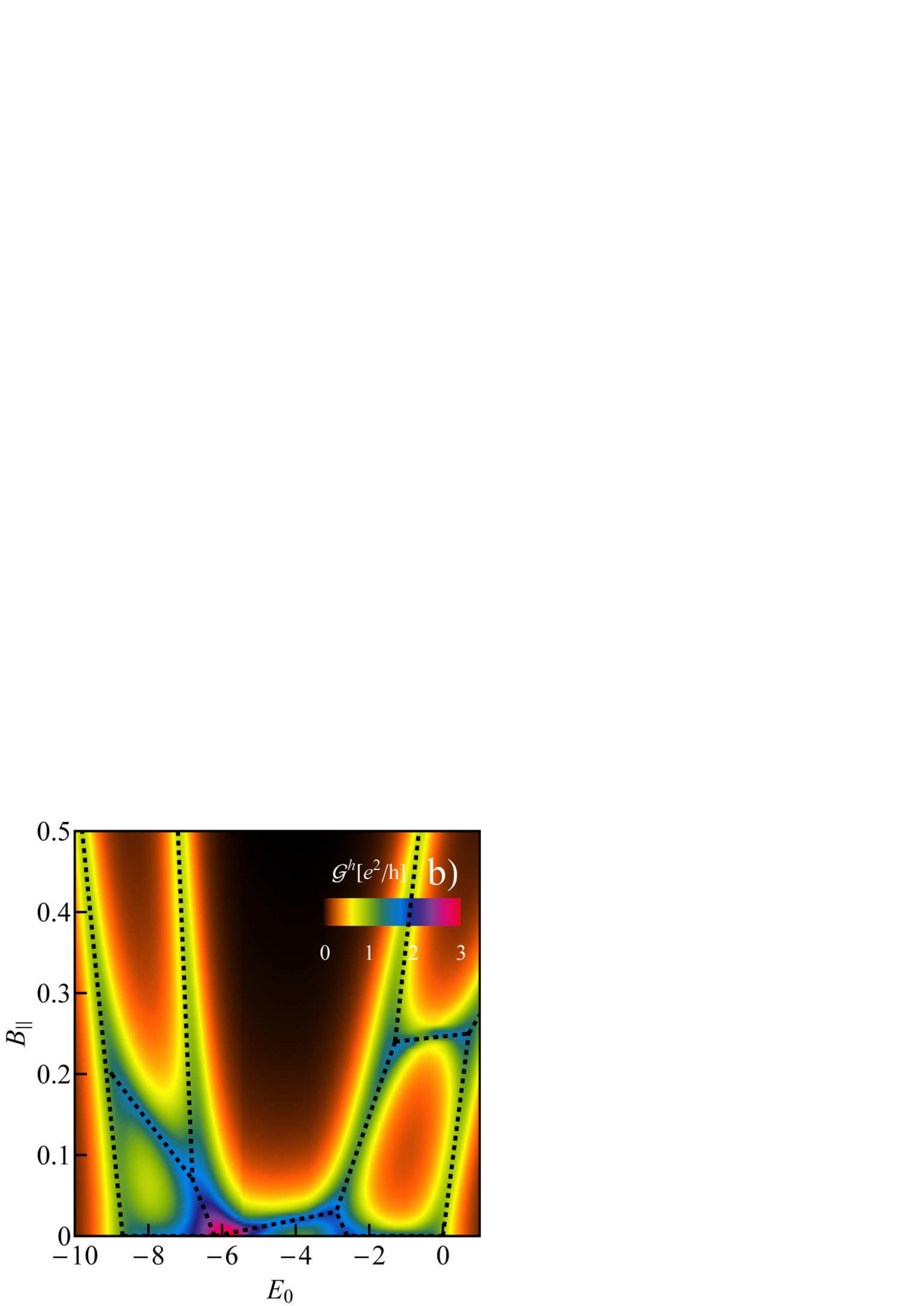}
\includegraphics[width=0.48\linewidth,bb=0 0 439 463,clip]{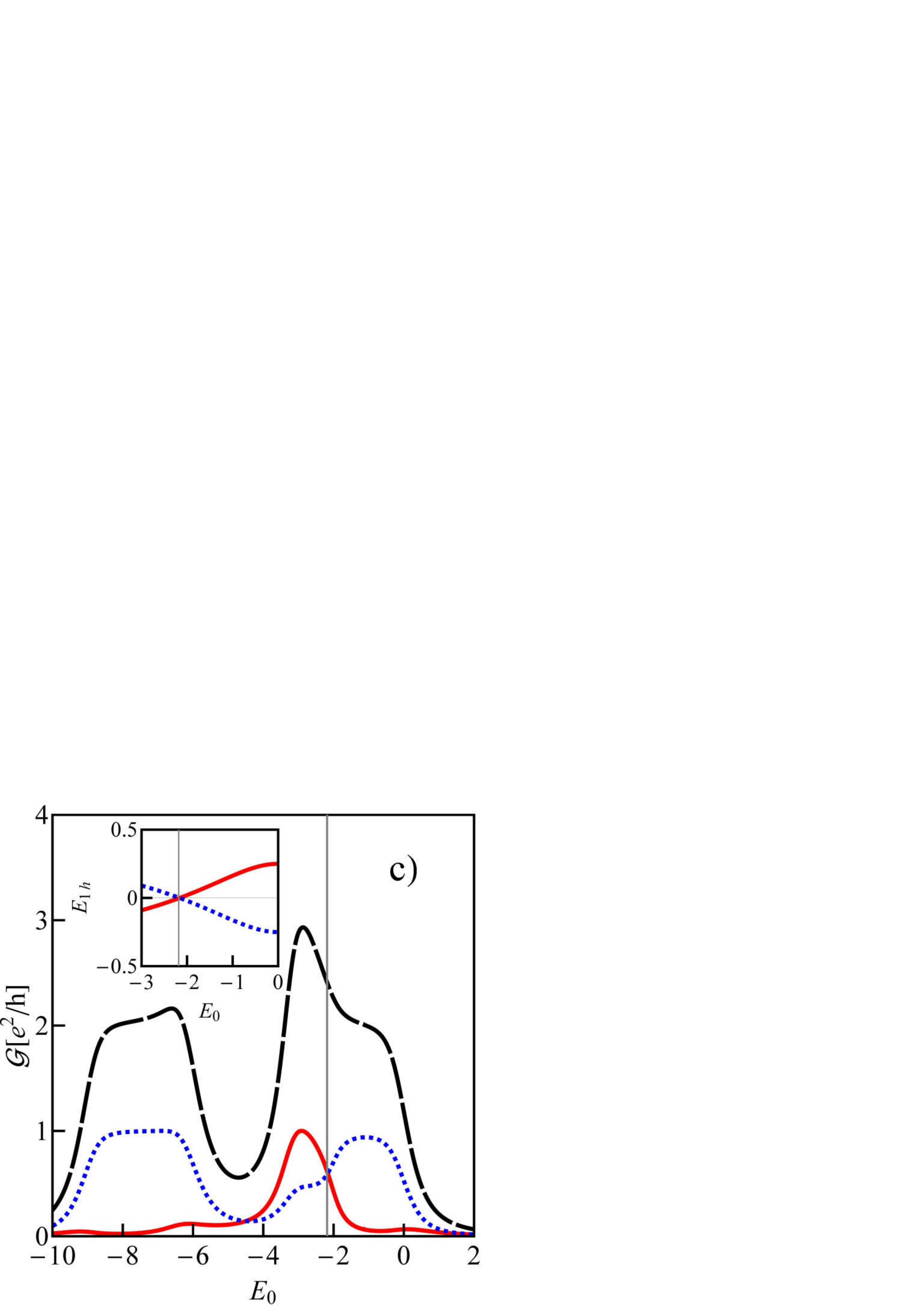}
\includegraphics[width=0.48\linewidth,bb=0 0 439 463,clip]{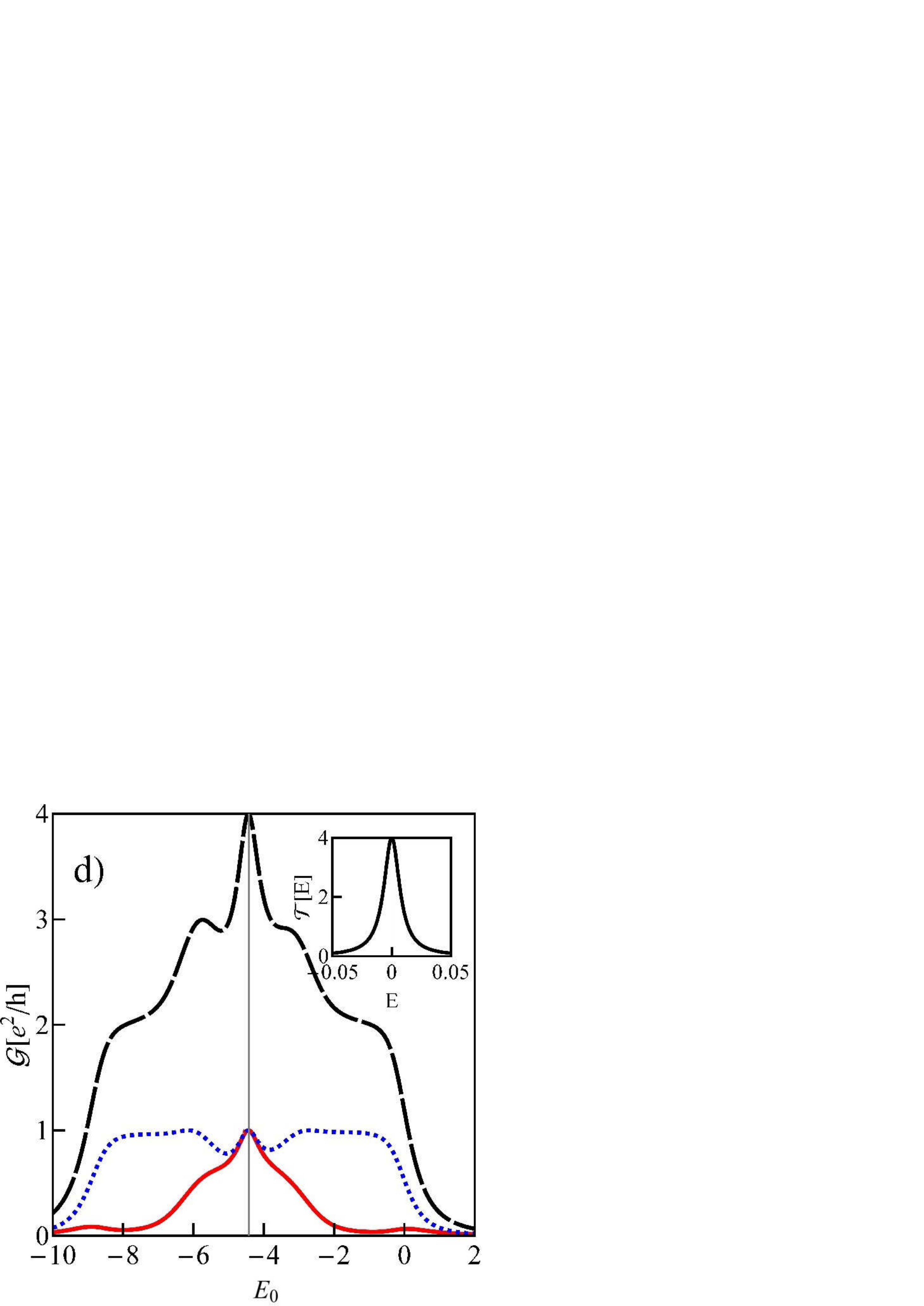}
\includegraphics[width=0.48\linewidth,bb=0 0 439 463,clip]{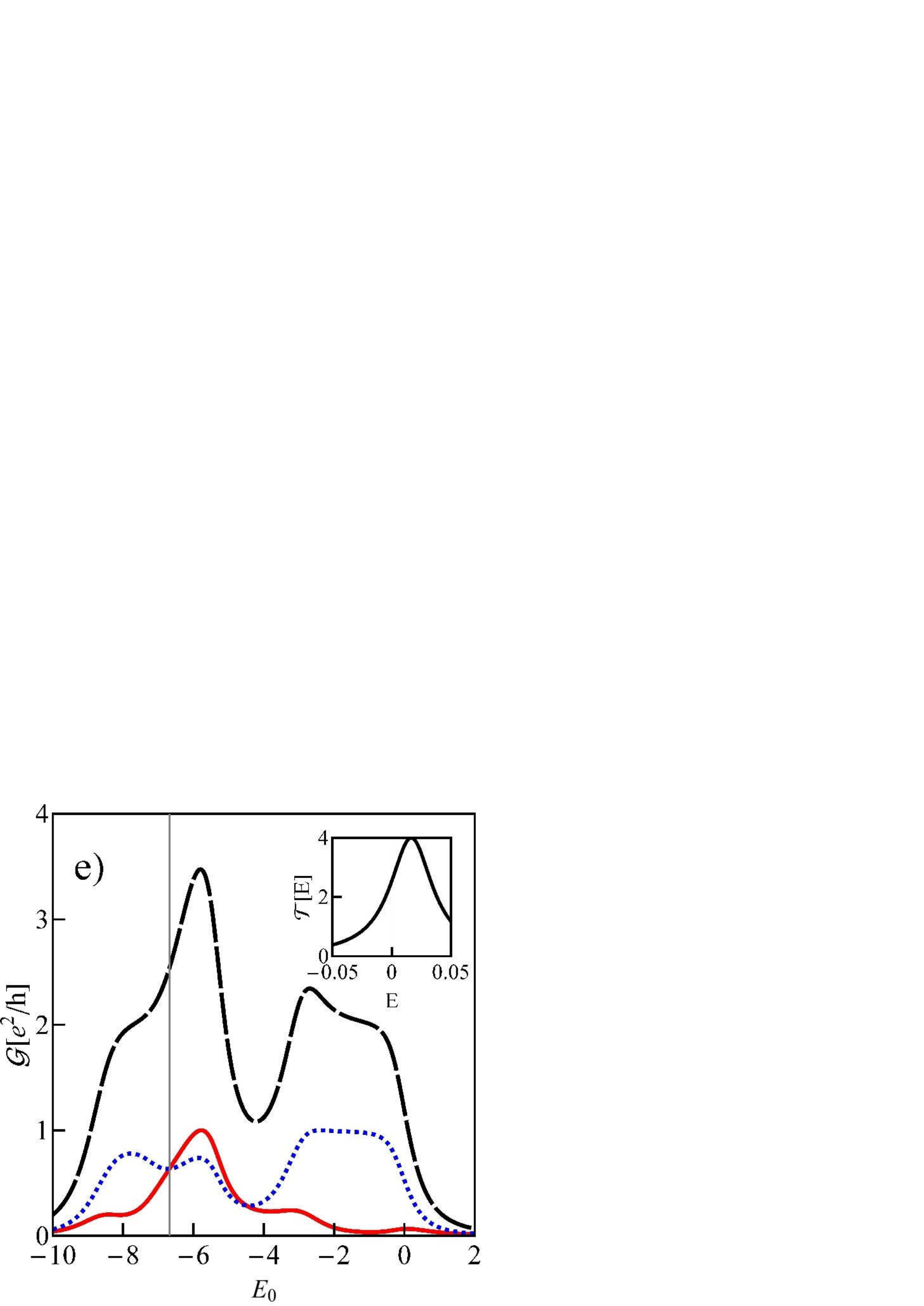}
\caption{\label{fig9} (Color online) Small bandgap CNTQD. (a,b) Hole conductance map ${\cal{G}}^{h}(E_{0}, B_{\parallel})$ for $\Delta_{g}=2$ (a) and $\Delta_{g}=5$ (b) ($\Delta_{Z}=-1/2$, $\Delta_{O}=3/4$, $\mu_{o}=5$). Inset presents field dependencies of single-hole states. (c,d,e) Zero field conductances for $N_{h} = 1$, $\Delta_{g} =2$   (b), $N_{h} = 2$, $\Delta_{g} =4$ (c), $N_{h} = 3$, $\Delta_{g} =6$ (d) (solid lines). Dotted and dashed lines present partial conductances corresponding to the spin-orbital doublets. Inset of (a) illustrates crossing of two single-hole doublets and insets of (c,d) present the corresponding transmissions of SU(4) Kondo resonances for $N_{h}=2$ and $N_{h}=3$.}
\end{figure}
The  example for $N=1$ is presented in the  inset of Fig. 9c.  For a certain value of gate voltage the curves cross, what means vanishing of the effective spin-orbit splitting and SU(4) symmetry is recovered. Similar crossings of the states, for different values of SO coupling are also possible in $N=2$ and $N=3$ valleys.  Figures  9c, e illustrate, how crossover to Kondo effect of higher symmetry in odd valleys  (SU(2)$\rightarrow$ SU(4)) manifest in conductance and similarly Fig. 9d presents answer of conductance on the  rebuild of Kondo correlations  destroyed by SO coupling in $N=2$ region. It is hard to keep track of the  crossover in odd valleys observing  only the evolution of total conductance with the change of the gap, because linear conductance for SU(4) symmetry coincides with that for SU(2), but SU(4) Kondo effect reflects in equal values of  partial spin-orbital  conductances reaching values of $1/2(e^{2}/h)$ each (Figs. 9c, e).  In even valley the gate induced rebuild of SU(4)  Kondo correlations manifests not only  in partial conductances, which have equal values $e^{2}/h$, but in this case it reflects also in total conductance  by  the emergence of a distinct peak. The described phenomena allow switching between transport Kondo regimes of different symmetries or even between Kondo and non-Kondo states by small changes of gate  voltage. Having in mind the property that  strain can modify the gap in nearly metallic nanotubes \cite{Minot2} and thus can in some cases  induce  Kondo correlations in $2e$ valley and opens new transport channels, one can think about possibility to use this  mechanism for  nano-mechanical switching of the current.

\section{Intershell Kondo Effects}
The electron energies of CNTQD are quantized by  periodic boundary  conditions along the circumference (quantization of $k_{\perp}$) and by longitudinal confinement (quantization of  $k_{\parallel}$).
\begin{figure}
\includegraphics[width=0.6\linewidth,bb=0 0 439 469,clip]{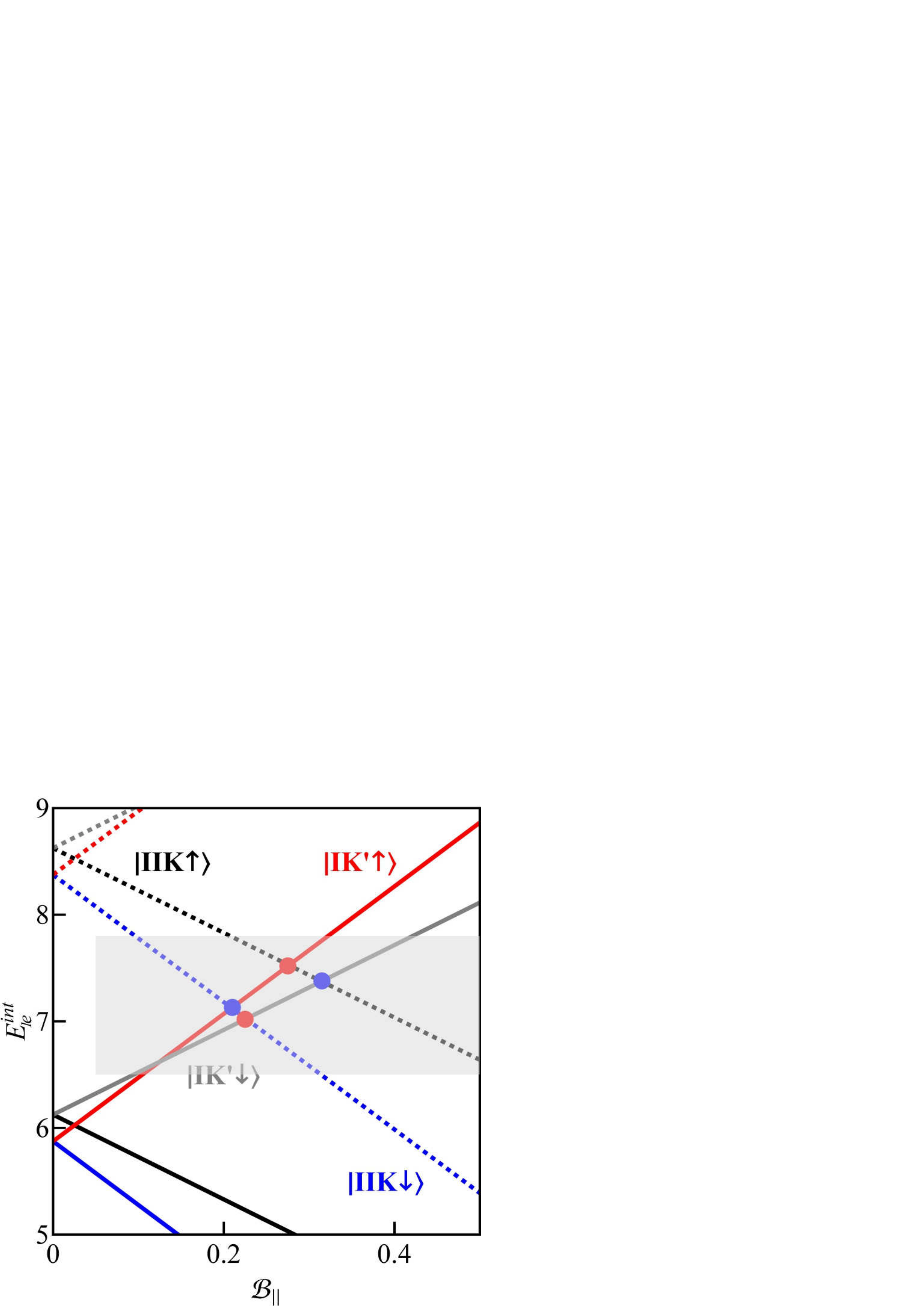}
\caption{\label{fig10} (Color online) Schematic  diagram of energy levels of two consecutive CNTQD shells ($E^{int}_{1e}=E_{1e}-2{\cal{U}}$ , where $E_{1e}$ is the shifted dot energy in the lower shell). The solid lines present energies of the lower shell and the dotted lines  the energies of the  upper. The blue dots mark valley crossing for the same spin channel and red dots indicate spin-valley crossing.}
\end{figure}
For small diameter the level spacing  corresponding to  the circumference quantization is large, of order of $0.5-1$ eV \cite{Dekker}, and it is enough  to restrict, similarly as we have done so far, to a single pair of clockwise and anticlockwise modes. In this approximation the only quantum number reflecting circumferential quantization is  valley pseudospin. In most cases a similar argument on large level spacing applies also to longitudinal modes.  In general, the energies of the longitudinal modes depend on the bandgap and on the  details of confining potential. In the case of a sharp confining potential and for states lying  not to close to the gap,  the resulting quantized energies are inversely proportional to the dot length.  Typically they are of order of several meV \cite{Reich} and usually these  energies  are much larger than the rest of energies relevant for the issues discussed here.  Restriction  to a single shell i.e. to a manifold of four states ({$|K\sigma\rangle$}) corresponding to a single value of $k_{\parallel}$ is then  justified. All our previous discussion was limited to this case. In this section  we abandon this assumption and  focus on   the intershell effects, that are relevant for higher  magnetic fields or for longer dots.
\begin{figure}
\includegraphics[width=0.48\linewidth,bb=0 0 439 454,clip]{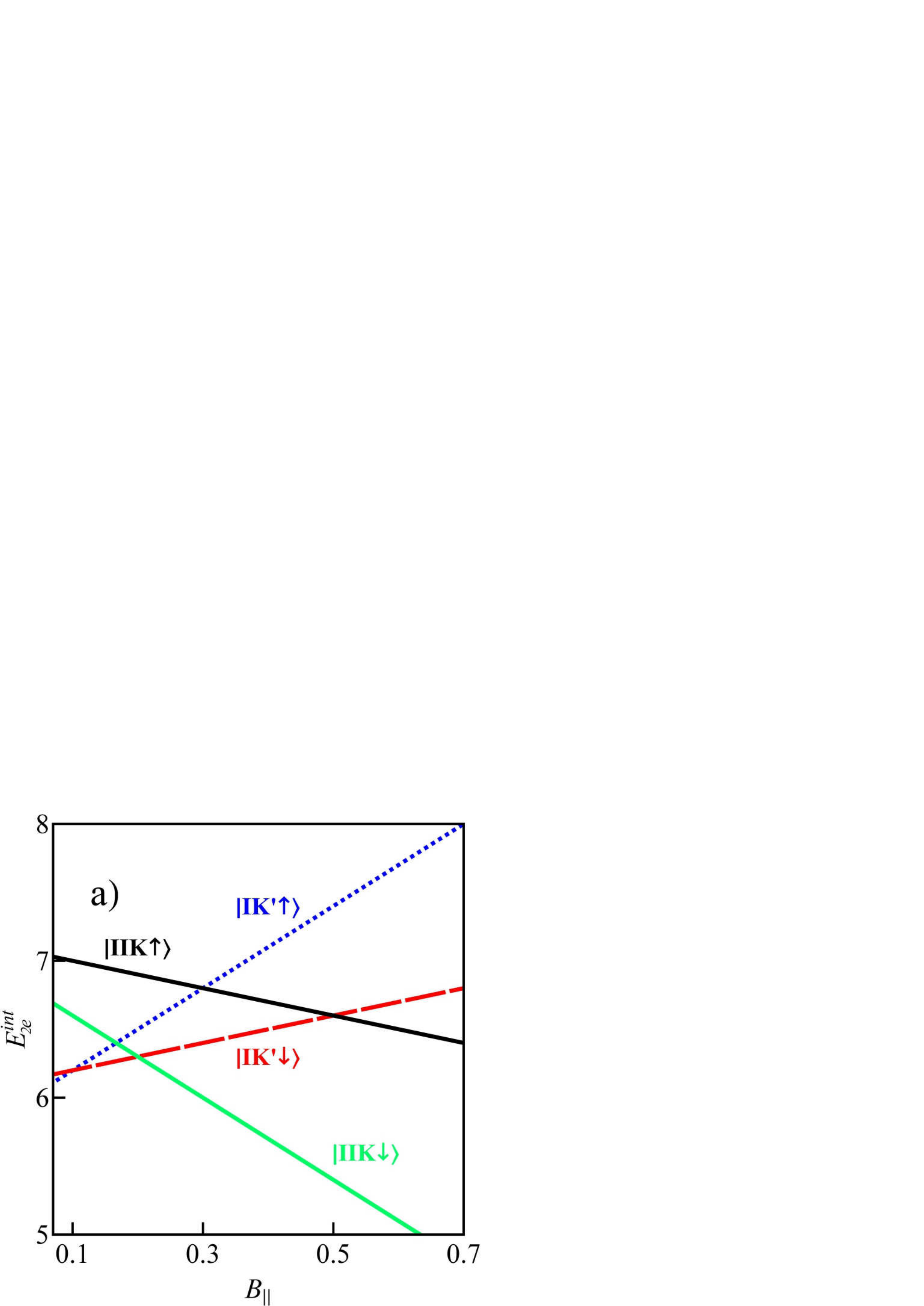}
\includegraphics[width=0.48\linewidth,bb=0 0 439 454,clip]{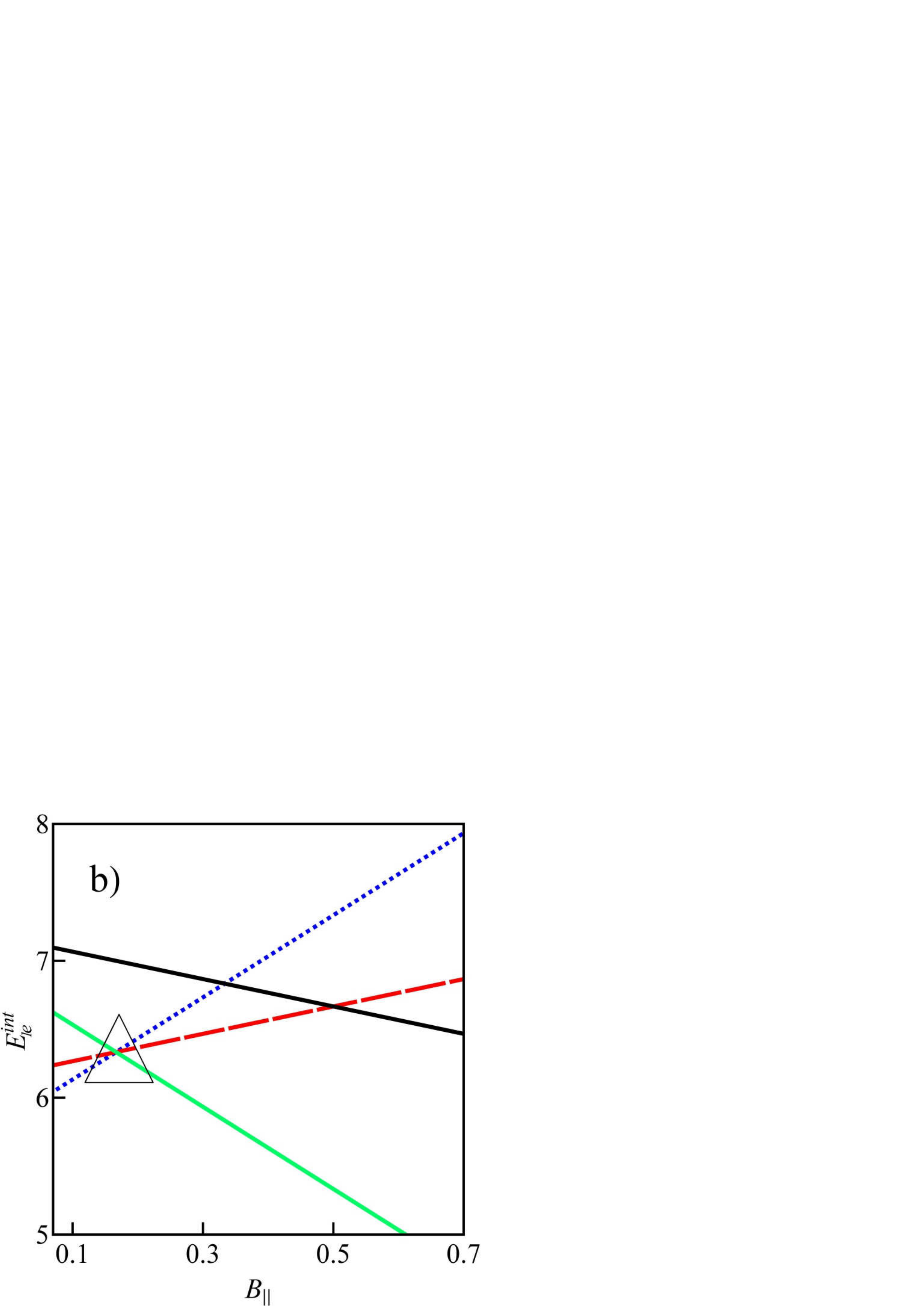}
\includegraphics[width=0.48\linewidth,bb=0 0 439 429,clip]{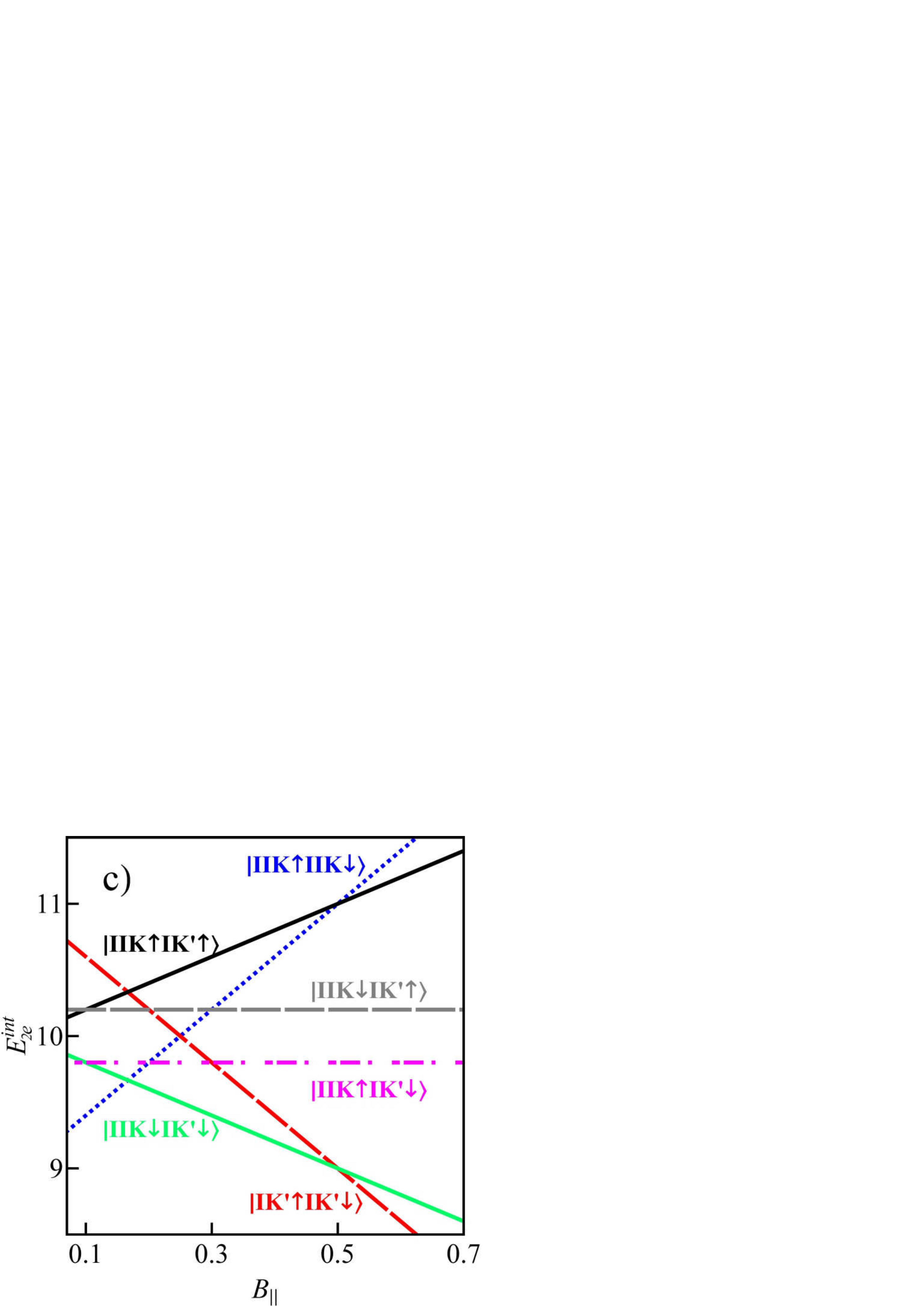}
\includegraphics[width=0.48\linewidth,bb=0 0 439 429,clip]{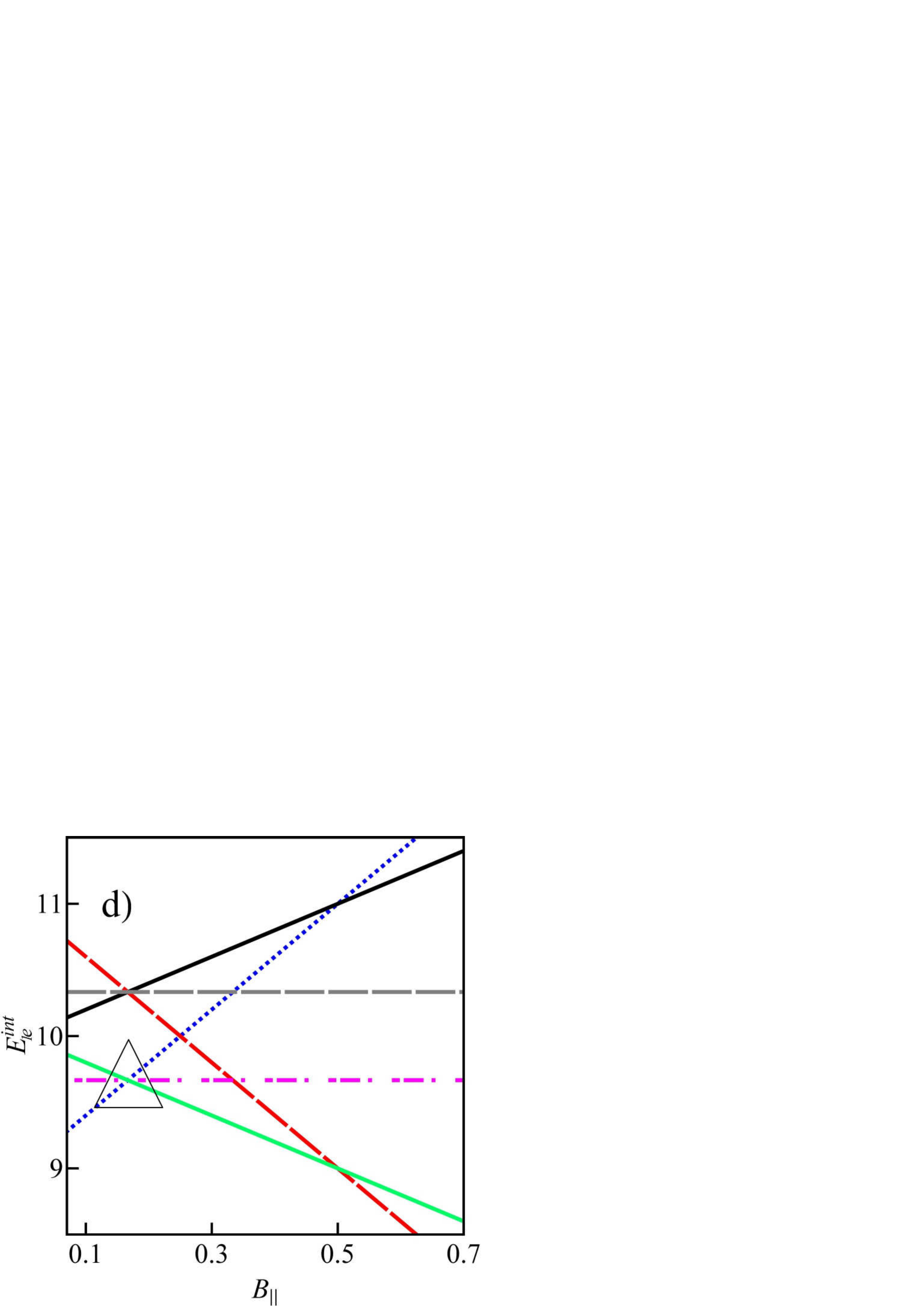}
\caption{\label{fig11} (Color online) Energies in intershell manifold. (a,b) Field dependencies of single-electron energies for $\Delta=1/10$ (a) and $\Delta=1/6$ (b). (c,d) Two-electron energies for $\Delta=1/10$  and $\Delta=1/6$ correspondingly ($E^{int}_{2e}=E_{2e}-2{\cal{U}}$). Energy separation of the shells $\Delta E=1$. Triangles on Figs. b,d mark SU(3) cases.}
\end{figure}
Fig. 10 schematically illustrates spectrum of two neighboring shells versus parallel magnetic field. In the following we concentrate on  the description of many-body processes only for the high fields, where intershell crossings of the states occur and the states from the lower doublet of the lower shell and similarly  the states form higher  doublet of the higher shell are well separated from the rest and with a good approximation it is enough to restrict to the intershell manifold (IM) of the states ($|IM\rangle = |IK'\downarrow\rangle,|IK'\uparrow\rangle, |IIK\downarrow\rangle, |IIK\uparrow\rangle$) (the first quantum  numbers  label the shells). In other words, instead of analyzing more generally all occupation regions from both shells,  we focus on the  high field regions of   $N_{I} = 3,4$ occupations of lower shell and $N_{II} = 1,2$ for higher shell.
\begin{figure}
\includegraphics[width=0.48\linewidth,bb=0 0 439 442,clip]{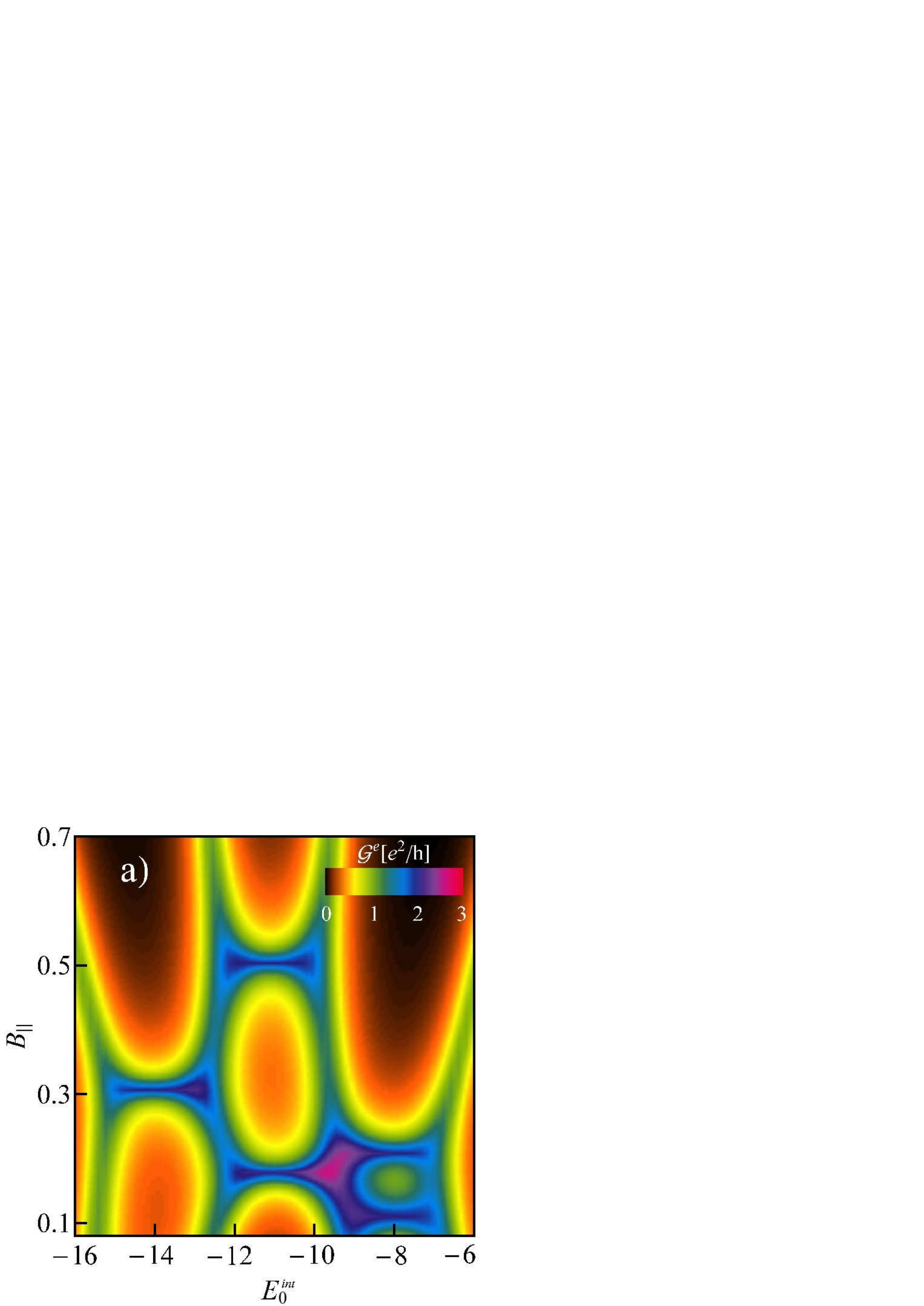}
\includegraphics[width=0.48\linewidth,bb=0 0 439 442,clip]{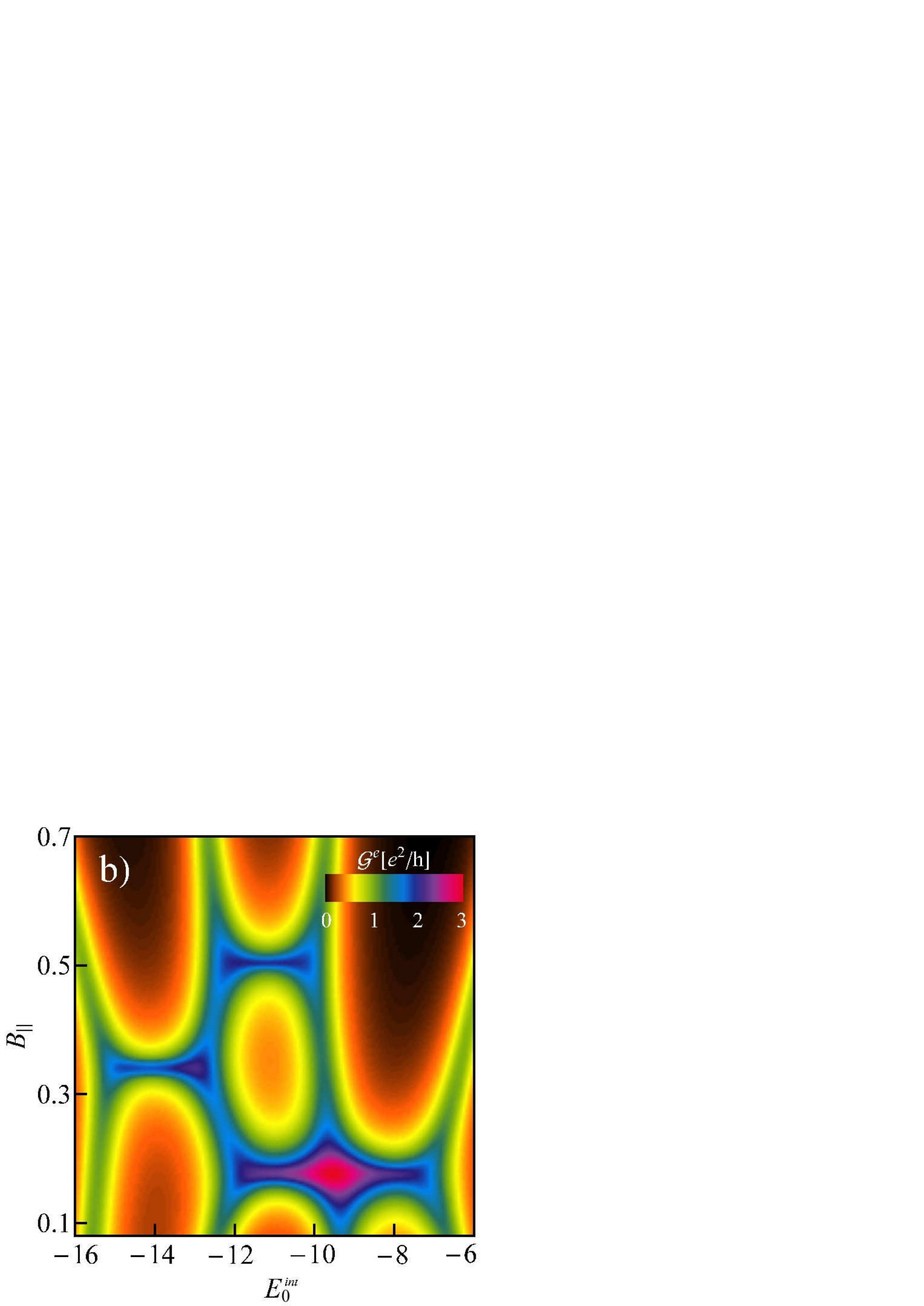}
\includegraphics[width=0.48\linewidth,bb=0 0 439 442,clip]{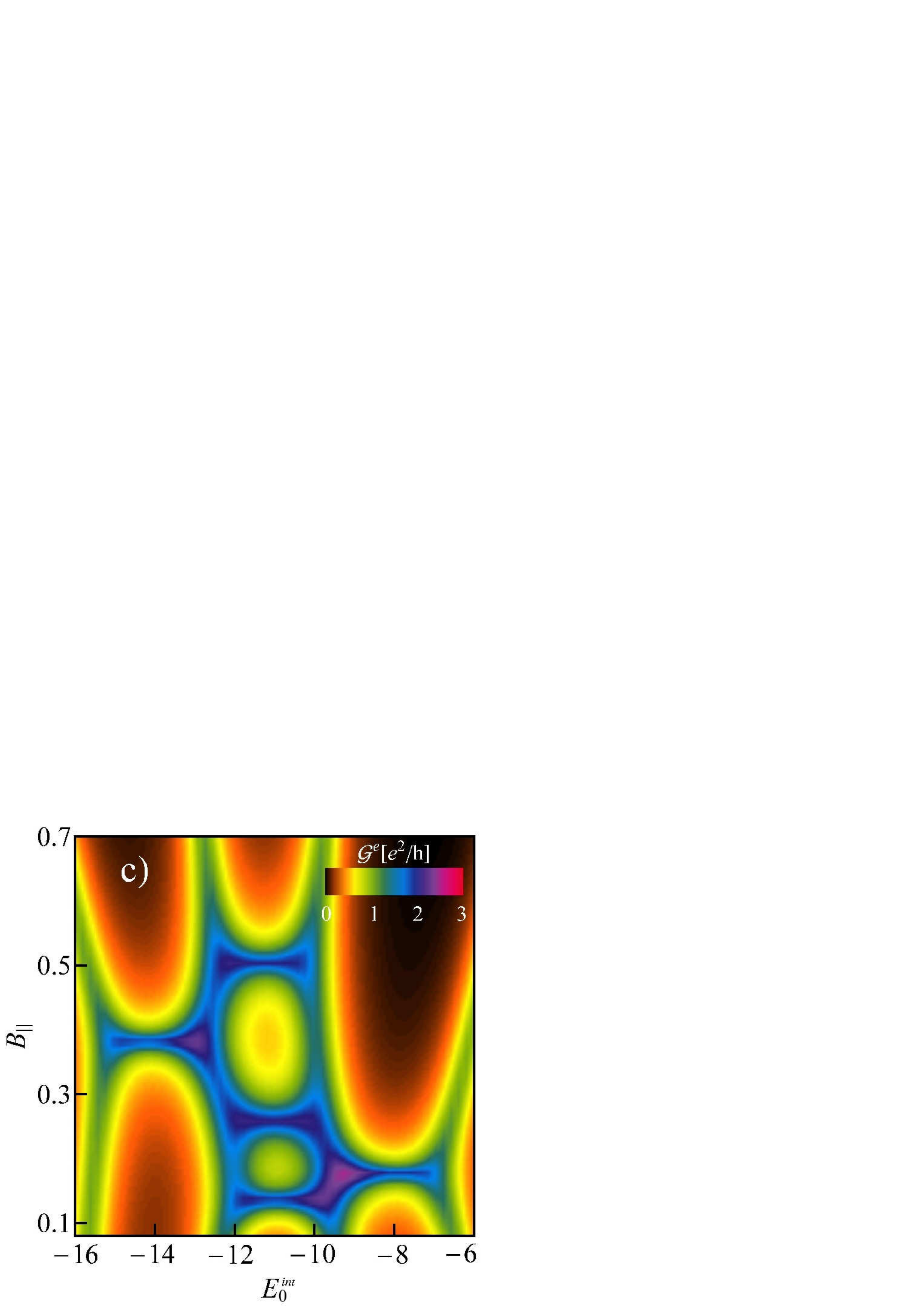}
\includegraphics[width=0.48\linewidth,bb=0 0 439 442,clip]{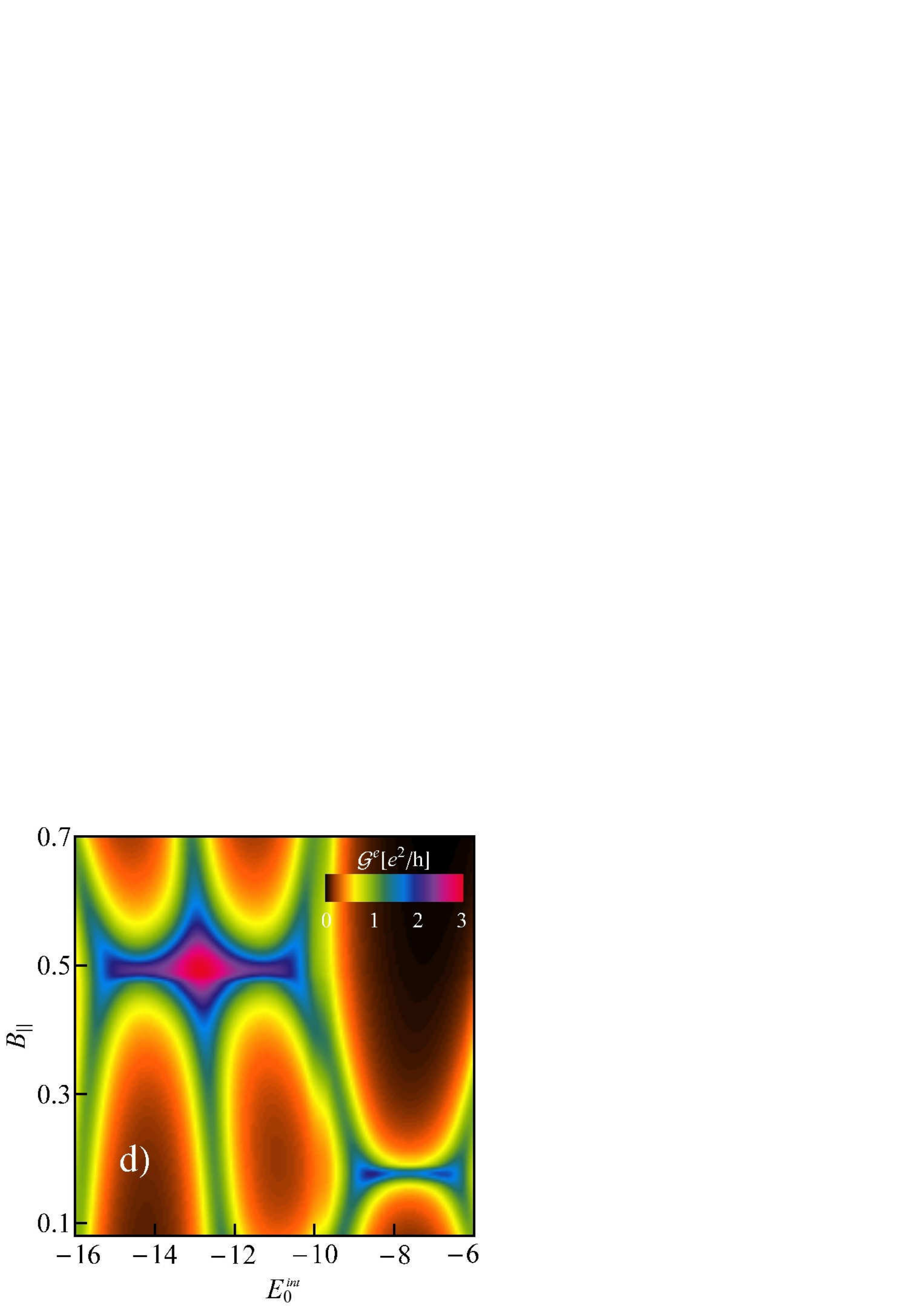}
\includegraphics[width=0.48\linewidth,bb=0 0 439 468,clip]{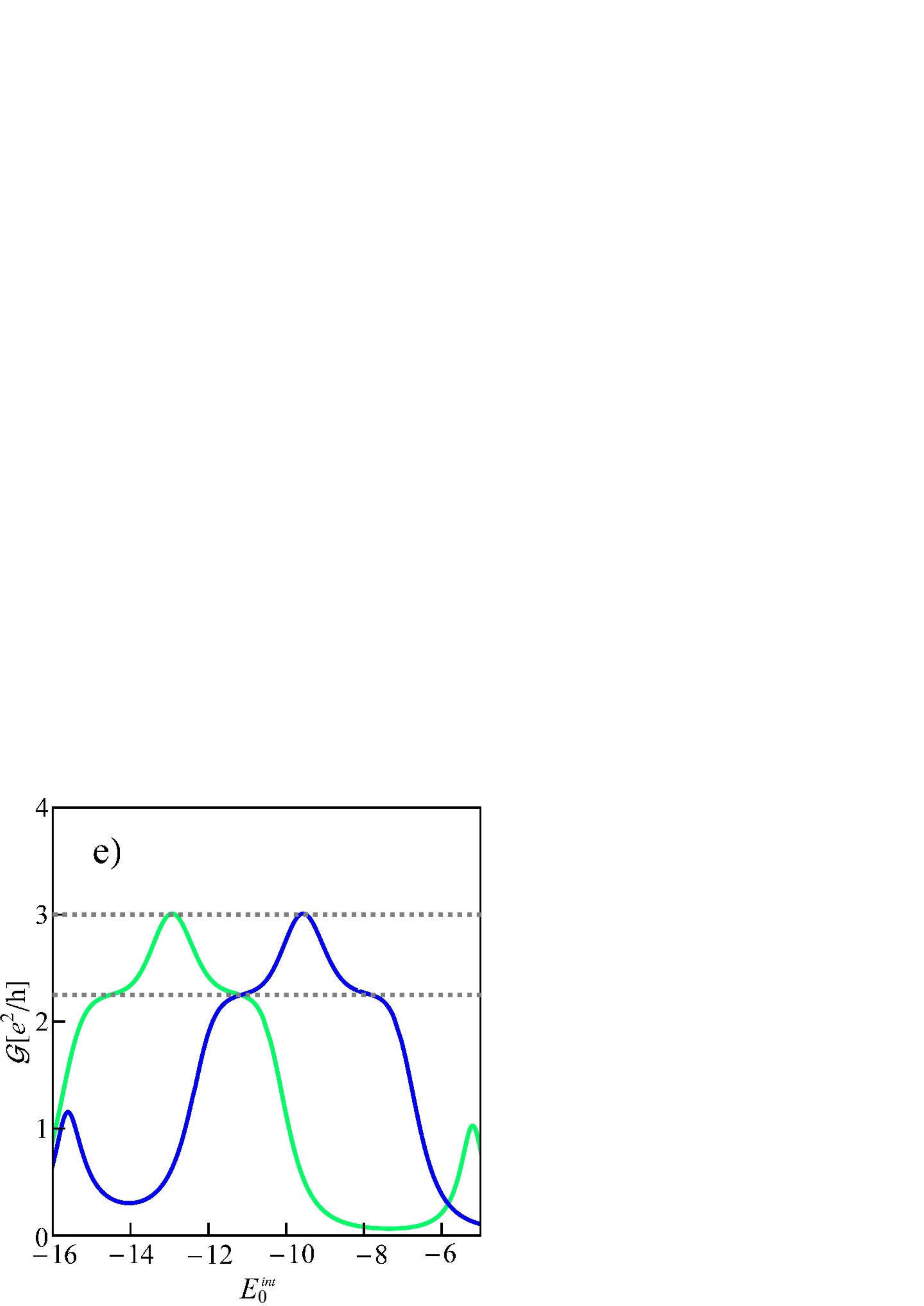}
\includegraphics[width=0.48\linewidth,bb=0 0 439 442,clip]{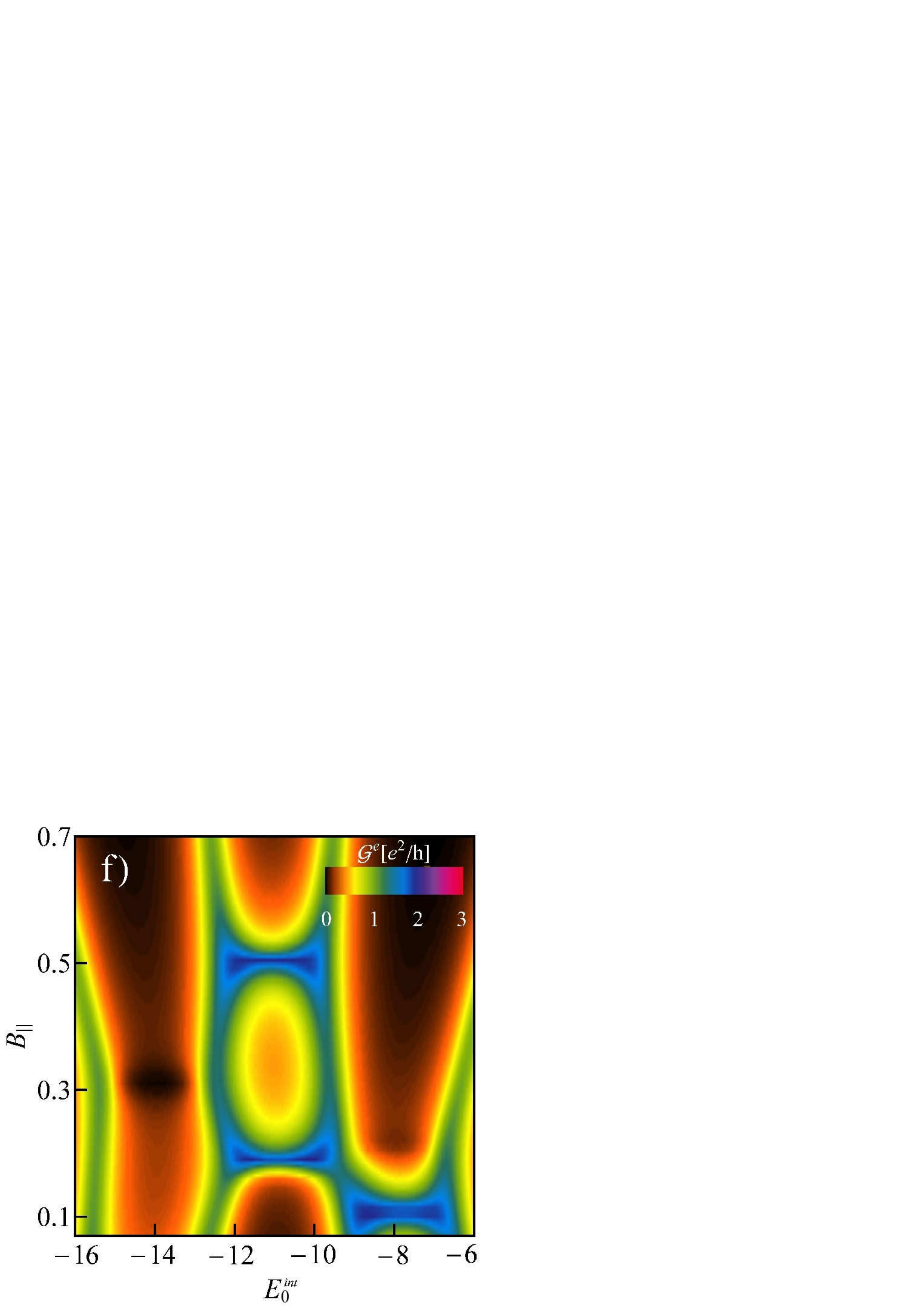}
\caption{\label{fig12} (Color online) Intershell gate-magnetic field conductance maps ($E^{int}_{0}=E_{0}-2{\cal{U}}$). (a)  $\Delta = 1/10$, (b), $\Delta = 1/6$ (SU(3) Kondo for $N = 1,2$)  (c), $\Delta = 1/4$, (d)  $\Delta = 1/2$ (SU(3) Kondo for $N = 2,3$). (e)  Conductances  presented  for magnetic fields, for which  SU(3) Kondo effects  occur $\Delta = 1/6$,  $B_{\parallel} = 1/6$    (broken line), $\Delta = 1/2$,  $B_{\parallel} = 1/2$ (solid). (f) Conductance for $\Delta = 1/10$,  with  finite intershell intervalley mixing $\Delta^{int}_{KK'} = 1/10$.}
\end{figure}
  Alternatively we can look at this region as single occupied range of (IM)  $N_{int} =1$ ($N_{I} =3$), and similarly double occupied  $N_{int}=2$ ($N_{I}=4$), triple occupied  $N_{int} =3$ ($N_{II}=1$) and full occupied $N_{int}=4$ ($N_{II}=2$). The considered region of energies and fields is shown in Fig. 10 (box  bounded by dotted lines).  Formally the many body correlations can be  described within SB formalism, similarly to the cases discussed in the previous paragraphs, by $16$ slave bosons. There are two different types of intershell line crossings: spin conserving crossing (valley SU(2)) or crossing of the lines of opposite spins (spin-valley SU(2)). Fig. 12 shows conductance maps for several values of spin-orbit splitting. The number of Kondo enhanced conductance  lines appearing in each occupancy region and the characteristic fields, for which the lines appear,  depend on the value of $\Delta$ and they correspond to degeneracy recovery in the  presented field range.
 As an example let us analyze conductance for $\Delta = 1/10$ (Fig. 12a).
 The lower line for $N_{int} = 1$  ($E^{int}_{0}\sim-7.5$) is the intershell Kondo resonance ($B_{\parallel}=0.1$) and the higher at $B_{\parallel}=0.2$ reflects Kondo revival due to cotunneling induced effective intershell valley quantum fluctuations between the states  $|IK'\downarrow\rangle$ and $|IIK\downarrow\rangle$ (see Fig. 11a). In $N_{int} = 3$ region ($E^{int}_{0}=-13.5$) intershell valley Kondo effect occurs at $B_{\parallel}=0.3$ due to effective fluctuations between the $|\underline{IIK\uparrow}\rangle$ and  $|\underline{IK'\uparrow}\rangle$ states (not presented). In both cases significant conductance spin polarization is observed  (Figs. 12a and 13c), because Kondo processes occur within the single spin channels (for $N_{int}=1$ $PC_{s}\approx-1$ and for $N_{int}=3$ $PC_{s}\approx1$).  In $N_{int} =2$ region two Kondo lines correspond to intershell spin- valley effects, the lower occurring for $B_{\parallel}=1/6$ results from  $|IIK\uparrow IIK\downarrow\rangle$  and $|IIK\downarrow IK'\downarrow\rangle$ fluctuations and the higher for $B_{\parallel}=1/2$ from $|IIK\uparrow IIK\downarrow\rangle$ and $|IK'\uparrow IK'\downarrow\rangle$ fluctuations (Fig. 11c). In both cases the corresponding conductances are unpolarized (Figs. 12a and 13c).
Apart from the field induced SU(2) Kondo revivals also higher symmetry effects are possible. For the assumed parameters it holds for $\Delta=1/6$  and $\Delta=1/2$. As it is illustrated for $\Delta =1/6$ on Fig. 11b (single electron states) and Fig. 11d (two-electron states) magnetic field brings three states to degeneracy.  Each of these states is coupled with equal strength to the corresponding state in electrode and in the considered strong coupling limit  Kondo SU(3) effect appears at half filling ($N=2$) and in one of odd occupied regions ($N=3$ for $\Delta=1/6$,  or  $N=1$ for $\Delta = 1/2$). It manifests by the  enhanced conductance, which in the limit of small coupling to the electrodes approaches  ${\cal{G}}=(9/4)(e^{2}/h)$ value.  The conductance plotted for the  fields, for which the   threefold degeneracy occurs exhibit  clear plateaus  evidencing  the appearance of the mentioned Kondo effects (Fig. 12 e).  Let us close this section by a short remark on the impact of intershell intervalley mixing. Fig.  12f shows conductance map for $\Delta^{int}_{KK'}=0.1$,  corresponding map for $\Delta^{int}_{KK'}=0$ is presented in Fig. 12a. In the $N=3$ region,  instead of Kondo enhanced line, line  of strongly reduced  conductance is observed for finite valley mixing and in region of single occupancy instead of two Kondo lines only one broadened line is seen. To understand these changes we show on Figures 13a, b the evolution of conductances with increasing valley mixing drawn for the  gates corresponding to the centers of $N_{int}=3$ and $N_{int}=1$ areas.
\begin{figure}
\includegraphics[width=0.48\linewidth,bb=0 0 439 455,clip]{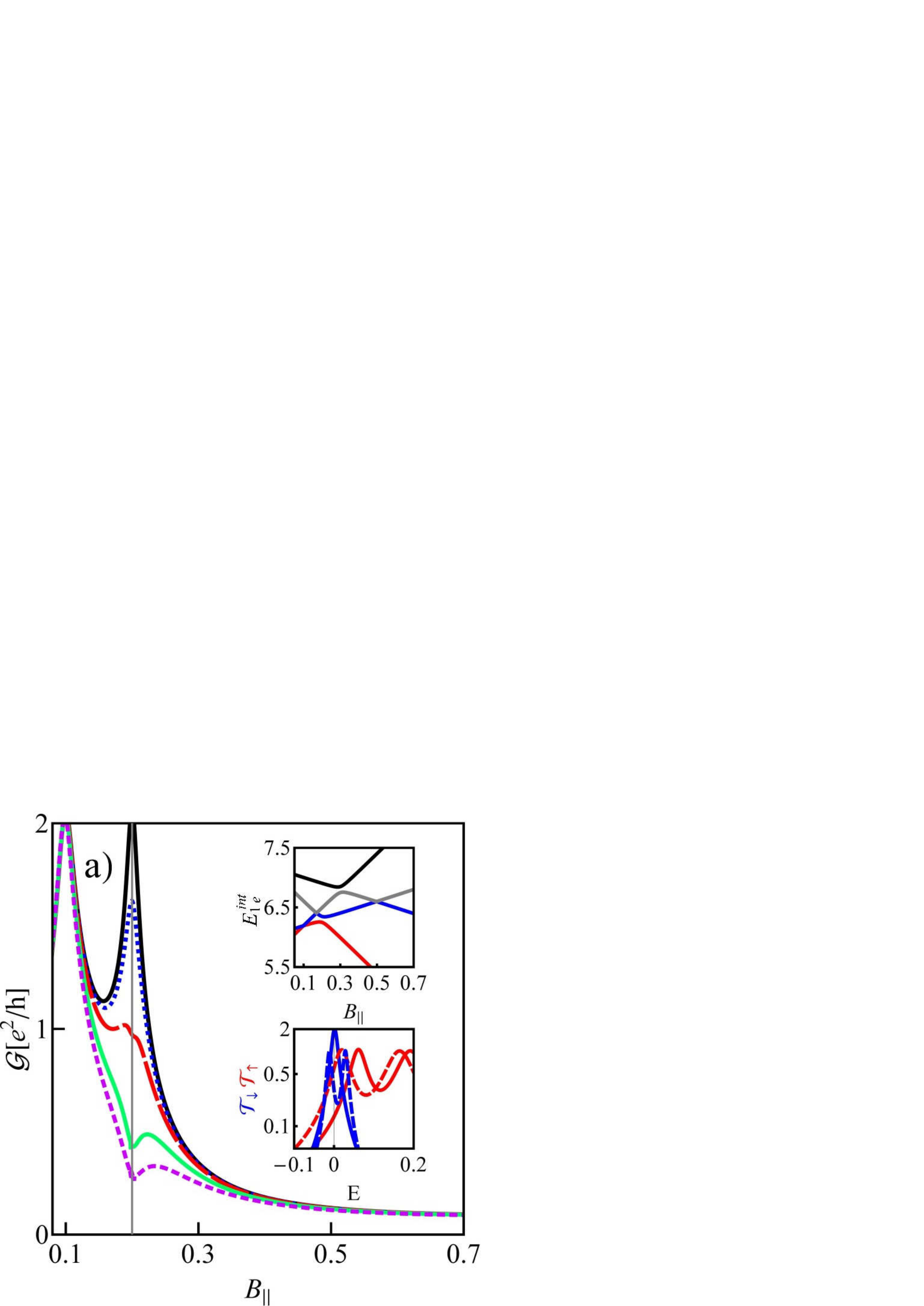}
\includegraphics[width=0.48\linewidth,bb=0 0 439 455,clip]{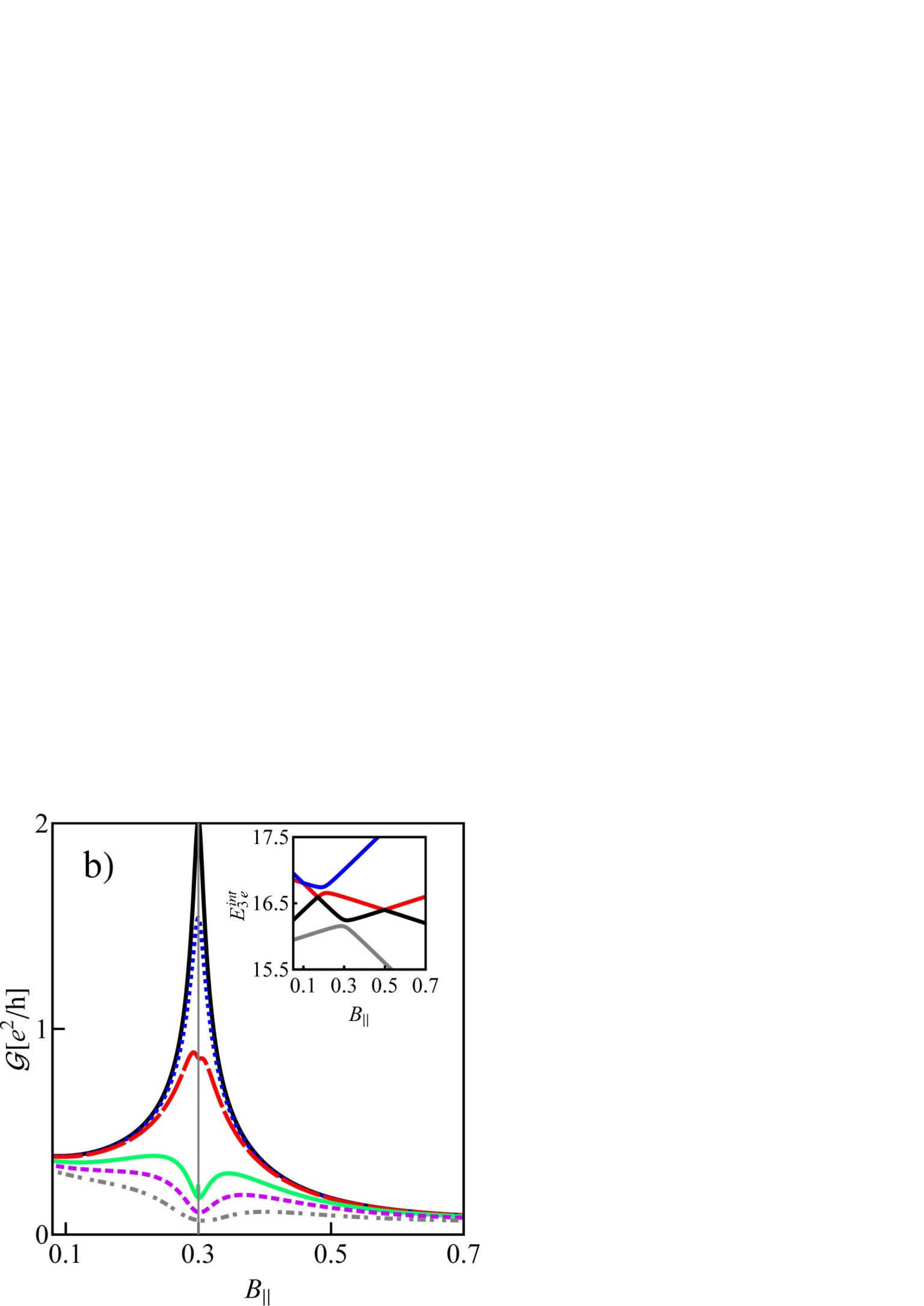}
\includegraphics[width=0.48\linewidth,bb=0 0 439 442,clip]{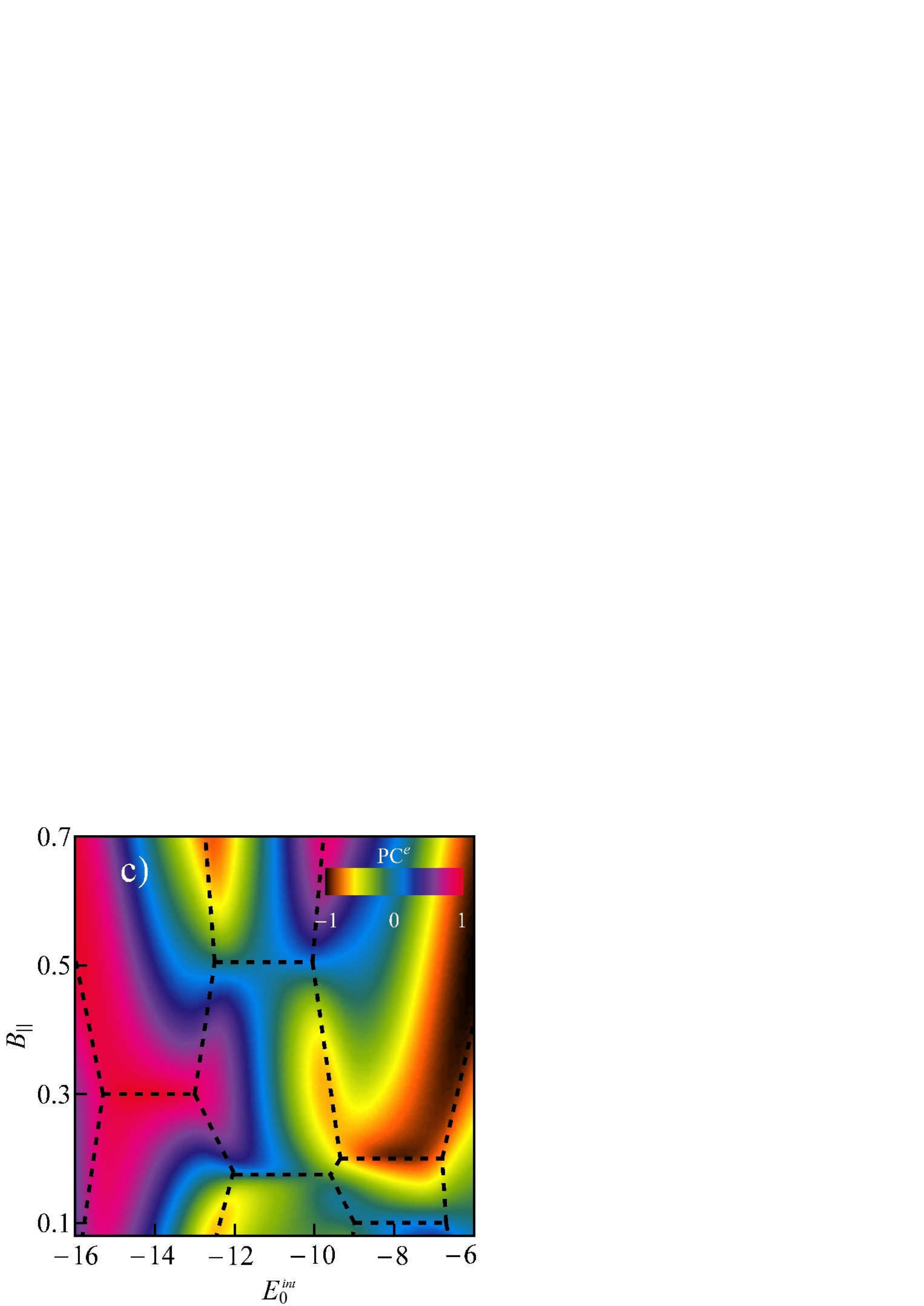}
\includegraphics[width=0.48\linewidth,bb=0 0 439 442,clip]{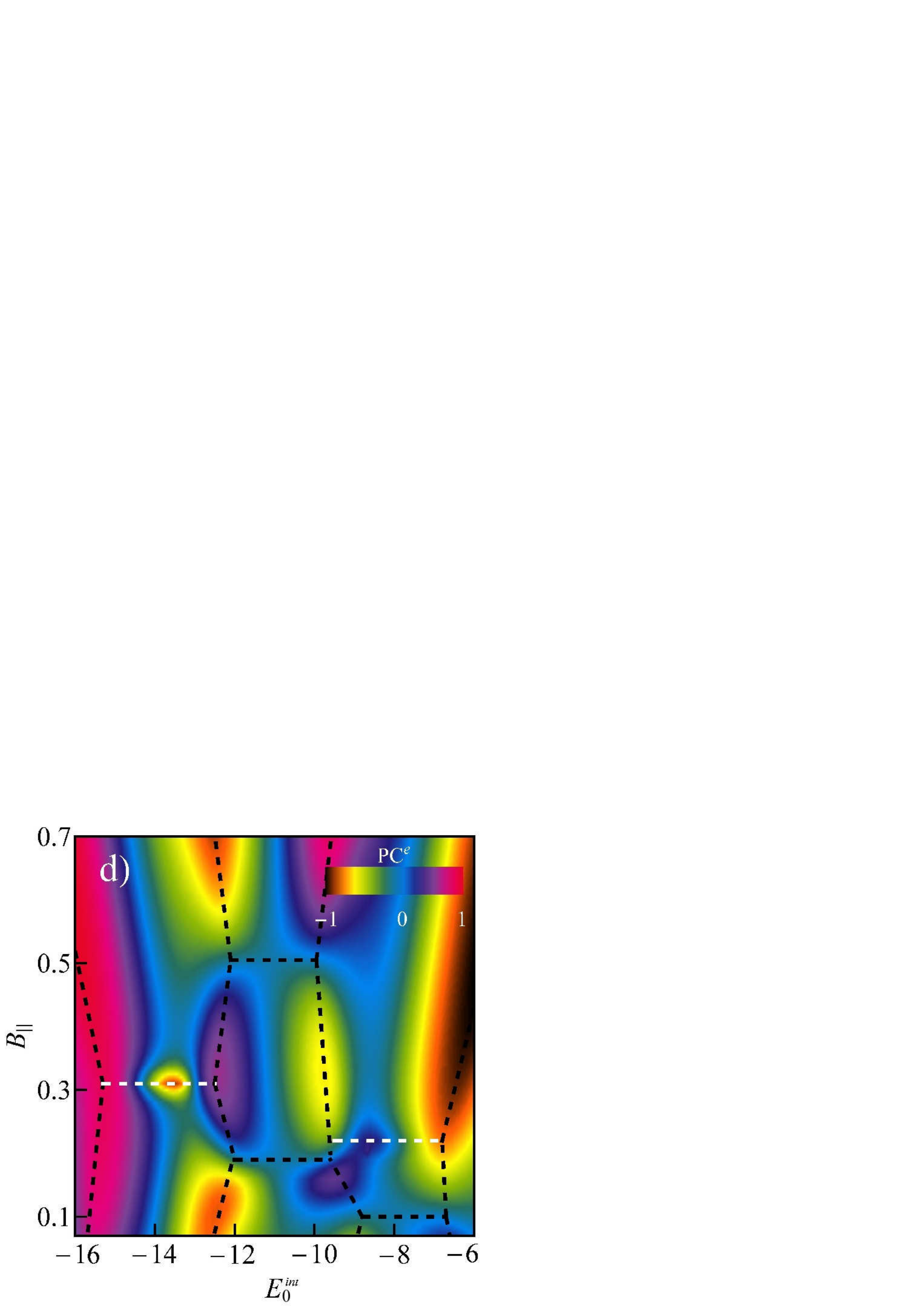}
\caption{\label{fig13} (Color online) Conductances for $\Delta= 0.1$  presented  for (a)  the centre of intershell occupation region  $N = 1$ ($E^{int}_{0} =-7.5$) and (b)  centre of intershell occupation $N = 3$ ($E^{int}_{0} =-13.5$)  for different intershell valley mixings: $\Delta^{int}_{KK'} = 0$ (solid black line), $\Delta^{int}_{KK'} = 0.005$ (dotted blue), $\Delta^{int}_{KK'} = 0.01$ (dashed red), $\Delta^{int}_{KK'} = 0.02$ (green) and  $\Delta^{int}_{KK'} = 0.03$ (short dashed magenta). Insets shows the field evolutions of single- electron energies (a) and energies  of   three- electron states of intershell manifold (b)($E^{int}_{3e}=E_{3e}-2{\cal{U}}$). Spin  polarization map of conductance for $\Delta = 0.1$ and $\Delta^{int}_{KK'} = 0$ (solid black line) , $\Delta^{int}_{KK'} = 0.005$ (dotted blue), $\Delta^{int}_{KK'} = 0.01$ (dashed red), $\Delta^{int}_{KK'} = 0.02$ (green), $\Delta^{int}_{KK'} = 0.03$ (short dashed magenta) and $\Delta^{int}_{KK'} = 0.05$ (dashed dotted gray). (c,d) Spin  polarization map of conductance for $\Delta = 0.1$, $\Delta^{int}_{KK'} = 0$ and $\Delta^{int}_{KK'} = 0.01$.}
\end{figure}
We additionally present in the insets  schematic views of the field dependencies of  pure dot electron states perturbed by valley mixing. For $N_{int}=3$ anticrossing of the ground state with one of the states  from the higher shell occurs and consequently the Kondo peak splits and conductance maximum first lowers for small $\Delta^{int}_{KK'}$ and for stronger valley mixing a dip is observed. For $N_{int}=1$ a similar anticrossing of the states from different shells  is observed with correspondingly reduced conductance ($B_{\parallel}=0.2$), but for slightly lower field  ($B_{\parallel}=0.1$) crossing of the states from the same (lower) shell occurs leading to a modified Kondo line.  The latter problem of restoring of the Kondo resonance within single shell has been already discussed in section III C. Comparing  the spin  polarizations of conductance presented on Fig. 13c for $\Delta^{int}_{KK'}=0$ and $\Delta^{int}_{KK'}\neq0$ (Fig. 13d) it is seen that the valley mixing induced  splitting of the Kondo resonance results in the change of sign of polarization  for  $N_{int} = 1,3$.  To get insight into this effect we present in the lower inset of Fig. 13 a example for transmission for $N_{int}=1$.
 ${\cal{T}}_{\downarrow}$   represents Kondo resonance, which for $\Delta_{KK'}=0$ is an unsplit peak  ( compare Fig. 11a), whereas ${\cal{T}}_{\uparrow}$  reflects excited processes involving spin up states lying higher in energy and this line is split already for vanishing valley mixing. For  $\Delta_{KK'}\neq0$ the energies of Kondo active states differ and instead of location of the Fermi level at the peak of ${\cal{T}}_{\downarrow}$  ($\Delta_{KK'}=0$), for finite mixing $E_{F}$ places close to the dip. ${\cal{T}}_{\uparrow}$  shifts towards lower energies with the increase of $\Delta_{KK'}$  and continues to split.  For sufficiently large valley mixing ${\cal{T}}_{\downarrow}$   dominates at the Fermi level and reverse of polarization results.

\section{Concluding remarks}
We have studied the interplay of different symmetry breaking perturbations on  transport through CNTQD in the range of strong correlations. Spin-orbit interaction or valley mixing breaks the  spin-orbital symmetry in carbon  nanotubes and in odd occupied valleys lifts the fourfold degeneracy of the states  leaving Kramers  double degeneracy. A crossover from highly symmetric SU(4) Kondo effect into lower symmetry SU(2)  then results.  At half filling the sixfold degeneracy is removed under these perturbations  and a quartet and two singlets appear. For strong perturbation  Kondo correlations are destroyed in this case, because the dot state of the lowest energy is  singlet.  Magnetic field breaks   time-reversal symmetry and   degeneracy is  completely lifted and consequently Kondo resonances disappear for fields exceeding  Kondo energy scale.  Magnetic field acts both on spin and orbital (valley) pseudospin, in the case of wide gap nanotubes field linearly increases or decreases the energies  depending on the orientations of magnetic moments associated with  the dot states. At some fields the degeneracy of some states might be recovered and revival of Kondo effect can occur.  Depending which  states come to degeneracy, the screened  isospin in Kondo processes is a pure spin or orbital pseudospin or a mixture of both, what  reflects in complete or only partial vanishing of spin or valley polarizations. By applying gate voltage one can change the region of the dot occupations, but for wide gap tubes it is not possible electrically recover the  degeneracy within the same occupation range.  In nearly metallic nanotubes the field evolution of state energies is not linear and then not only magnetic field but also gate voltage can recover degeneracy within a valley of a given occupation,  opening the conditions for building up of Kondo correlations. Interesting, in these narrow gap systems the gate induced reconstruction of dot states can compensate in specific conditions the changes induced by spin-orbit coupling  and revival of SU(4) Kondo effect can result at zero magnetic field.  Kondo physics in CNTQDs is even richer if one includes the impact of other perturbations e.g. local  Coulomb interaction induced  valley back scattering, important for short QDs,  or intershell effects, which  may play the role for longer dots and in high magnetic fields. In these cases  the possibility of the occurrence of SU(3) Kondo effect is foreseen by us. To get more comprehensive description of intra- and intershell many-body effects on equal footing extended basis of states from both shells have to be considered requiring a use of  more slave boson operators. This problem will be discussed elsewhere \cite{Flokow}.
Also the presented discussion of SU(3) Kondo effect is only preliminarily, the more detailed analysis is under our investigation.
Theoretical challenge which we encounter considering many-body processes in the presence of  VBS perturbation is the impact of VBS induced intradot entanglement of the dot states on the cotunneling processes in which participate also  unentangled states of electrodes. Even in the case of the full geometrical symmetry,  the mentioned  distinction  of electrode and dot states  reflects in Kondo resonance.  Zero frequency transmissions of the highly symmetric  many-body resonances (SU(N)) are identical for the degenerate states,  but as suggested by our preliminary results they slightly differ (depending on the degree of entanglement)  for finite frequencies. This differentiates this case from the situation when only  product dot and electrode states participate in many-body processes, then  all N partial  transmissions are identical. It is also worth pointing out that taking into account states from  more than two shells degeneracy  of higher order than three can occur at high fields.
Rapid progress in technology allows production  of ultraclean CNTQDs  making them the ideal objects for studying different strongly correlated regimes. The richness of many-body states occurring in these systems results from a competition of different interactions and from  different degrees of freedom involved in formation of resonances in the strongly correlated CNTQDs. This is promising for information processing, because it opens the path for magnetic, electric or mechanical manipulation not only of the spin, but also valley degree of freedom, or both of them.

\begin{acknowledgments}
This project was supported by the Polish National Science Centre from the funds awarded through the decision No. DEC-2013/10/M/ST3/00488.
\end{acknowledgments}

\def\refname{References}


\begin{thebibliography}{99}

\bibitem{Saito}
R. Saito, G. Dresselhaus and M. S. Dresselhaus,  \emph{Physical properties of carbon nanotubes} (Imperial College Press, U.K., 1998).
\bibitem{Cottet}
A. Cottet, T. Kontos, S. Sahoo, H. T. Man, M. -S. Choi, W. Belzig, C. Bruder, A. F. Morpurgo, and C. Sch\"{o}nenberger, Semicond. Sci. Technol. \textbf{21}, S78, (2006).
\bibitem{Javey}
\emph{Carbon Nanotube Electronics}, edited by A. Javey and J. Kong, Springer (2009).
\bibitem{Avouris}
P. Avouris, Phys. Tod. \textbf{62}, 34 (2009).
\bibitem{Shulaker}
M. Shulaker, G. Hills, N. Patil, H. Wei, H.-Y. Chen, H.-S. P. Wong and S. Mitra, Nature \textbf{501}, 526 (2013).
\bibitem{Laird}
A. Laird, E. Kuemmeth, Ferdinand, G. Steele, K. Grove-Rasmussen, J. Nyg{\aa}rd, K. Flensberg, L. P. Kouwenhoven, Rev. Mod. Phys. \textbf{87}, 703  (2015).
\bibitem{Hamada}
N. Hamada, S. I. Sawada and A. Oshiyama, Phys. Rev. Lett. \textbf{68}, 1579 (1992).
\bibitem{Saito2}
R. Saito, M. Fujita, and G. Dresselhaus, Appl. Phys. Lett. \textbf{60}, 2204 (1992).
\bibitem{Steele}
G. A. Steele, F. Pei, E.M. Larid, J. M. Jol, H. B. Meerwaldt and L. P. Kouvenhoven, Nat. Comm. \textbf{4}, 1573 (2013).
\bibitem{Cao}
J. Cao, Q. Wang, and H. J. Dai, Nat. Mater. \textbf{4}, 745 (2005).
\bibitem{Kuemmeth}
F. Kuemmeth, S. Ilani, D. C. Ralph, and P. L. McEuen, Nature  \textbf{452}, 448 (2008).
\bibitem{Minot}
E. D. Minot, Y. Yaish, V. Sazanova, and P. L. McEuen, Nature \textbf{428}, 536 (2004).
\bibitem{Cobden}
D. H. Cobden and J. Nyg{\aa}rd, Phys. Rev. Lett. \textbf{89}, 046803 (2002).
\bibitem{Jarillo}
P. Jarillo-Herrero, J. Kong, H. S. J. Van der Zant, C. Dekker, L. P. Kouvenhoven, and S. De Franceschi, Nature \textbf{434}, 484 (2005).
\bibitem{Makarovski}
A. Makarovski, J. Liu, and G. Finkelstein, Phys. Rev. Lett. \textbf{99}, 066801 (2007).
\bibitem{Grove}
K. Grove-Rasmussen, H. J. Jorgensen, and P. E. Lindelof, Physica E \textbf{40}, 92 (2007).
\bibitem{Wu}
F. Wu, R. Danneau, P. Queipo, E. Kauppinen, T. Tsuneta, and P. J. Hakonen, Phys. Rev. B \textbf{79},  073404 (2009).
\bibitem{Pohjola}
T. Pohjola, H. Schoeller, and G. Sch\"{o}n, Europhys. Lett. \textbf{54}, 241 (2001).
\bibitem{Borda}
L. Borda, G. Zar\'{a}nd, W. Hofstetter, B. I. Halperin, and J. von Delft, Phys. Rev. Lett. \textbf{90}, 026602 (2003).
\bibitem{Chudnovskiy}
A. I. Chudnovskiy, Europhys. Lett. \textbf{71}, 672 (2005).
\bibitem{Choi}
M.-S. Choi, R. L\'{o}pez, and R. Aguado, Phys. Rev. Lett. \textbf{95}, 067204 (2005).
\bibitem{Lipinski}
S. Lipi\'{n}ski and D. Krychowski, Phys. Stat. Sol. (b) \textbf{243}, 206, (2006); J. Alloys Compd. \textbf{423}, 215 (2006).
\bibitem{Lopez}
R. L\'{o}pez, D. Sanchez, M. Lee, M.-S. Choi, P. Simon, and K. Le Hur, Phys. Rev. B \textbf{71}, 115312 (2005).
\bibitem{Lim}
J. S. Lim, M.-S. Choi, M. Y. Choi, R. L\'{o}pez, and R. Aguado, Phys. Rev. B \textbf{74}, 205119 (2006).
\bibitem{Zarand}
G. Zar\'{a}nd, Philos. Mag. \textbf{86}, 2043 (2006).
\bibitem{Galpin}
M. R. Galpin, D. E. Ogan, and H. R. Krishnamurthy, J. Phys.:Condens. Matter \textbf{18}, 6571 (2006).
\bibitem{Sakano}
R. Sakano and N. Kawakami, Phys. Rev. B 73, 155332 (2006).
\bibitem{Mravlje}
J. Mravlje, A. Ram\v{s}ak, and T. Rejec, Phys. Rev. B \textbf{73}, 241305R (2006).
\bibitem{Le Hur}
K. Le Hur, P. Simon, and D. Loss, Phys. Rev. B \textbf{75}, 035332 (2007).
\bibitem{Busser}
C. A. B\"{u}sser and G. B. Martins, Phys. Rev. B \textbf{75}, 045406 (2007).
\bibitem{Fang}
T.-F. Fang, W. Zuo, and H.-G. Luo, Phys. Rev. Lett. \textbf{101}, 246805 (2008).
\bibitem{Anders}
F. B. Anders, D. E. Logan, M. R. Galpin, and G. Finkelstein, Phys. Rev. Lett. \textbf{100}, 086809 (2008).
\bibitem{Krychowski}
S. Lipi\'{n}ski, D. Krychowski, Phys. Rev. B \textbf{81}, 115327 (2010).
\bibitem{Makarovski2}
A. Makarovski, A. Zhukov, J. Liu, and G. Finkelstein, Phys. Rev. B \textbf{75}, 241407 (2007).
\bibitem{Keller}
A. J. Keller, S. Amasha, I. Weymann, I. G. Rau, J. A. Katine, H. Shtrikman, G. Zarand, and D. Goldhaber-Gordon, Nat. Phys. \textbf{10}, 145 (2013).
\bibitem{Filippone}
M. Filippone, C. P. Moca, G. Zar\'{a}nd, and C. Mora, Phys. Rev B \textbf{90}, 121406 (2014).
\bibitem{Schmid}
D. R. Schmid, S. Smirnov, M. Marga\'{n}ska, A. Dirnaichner, P. L. Stiller, M.  Grifoni, A. K. H\"{u}ttel, and C. Strunk, Phys. Rev. B \textbf{91}, 155435  (2015).
\bibitem{Krychowski2}
D. Krychowski and S. Lipi\'{n}ski, Phys. Rev. B \textbf{93}, 075416 (2016).
\bibitem{Jespersen}
T. S. Jespersen, K. Grove-Rasmussen, J. Paaske, K. Muraki, T. Fujisawa, J. Nyg{\aa}rd, and K. Flensberg, Nature Phys. \textbf{7}, 348 (2011).
\bibitem{Grove2}
K. Grove-Rasmussen, S. Grap, J. Paaske, K. Flensberg, S. Andergassen, V. Meden, H. I. Jorgensen, K. Muraki, and T. Fujisawa, Phys. Rev. Lett. \textbf{108}, 176802 (2012).
\bibitem{Churchill}
H. O. H. Churchill, F. Kuemmeth, J. W. Harlow, A. J. Bestwick, E. I. Rashba, K. Flensberg, C. H. Stwertka, T. Taychatanapat, S. K. Watson, and C. M. Marcus, Phys. Rev. Lett. \textbf{102}, 166802 (2009).
\bibitem{Bulaev}
D. V. Bulaev, B. Trauzettel, and D. Loss, Phys. Rev. B \textbf{77}, 235301 (2008).
\bibitem{Marcus}
K. Flensberg, and C. M. Marcus, Phys. Rev. B \textbf{81}, 195418 (2010).
\bibitem{Jhang}
S. H. Jhang, M. Marganska, Y. Skourski, D. Preusche, B. Witkamp, M. Grifoni, H. van der Zant, J. Wosnitza, and C. Strunk, Phys, Rev. B \textbf{82}, 041404(R) (2010).
\bibitem{Pei}
F, Pei, E. A. Larid, G. A. Steele, and L. Kouvenhoven, Nat. Nanotechnol. \textbf{7}, 630 (2012).
\bibitem{Cleuziou}
J. P. Cleuziou, N. V. N'Guyen, S. Florens, and W. Wernsdorfer, Phys. Rev. Lett. \textbf{111}, 136803 (2013).
\bibitem{Ando}
T. Ando, J. Phys. Soc. Jpn. \textbf{69}, 1757 (2000).
\bibitem{Huertas}
D. Huertas-Hernando, F.  Guinea, and A. Brataas, Phys. Rev. B \textbf{74}, 155426 (2006).
\bibitem{Izumida}
W. Izumida, K.  Sato, and R. Saito, J. Phys. Soc. Jpn. \textbf{78}, 074707 (2009).
\bibitem{Fang2}
T. F. Fang, W. Zuo, and H. G. Luo, Phys. Rev. Lett. \textbf{104}, 169902(E) (2010).
\bibitem{Galpin2}
M. R. Galpin, F. W. Jayatilaka, D. E. Logan, and F. B. Anders, Phys. Rev. B \textbf{81}, 075437 (2010).
\bibitem{Mantelli}
D. Mantelli, C. P. Moca, G. Zarand, and M. Grifoni, Physica E \textbf{77}, 180 (2016).
\bibitem{Carmi}
A. Carmi, Y. Oreg, and M. Berkooz, Phys. Rev. Lett. \textbf{106}, 106401 (2011).
\bibitem{Moca}
C. P.  Moca, A. Alex, J. von Delft, and G. Zar\'{a}nd, Phys. Rev. B \textbf{86}, 195128 (2012).
\bibitem{Lopez2}
R.  L\'{o}pez, T. Rejec, J. Martinek, and R. Zitko, Phys. Rev. B \textbf{87}, 035135 (2013).
\bibitem{Jeong}
J. S. Jeong, and H. W. Lee, Phys. Rev. B \textbf{80}, 075409 (2009).
\bibitem{Kotliar}
G.  Kotliar, and A. E. Ruckenstein, Phys. Rev. Lett. \textbf{57}, 1362 (1986).
\bibitem{Dong}
B. Dong, and X. L. Lei, J. Phys. Condens. Matt. 13, 9245 (2001); Phys. Rev. B \textbf{63}, 235306 (2001).
\bibitem{Bulka}
B. R. Bu{\l}ka, S. Lipi\'{n}ski, Phys. Rev \textbf{67}, 024404 (2003).
\bibitem{Trocha}
P. Trocha, Phys. Rev. B \textbf{82}, 125323 (2010).
\bibitem{Herrero2}
P. Jarillo-Herrero, S. Sapmaz, C. Dekker, L. P. Kouvenhoven, and H. S. J. Van der Zant, Nature \textbf{429}, 389 (2004).
\bibitem{Babic2}
B. Babic, T. Kontos, and C. Sch\"{o}nenberger, Phys. Rev. B \textbf{70}, 235419 (2004).
\bibitem{Kouvenhoven2}
L. P. Kouwenhoven, C. M. Marcus, P. L. McEuen, S. Tarucha, R. M. Westerwelt, and N. S. Wingreen, in \emph{Proceedings of Advanced Study Institute on Mesoscopic Electron Transport}, edited by L. Sohn, L. P. Kouvenhoven, and G. Sch\"{o}n (Kluwer, Dordrecht, 1997).
\bibitem{Vituschinsky}
P. Vitushinsky, A. A. Clerk, and K. Le Hur, Phys. Rev. Lett. \textbf{100}, 036603 (2008).
\bibitem{Mora}
C. Mora, X. Leyronas, and N. Regneult, Phys Re. Lett. \textbf{100}, 036604 (2008).
\bibitem{Hewson}
A. C. Hewson, \emph{The Kondo Problem to Heavy Fermions} (Cambridge University Press, Cambridge, U.K., 1997).
\bibitem{Kubo}
T. Kubo, Y. Tokura, T. Hatano, and S. Tarucha, Phys. Rev. B \textbf{74}, 205310 (2006).
\bibitem{Karwacki}
{\L}. Karwacki, and P. Trocha, Phys. Rev. B \textbf{94}, 085418 (2016).
\bibitem{Makarovski3}
A. Makarovski, J. Liu, and G. Finkelstein, Phys. Rev. Lett. \textbf{99}, 066801 (2007).
\bibitem{Ando2}
T. Ando, J. Phys. Soc. Jpn. \textbf{75}, 024707 (2006).
\bibitem{Wunsch}
B. Wunsch, Phys. Rev. B \textbf{79}, 235408 (2009).
\bibitem{Secchi}
A. Secchi, and M. Rontani, Phys. Rev. B \textbf{88}, 125403 (2013).
\bibitem{Pecker}
S. Pecker, F. Kuemmeth, A. Secchi, M. Rotani, D. C. Ralph, P. L. McEuen, and S. Ilani, Nat. Phys. \textbf{9}, 576 (2013).
\bibitem{Deshpande}
W. Deshpande, B. Chandra, R. Caldwell, D. S. Novikov, J. Hone, and M. Bockrath, Science \textbf{323}, 106 (2009).
\bibitem{Minot2}
E. D. Minot, Y. Yaish, V. Sazonova, J.-Y. Park, M. Brink  and P. L. McEuen, Phys. Rev. Lett. \textbf{90}, 156401 (2003).
\bibitem{Dekker}
C. Dekker, Phys. Tod. \textbf{52}, 22 (1999).
\bibitem{Reich}
S. Reich, C. Thomsen, and J. Menltzsch, \emph{Carbon Nanotubes Basic Concepts and Physical Properties} (Wiley-VCH Verlag GmbH \& Co. KGaA Weinheim, 2004).
\bibitem{Flokow}
P. Flork\'{o}w, D. Krychowski, S. Lipi\'{n}ski, (to be published).

\end{thebibliography}
\end{document}